\documentclass[prx,twocolumn,secnumarabic,balancelastpage,amsmath,amssymb,superscriptaddress]{revtex4-2}
 
\usepackage{amsfonts}
\usepackage[dvips]{graphicx}
\usepackage{mathrsfs}
\usepackage{amsthm,amssymb}
\usepackage[colorlinks, citecolor=blue, linkcolor=blue, linktocpage]{hyperref}
\usepackage{graphicx}  %
\usepackage{dcolumn}   %
\usepackage{bm}        %
\usepackage{amssymb}   %
\usepackage{floatrow}
\usepackage{mathtools} 
\usepackage{braket}
\usepackage{svg}
\usepackage[caption=false]{subfig} %
\usepackage{algorithm,algpseudocode}
\usepackage{boxedminipage}
\usepackage{varwidth} %
\usepackage{todonotes}
\usepackage{physics}
\usepackage[english]{babel}
\usepackage{blindtext}
\usepackage{makecell}
\usepackage{xcolor}
\usepackage{soul}
\usepackage [autostyle, english = american]{csquotes}
\MakeOuterQuote{"}

\usepackage{enumerate}
\usepackage{enumitem}

\newif\ifptitle
\newif\ifpnumber
\newcounter{para}

\ptitletrue  %
\pnumbertrue  %

\makeatletter 
\renewcommand{\fnum@figure}{\textbf{Figure~\thefigure}}
\makeatother

\makeatletter
\renewcommand{\thetable}{\arabic{table}} 
\renewcommand{\fnum@table}{\textbf{Table~\thetable}}
\makeatother

\usepackage{makecell}
\usepackage{array}
\newcolumntype{?}{!{\vrule width 2\arrayrulewidth}}

\makeatletter
\def\blfootnote{\xdef\@thefnmark{}\@footnotetext}
\makeatother

\newcommand{\tot}{\text{tot}}

\newcommand{\eff}{\text{eff}}

\renewcommand{\selectlanguage}[1]{}

\begin{document}

\title{A Dipolar Chiral Spin Liquid on the Breathed Kagome Lattice} 

\author{Francisco Machado}
\thanks{These authors contributed equally to this work.}
\affiliation{ITAMP, Harvard-Smithsonian Center for Astrophysics, Cambridge, MA 02138, USA}
\affiliation{Department of Physics, Harvard University, Cambridge, MA 02138, USA}
\affiliation{QuTech, Delft University of Technology, PO Box 5046, 2600 GA Delft, The Netherlands}

\author{Sabrina Chern}
\thanks{These authors contributed equally to this work.}
\affiliation{Department of Physics, Harvard University, Cambridge, MA 02138, USA}

\author{Michael P. Zaletel}
\affiliation{Department of Physics, University of California, Berkeley, CA 94720, USA}
\affiliation{Material Science Division, Lawrence Berkeley National Laboratory, Berkeley, CA 94720, USA}

\author{Norman Y. Yao}
\affiliation{Department of Physics, Harvard University, Cambridge, MA 02138, USA}
\affiliation{Harvard Quantum Initiative, Harvard University, Cambridge, MA 02138, USA}

\date{\today}

\begin{abstract}
  Continuous control over lattice geometry, when combined with long-range interactions, offers a powerful yet underexplored tool to generate highly frustrated quantum spin systems.
  By considering long-range dipolar antiferromagnetic interactions on a breathed Kagome lattice, we demonstrate how these tools can be leveraged to stabilize a chiral spin liquid. 
  We support this prediction with large-scale density-matrix renormalization group calculations and explore the surrounding phase diagram, identifying a route to adiabatic preparation via a locally varying magnetic field. 
  At the same time, we identify the relevant low-energy degrees of freedom in each unit cell, providing a complementary language to study the chiral spin liquid.
  Finally, we carefully analyze its stability and signatures in finite-sized clusters, proposing direct, experimentally viable measurements of the chiral edge mode in both Rydberg atom and ultracold polar molecule arrays.
\end{abstract}

\maketitle

The inability to simultaneously satisfy all local energetic constraints---often termed \emph{frustration}~\cite{anderson:1978}--- is responsible for a wealth of complex and delicate phenomena, ranging from protein folding and optimization problems to structural glasses and spin liquids~\cite{Mezard_2000, Onuchic_protein_folding}.
Within spin systems, frustration often arises from antiferromagnetic interactions on a non-bipartite lattice~\cite{savary:2017, Balents2010}.
The presence of quantum fluctuations provides a tunneling mechanism between different configurations, stabilizing a coherent macroscopic superposition as the true ground state~\cite{Rokhsar_Kivelson_1988}.
This defines a new zero temperature phase of matter that lacks any local order parameter, yet displays robust behavior such as long-range entanglement and anyonic excitations~\cite{savary:2017}.

While this mechanism appears straightforward, quantum spin liquids (QSLs) often inhabit a small portion of a much broader and complex phase diagram~\cite{Balents2010}.
This has motivated the investigation of a wide range of theoretical models, using beyond nearest neighbor interactions~\cite{he_chiral_2014, gong_global_2015}, complicated multi-body terms~\cite{bauer:2014, Kang_2024_DM_ints}, and different lattices~\cite{kitaev:2006}.
The identification of these models has laid out a blueprint for the search of quantum spin liquid candidate materials such as $\alpha$-RuCl$_3$, Herbertsmithite, other Kagome materials, and organic salts~\cite{Norman_2016_RMP_Herbertsmithite, Suzuki2021, li:2025a, Stahl2024}.
However, the lack of clear positive signatures has precluded the definite experimental identification of a spin liquid~\cite{Wen_2019}, motivating the search for new approaches and models for their stabilization and study.

\begin{figure}[h!]
    \centering
    \includegraphics[width=3.0in]{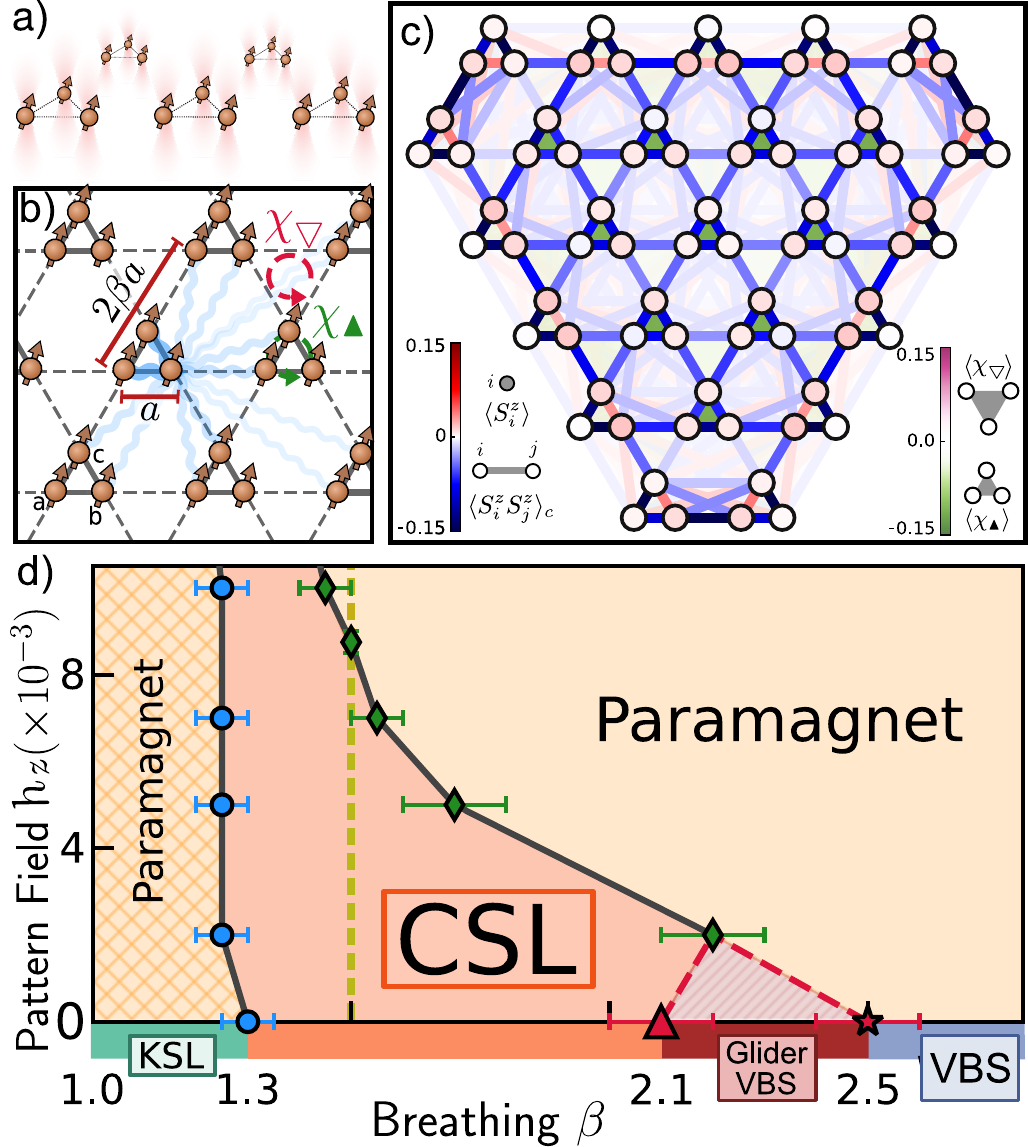}
    \caption{
    {\bf (a)} Schematic of a tweezer array platform where the position of the dipolar spins can be arbitrarily chosen. 
    {\bf (b)} Top view of the the two-dimensional breathed Kagome lattice, where the distance between the Kagome unit cells is scaled by the breathing parameter $\beta$.
    We label the small and large triangles by $\blacktriangle$ and $\bigtriangledown$, respectively.
    {\bf (c)} A cluster composed of $N=75$ spins that exhibits signatures of the chiral spin liquid (CSL):
    a lack of magnetic ordering (in single and two-site spin operators),
    time reversal symmetry breaking (i.e. non-zero chirality), 
    and additional static and dynamic responses (see Figs.~\ref{fig:Cluster_Dynamics} and \ref{fig:Cluster_Dynamics_Chiral}). 
    {\bf (d)} Phase diagram with respect to breathing $\beta$ and the strength $h_z$ of a local magnetic field pattern. 
    There is a direct, second order transition between the paramagnet and the CSL at $h_z \sim 0.008$ for $\beta = 1.5$ (yellow dashed line, details in Fig.~\ref{fig:OnsiteField}).
    }
    \label{fig:schematic}
\end{figure}

The emergence of modern atomic, molecular, and optical (AMO) experiments has provided a complementary perspective for the preparation and measurement of QSLs. 
This ``bottom-up'' approach allows for local control and measurement over individual degrees of freedom and their interactions. 
In addition, AMO experiments offer a natural platform to perform non-equilibrium dynamical experiments, a regime that is challenging in condensed matter experiments. 
These elements have already been leveraged to explore signatures of QSLs in both Rydberg tweezer arrays~\cite{semeghini:2021, Evered2025} and atoms in optical lattices~\cite{sun_engineering_2023}.

Here, we demonstrate how this approach can be further extended by focusing on how the \textit{continuous} control of geometry serves as a powerful and simple tuning parameter.
We demonstrate that two simple ingredients, continuous control over lattice geometry and long-range interactions, offer a simple route to stabilize quantum spin liquids.
More specifically, we show how a chiral spin liquid (CSL) state can emerge from dipolar interacting spins in a deformed Kagome lattice.
These ingredients naturally arise within AMO platforms, in particular dipolar Rydberg atom and molecular arrays, offering a new way to prepare and probe a CSL. 

\vspace{2mm}
After a brief introduction to the CSL phase in Sec.~\ref{sec:intro_CSL}, our work is organized across two sections.
In Sec.~\ref{sec:Model}, we introduce the central model in our work---dipolar interacting spin-1/2 particles on a \emph{breathed} Kagome lattice---and characterize the resulting CSL state.
We present a careful study of the CSL state (both TRS symmetry breaking and topological order), the phase diagram of our model as a function of breathing, and introduce an effective model utilizing the low-energy degrees of freedom within each unit-cell.

In Sec.~\ref{sec:ExperimentalConsiderations}, we discuss how the tools of AMO platforms enable us to better stabilize, prepare, and study the CSL in experiments.
We begin by studying the CSL state under different finite-sized geometries and using additional Ising interactions, highlighting how geometry control and Floquet engineering can enhance the stability of the CSL phase.
We then investigate how additional fields can be utilized to engineer an adiabatic path into the CSL state.
While global fields lead to a complex ladder of magnetically ordered phases, a local pattern enables a single, direct transition between a trivial paramagnetic and the CSL state.
Finally, we study how direct access to the system's local degrees of freedom enables the characterization of CSL state through a transverse spin response, a robust chiral edge mode, and signatures of the trapping of a semion. 
We conclude by remarking on how our protocol can be implemented in state-of-the-art experimental platforms in Rydberg atom and molecular tweezer arrays, as well as neutral atoms in optical lattices.

\section{Background}
\label{sec:intro_CSL}

The chiral spin liquid (CSL) first emerged as a candidate ground state of frustrated Heisenberg systems in a 2D triangular lattice~\cite{kalmeyer:1987}.
Its theoretical description corresponds to the bosonic analog of the fractional Quantum Hall (FQH) state at filling $\nu=1/2$~\cite{kalmeyer_laughlin_1989, wen_chiral_1989}.
As such, the CSL is characterized by gapped excitations in the bulk that exhibit anyonic statistics (i.e. semions), while the edge hosts a topologically protected gapless mode~\cite{wen_chiral_1989, wen_chiral_1990}. 
The chirality of the gapless mode is determined by either the orientation of a time-reversal symmetry (TRS) breaking field (i.e. magnetic field in the FQH picture), or the orientation of spontaneous TRS-breaking.

Since then, different theoretical studies connected the CSL to both a U(1) lattice gauge theory~\cite{he_kagome_2015,he_kagome_2016}, as well as to the fractional Chern Insulator (FCI), the lattice analog of the FQH state~\cite{yao_realizing_2013, bauer:2014}. 
While these investigations offer important theoretical frameworks for understanding this topological state, a clear path towards the observation of the CSL in a real system remains an outstanding question. 

Progress towards this goal followed the subsequent development of numerical methods.
This has enabled the study of a broader landscape of models, and provided evidence for the CSL in three distinct settings:
(i) the Kagome antiferromagnetic Heisenberg model with an explicit TRS-breaking field~\cite{bauer:2014}, 
(ii) the extended $J_1-J_2-J_3$ Kagome antiferromagnetic Heisenberg model~\cite{gong_global_2015}, 
and (iii) the triangular Hubbard model~\cite{szasz_chiral_2020}.

While these models better capture real physical systems, they either require engineering complicated spin interactions (i and ii) or a spinful itinerant electron system (iii). 
In addition, the spin liquid phase only appears for a small sliver of the theoretical phase diagram, requiring precise control over the interactions.
Recent progress towards frustrated Hubbard models in stacked 2D materials offer a promising route to observe the CSL~\cite{motruk:2023,Kuhlenkamp2024}, yet the realization of strong TRS-breaking fields or controllable next-neighbor interactions remains an important open challenge in both solid-state and AMO systems.

Recent work has suggested that models with simple two-body, long-range dipolar interactions can host spin liquid states~\cite{Yao_dipolar_2018,bintz2024DSL}. 
We build on this idea to demonstrate that these simple interactions, when coupled to a continuous control over the geometry, can stabilize highly frustrated quantum phases of matter in large swaths of the phase diagram.
This offers a straightforward path to prepare and probe the CSL state.

\section{Dipolar Spins in a breathed Kagome Lattice}

\subsection*{Theoretical Model} 
\label{sec:Model}

Throughout our work we will be considering an antiferromagnetic dipolar XY model [Fig.~\ref{fig:schematic}(a)] described by the following Hamiltonian: 
\begin{equation}
    H_{\text{XY}} = J \sum_{i,j} \frac{S^x_iS^x_j + S^y_iS^y_j}{|\bm{r}_i -\bm{r}_j|^3} 
    \label{eq:H_XY}
\end{equation}
where $S_j^\alpha$ is the $\alpha$ component of the $j$-th spin-$1/2$ located at $\bm{r}_j$, and $J>0$ is the coupling strength.
We will work in units where the closest spin-spin distance and interaction strength are both normalized to unity, $\hbar=a=J=1$.

Our starting point is the conventional Kagome lattice, which is composed of corner sharing small and large triangles.
We continuously tune this geometry using a breathing parameter $\beta \ge 1$, which controls the relative size of the $\bigtriangledown$ and $\bigtriangleup$ triangles. Specifically, if the large triangle has lattice constant $1$, the small triangle has lattice constant $2 \beta - 1$, so that the lattice constant of the full unit cell is $2 \beta$ as shown in Fig.~\ref{fig:schematic}(b).
While this perturbation does not directly change the connectivity of the lattice, the interplay between the deformed geometry and the long-range dipolar terms controls the relative strength of the spin-spin interactions.

\subsubsection*{Numerical Methods}

The main numerical method used in our investigation is the Density Matrix Renormalization Group (DMRG) class of algorithms applied to Matrix Product States (MPS).
First utilized to study ground states in strongly correlated one-dimensional systems \cite{White1992,White1993,WhiteHuse_Heisenberg_1993,Hastings_2007}, DMRG has since been extended to infinite systems and two-dimensional geometries,~\cite{mcculloch2008iDMRG, Liang1994_2dDMRG}, as well as the study of dynamics~\cite{Haegeman_2011}.

When extending MPS methods to 2D, 
the linear structure of the MPS must be ``wrapped'' along the two-dimensional plane.
Using this approach, the 2D model is directly mapped onto a longer-range 1D model at the cost of higher computational complexity.
This limits the calculations to cylinders of moderate width: in this work we focus on widths of 4, 5, and 6 unit cells. 
Unless when discussing finite clusters, we use the infinite DMRG algorithm, which assumes the MPS to be translationally invariant along the cylinder axis, enabling a direct study of the cylinder's thermodynamic limit. 

Crucially, there is a choice of boundary conditions when wrapping the MPS around the cylinder. 
Formally, the cylinder is defined by identifying $\vec{x}$ with $\vec{x} + n \vec{a_1} + m \vec{a_2}$, where $\vec{a}_1$ and $\vec{a}_2$ are the basis vectors of the breathed Kagome lattice---this defines the geometry termed YC2n-2m~\cite{yan_2011, he_signatures_2017}, where $n$ is the cylinder circumference, and $m$ the ``shear'' of the boundary conditions. 
Throughout this work, we study the geometries YC8, YC10, and YC12  (i.e.~cylindrical boundary conditions with implicit $m=0$), and YC8-2, YC10-2, and YC12-2 (i.e.~helical boundary conditions).
The latter boundary conditions provide a significant reduction in the computational cost by making the model translationally invariant along the MPS direction at the expense of explicitly breaking the mirror symmetry due to the twist of the boundary conditions.

One important consideration is the treatment of the long-range power-law tail of our dipolar interactions, which cannot be exactly represented as a Matrix Product Operator (MPO).
As such, we must implement a cutoff range for the interactions, choosing to preserve all interactions between the three spins of a unit cell and the three spins of another unit cell.
More specifically, we consider three increasing range cutoffs (R1, R2, and R3) corresponding to the interactions across nearest, next-nearest and next-next-nearest unit cells, respectively (see SM for details)~\cite{SM}.

We emphasize that for all boundary conditions and interaction cutoffs, we observe robust signatures of the CSL, providing strong numerical evidence for the existence of this phase. 
Some choices of cutoffs and boundary conditions introduce nuanced effects arising from symmetry considerations that we will discuss as they emerge in our work.

\subsection{Characterizing the Chiral Spin Liquid}
\label{sec:CSL_beta1.5}

\begin{figure*}[t]
    \includegraphics[width = 7.0in]{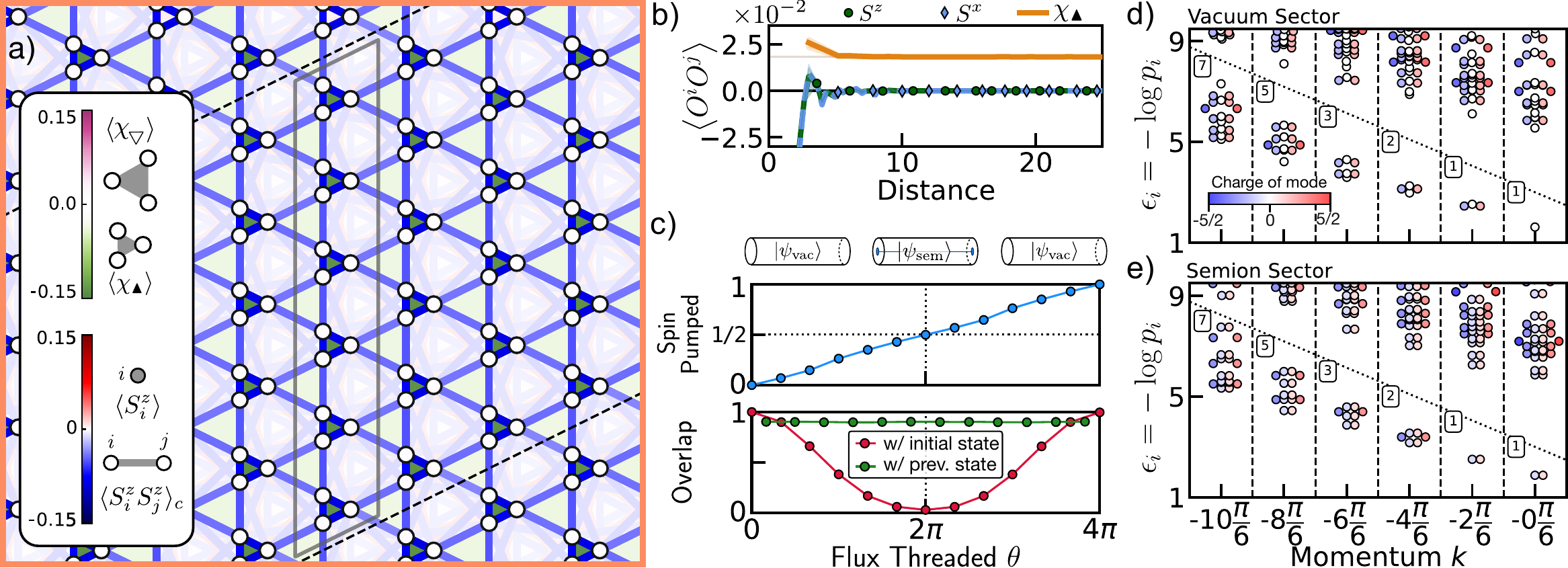}  
    \caption{
    Full characterization of the CSL state at breathing $\beta = 1.5$ on a YC12=0 cylinder with R2 interaction cutoff. 
    {\bf (a)} Local spin and chiral correlations for the iMPS ground state.
    Black rhombus corresponds to the iMPS unit cell, with dashed lines indicating the periodic boundary condition of the cylinder.
    Each circle encodes the single-site magnetization $\langle \sigma^z \rangle$, while the lines encode the spin-spin correlations $\langle \sigma_i^z \sigma_j^z \rangle$. 
    The system does not display any magnetic ordering, up to small finite-size and convergence effects ($\sim 1\% $ of correlation.) .
    The color of the small ($\blacktriangle$) and large ($\bigtriangledown$) triangles corresponds to their chiralities---the homogeneous value indicates time reversal symmetry breaking.
    {\bf (b)} Two-point correlation functions for $\sigma_z, \sigma_x, \text{ and } \chi_\blacktriangle$. 
    All spin correlations decay to zero at long distances, but the chiral-chiral correlations approach a non-zero value, indicating TRS-breaking. 
    {\bf (c)} Spin pumping along the cylinder and state overlap per cylinder ring upon the adiabatic threading of $\theta$ flux through the cylinder (top and bottom, respectively).
    At $\theta = 2 \pi$, a total of $1/2$ spin magnetization has been pumped, and the resulting ground state is orthogonal to the original state, indicating that the ground state is in a different \textit{topological} sector (top schematic).
    At $\theta=4\pi$, the final and the initial states are the same, and an integer total magnetization has been pumped.
    Adiabaticity of our protocol is ensured by the large overlap between $\ket{\psi(\theta)}$ with the previous state $\ket{\psi(\theta - \Delta \theta)}$. 
    {\bf (d)[(e)]} The entanglement spectrum of the vacuum [semion] as a function of momentum $k$ around the cylinder and the total magnetization quantum number $S_z$ (color).
    Both sectors display a low-energy chiral edge mode. 
    The number of modes of lowest $|S_z|$ states (boxes) agrees with the expected SU$_1$(2) WZW edge theory.
    }
    \label{fig:CSL} 
\end{figure*}

We begin by fully characterizing the chiral spin liquid ground state at breathing $\beta=1.5$~[Fig.~\ref{fig:CSL}].

One characteristic feature of gapped spin liquids is the lack of spin ordering all the way down to zero temperature.
This can be understood through the lens of the Hastings-Oshikawa-Lieb-Schultz-Mattis (HOLSM) theorem.
The HOLSM theorem states that a spin system with half-integer spin per unit cell, as well as translation and spin rotation symmetries, must either exhibit spontaneous symmetry breaking, topological order, or be gapless~\cite{Hastings_2004}.
As a result, the characterization of the decay of correlations in the ground state can offer strong support for the existence of underlying topological order.

Because our model, Eq.~\ref{eq:H_XY}, hosts both $U(1)$ spin-rotation symmetry (around $\hat{z}$) and a $\mathbb{Z}_2$ spin reflection symmetry, a spin liquid candidate would not display spontaneous symmetry breaking in either $S_x$ or $S_z$ spin observables.

Indeed, for both $S_x$ and $S_z$, the onsite expectation value is zero [Fig.~\ref{fig:CSL}(a)], and it exhibits neither long-range nor algebraic order~[Fig~\ref{fig:CSL}(b)]. %
We can complement this analysis by directly computing the MPS's correlation length~\cite{SM}, observing that it does not increase with increasing bond dimension: this feature highlights the presence of a short-range correlated state as expected for a gapped phase of matter.
In addition, the translation and rotation symmetries of the Kagome unit cell are also preserved; 
the pattern of two-point correlation functions (shading of the bonds) is symmetric across the entire lattice (up to small corrections of magnitude $\lesssim 2\times 10^{-2}$). 
Curiously, even though our model is highly anisotropic, the state's spin-spin correlations exhibit a near SU(2) symmetry with $\expval{S^x_iS^x_j}$ and $\expval{S^z_iS^z_j}$ differing at most by $6.5\times 10^{-3}$.
We return to this point later in our work when discussing our effective two-spin model, where the origin of this feature becomes more transparent [Sec.~\ref{sec:PhaseDiagram}].

The lack of any ordering pattern and bounded correlation length suggests that, via the HOLSM theorem, the ground state at $\beta=1.5$ is, indeed, a topological spin liquid.
To investigate the nature of the spin liquid state, we leverage the fact that the CSL is characterized by the presence of spontaneous TRS-breaking.
Formally, TRS acts as an anti-unitary operator $\Theta$ that flips the orientation of each spin: $\Theta = K e^{i \pi S_y}$, where $K$ is complex conjugation.
As such, TRS-breaking manifests itself by the long-range order of a local operator composed of an \emph{odd} number of spins, e.g. $S^z_i$ or $S^z_i S^x_j S^y_k$.

One way to diagnose TRS without being sensitive to spin rotation symmetry breaking is via the chirality observable~\cite{wen_chiral_1989}:
\begin{equation}
    \chi_{ijk} = \expval{\Vec{S}_i \cdot (\Vec{S}_j \times \Vec{S}_k)}
    \label{eqn:chirality}  
\end{equation}
for spins $i,j,k$.
While $i,j$ and $k$ can be any spins in the system, we will focus on the chirality around the small and the large triangles of our breathed Kagome lattice, which we denote by $\chi_\blacktriangle$ and $\chi_{\bigtriangledown}$ respectively~[Fig.~\ref{fig:schematic}(b)].

In Fig~\ref{fig:CSL}(b), we compute the chiral-chiral correlation function $\expval{\chi_{i,\blacktriangle}\chi_{j,\blacktriangle}}$ as we increase the distance between triangles $i$ and $j$.
This demonstrates that the ground state spontaneously breaks TRS \emph{without} breaking spin rotation symmetry.
This conclusion can also be directly deduced from the single chirality expectation values $\langle \chi_\blacktriangle\rangle $ and $\langle \chi_{\bigtriangledown} \rangle$~[green shading in Fig.~\ref{fig:CSL}(a)].
\footnote{The ground state manifold contains both chirality states and thus any superposition of the two is also a minimum energy state.
However, representing such superposition in the tensor network requires twice the bond dimension.
As a result, the iDMRG optimization converges to the symmetry broken state where the bond dimension can be better utilized to better capture many-body correlations within a specific symmetry sector.}
Noticeably, the magnitude of the chirality of the small triangles $\langle \chi_\blacktriangle\rangle $ is much larger than the chirality of the larger triangles $\langle \chi_{\bigtriangledown} \rangle$.
While expected from the lack of inversion symmetry in the breathed Kagome lattice, this feature follows from the larger interaction strength between spins in the same unit cell.

\subsubsection*{Detecting Topological Order}

While the previous considerations suggest that the ground state is a CSL, local correlations cannot alone unambiguously determine the presence of topological order~\cite{Wen_2013}. 
Such confirmation requires the computation of non-local quantities.
In the following sections we study four such probes: 
the fractionally quantized conductivity, 
the entanglement spectrum,
and the modular matrices $\mathcal{S}$ and $\mathcal{U}$.

\emph{Conductivity and ground state degeneracy}---%
The CSL is predicted to feature a quantized spin-Hall response, in which a gradient in a Zeeman field drives spin-current in the transverse direction. For numerical purposes, the spin-Hall response is best detected via Laughlin's ``flux-threading'' argument~\cite{LaughlinFlux1981}. By adiabatically threading spin-flux (via a twisted boundary condition), the amount of charge pumped through the system indicates its conductivity.

To thread flux through the system, we change the hopping term \emph{around} the cylinder periodic boundary condition to impart a phase of $\theta$---those hoppings are modified to $S_i^+S_j^-e^{i \theta} + h.c.$~\cite{SM} (see SM for diagram). 
To adiabatically introduce phase, we start from the original ground state at $\theta=0$, increase the phase $\theta \to \theta + \delta\theta$, and then reoptimize the ground state of the system.
This is performed until $\theta = 2\pi$ and the Hamiltonian returns back to itself (Eq.~\ref{eq:H_XY}). 

The spin pumped \emph{along} the cylinder can be computed using the charge resolved spectrum of the MPS state~\cite{zaletel_flux_2014}, with the conductivity being equal to the total spin pumped.
In Fig.~\ref{fig:CSL}(c), we observe that there is exactly half a spin pumped across the system when a $2 \pi$ flux is threaded, in agreement with our expectation for a conductivity of $\sigma = 1/2$ in the bosonic $\nu=1/2$ fractional quantum Hall~\cite{kalmeyer:1987}.

The presence of a fractionalized conductivity immediately indicates the presence of a degenerate ground state manifold~\cite{WenNiu1990}.
This can be verified by computing the overlap between the states for $\theta =0$ and $\theta=2\pi$: although the corresponding Hamiltonians are the same, the computed ground states are orthogonal, as measured by the local overlap between the two states~\footnote{Since we are considering an infinite cylinder geometry, we consider the overlap of the states \emph{per unit length} rather than the total overlap between the states, which would immediately be zero unless the two states are exactly proportional. 
To this end, the fact that the local overlap is almost zero, highlights that the states are already distinct within a single unit-cell of the iMPS.
}. 
At $\theta=4\pi$, the overlap returns to $1$, highlighting that there is, at least, a two-fold degeneracy of the ground state manifold (within this particular TRS-breaking sector).

The physical picture for the threading procedure can be understood in terms of the generation and movement of a \emph{semion} through the system.
Starting from the vacuum sector, threading $2\pi$ of magnetic flux moves one semion of a pair from $x\to-\infty$ to $x\to +\infty$ [Fig.~\ref{fig:CSL}(c)]. 
The resulting wavefunction holds no semions (i.e. remains in the ground state), but has a different anyonic flux through the cylinder---this is the semion sector ground state wavefunction.
When an additional $2\pi$ flux is threaded through the cylinder, the other semion is moved, and the anyonic flux is now zero; we return to the vacuum sector.

We conclude by noting that, due to the finite width of the cylinders considered, the topological sectors are not \emph{exactly} degenerate, but rather exhibit a small splitting, which is expected to decay exponentially with increasing width of the cylinder.
Indeed, upon increasing the cylinder width from $4$ to $6$ unit cells, the energy splitting between the two sectors decreases by a factor of approximately 15.
For YC12-0 geometry, this splitting amounts to a relative energy difference of $7\times 10^{-6}$ per site.

\emph{Determining the nature of the chiral edge theory}---%
One way of removing some of the sector ambiguity is by considering the entanglement spectrum of the system.
Besides providing a direct identification of the ground state sector, it captures the spectrum of the edge mode~\cite{Kitaev_2006, LiHaldane2008,qi_general_2012}. 
This enables us to utilize our iMPS to directly characterize the edge theory that would occur in a finite-sized boundary and extract the features of a \emph{gapless} edge spectrum from a \emph{gapped} bulk wavefunction.

More specifically, it has been noted that the low-lying spectrum of the entanglement Hamiltonian $H_{\text{ent}}$ (defined in terms of the half-system reduced density matrix $\rho_A = \text{Tr}_{\bar{A}} \ketbra{\psi}{\psi} = e^{-H_{\text{ent}}}$) is proportional to the low-lying spectrum of the edge theory~\cite{qi_general_2012}.
While in general hard to compute, the iMPS representation of the wavefunction enables us to calculate the spectrum~\cite{zaletel_topological_2013, cincio:2013}.

To properly analyze the resulting spectrum, we must take into account the presence of additional symmetries such as translational invariance around the cylinder and total $U(1)$ spin conservation.
First, in both entanglement spectra (associated with the $\theta=0$ and $\theta=2\pi$ wavefunctions), the approximate linear dispersion relation of the low-lying levels signifies the presence of a gapless chiral edge mode, as expected for the $\nu = 1/2$ FQH state.
Note that, depending on the direction of the TRS-breaking, we observe either left or right moving spectra---the symmetry breaking direction dictates the chirality of the edge mode.
Second, the degeneracy pattern of levels in the $S_z^{tot} = 0$ (white markers) manifold follows the expectation of a chiral bosonic mode $\{1,1,2,3,5,7,\ldots\}$ with total angular momentum $k_y$.
This identifies the edge theory as being an SU$_1$(2) Wess-Zumino-Witten conformal field theory~\cite{wen:1991}.  

While the XY interactions preserve only $S^z$, the entanglement spectrum nevertheless exhibits an approximate SU(2) symmetry evidenced by the (near) degeneracy of SU(2)-multiplets across $S_z^{tot}$ [different colors in Figs.~\ref{fig:CSL}(d,e)].
This splitting of the multiplets is observed to decrease with cylinder circumference \cite{SM}.

Furthermore, the entanglement spectrum enables us to unambiguously distinguish the vacuum and semion sectors by the degeneracy of the lowest energy modes. 
In the vacuum sector, the lowest energy mode at $k_y=0$ is non-degenerate and occurs in the $S_z^{tot} = 0$ sector. 
By contrast, the lowest energy modes in the semion sector are doubly degenerate at $S_z^{tot} = \pm 1/2$.
This serves as signature of the spin-1/2 moment carried by the anyonic spinons.

\emph{Characterizing anyonic statistics}---%
To this end, we compute the action of the modular transformations in the ground state manifold; these transformations directly connect the overlap of the ground states (under a suitable geometric transformation) to properties of the anyonic excitations~\cite{zaletel_topological_2013, zhang:2012, cincio:2013}.
The breathed Kagome lattice preserves the $\mathrm{C}_{3v}$ symmetry of the lattice, so the natural transformations to consider are $2\pi/3$ rotations, $\mathcal{R}_{2\pi/3}$, which can be cast as the product of the two generators of the modular transformations, $\mathcal{S}$ and $\mathcal{U}$~\cite{SM}:
\begin{equation}
  \mathcal{R}_{2\pi/3} =  \mathcal{U} \mathcal{S} \quad \Rightarrow \quad  \langle i | \mathcal{R}_{2\pi/3}|j \rangle = \left[ D^\dag U S D \right]_{ij} ~,
\end{equation}
where $i,j\in\{1,s\}$ label the two different degenerate ground states corresponding to the vacuum and semion sectors, $U$ and $S$ are the matrices encoding the action of $\mathcal{U}$ and $\mathcal{S}$ on the ground state manifold, and $D$ carries any additional, unphysical phase difference between the two ground states:
\begin{equation}
  U = e^{-i \frac{2 \pi}{24} c} \mqty(\theta_1 & 0 \\ 0 & \theta_s),\quad
  S = \mqty(S_{11} & S_{1s} \\ S_{s1} & S_{ss})~.
\end{equation}
The anyonic information is then included in the self-statistics of the quasi-particles $\theta_i$ and the mutual statistics and quantum dimension $S_{ij}$. 

To compute $U$ and $S$, we first build the wavefunction of the CSL on a $L\times L$ torus by concatenating our iMPS ground state with appropriate boundary conditions~\cite{zaletel_flux_2014}.
The subsequent calculation of the wavefunction overlap is, in general, exponentially hard; the physical rotation changes the relative orientation of the two MPS states, preventing their simple contraction.
We use a Monte Carlo procedure for computing the overlaps in a $6\times 6$ torus at bond dimension $4098$ to obtain \cite{cincio:2013, SM}:
\begin{align}
  &US = e^{-i\frac{2\pi}{24}c} \begin{bmatrix} \theta_1 S_{11} & \theta_1 S_{1s}\\ \theta_s S_{s1} & \theta_s S_{ss} \end{bmatrix}
  =\\
  & \frac{e^{-i\frac{2\pi}{24}(-1 {\scriptscriptstyle +0.37})}}{\sqrt{2}}  \begin{bmatrix}
    1 {\scriptstyle -0.006} & (1 {\scriptstyle +0.004})  {\scriptstyle e^{-i0.009}}\\
       -i (1 {\scriptstyle +0.039}) {\scriptstyle e^{i0.020}} & -i(-1 {\scriptstyle-0.002}) {\scriptstyle e^{-i0.002}}
    \end{bmatrix}\notag
\end{align}
While this analysis accurately captures the anyonic self-statistics and the quantum dimension of the theory (on the percent level), the central charge is significantly different from the expected value: $-0.63$ instead of $-1$ \footnote{The sign of both $\theta_s$ and $c$ are determined by the direction of the TRS spontaneous symmetry breaking. By considering the other symmetry sector, the symmetric values would be obtained.}.
Crucially, we find that while the other quantities converge faster with respect to the bond dimension of the wave functions, the central charge continues to exhibit a large flow (see SM for details~\cite{SM}).
In order to account for this finite size effect, as well as provide a better estimate of the system's central charge, we compute $[US]_{11}$ to larger bond dimension ($\chi=5120$) and perform a simple inverse bond dimensional scaling \cite{SM}.
Using this analysis, we extract a central charge of $c =  -1.07 \pm 0.08$, in much better agreement with the expected value for the CSL state.

\begin{figure*}[t]

    \includegraphics[width =\textwidth]{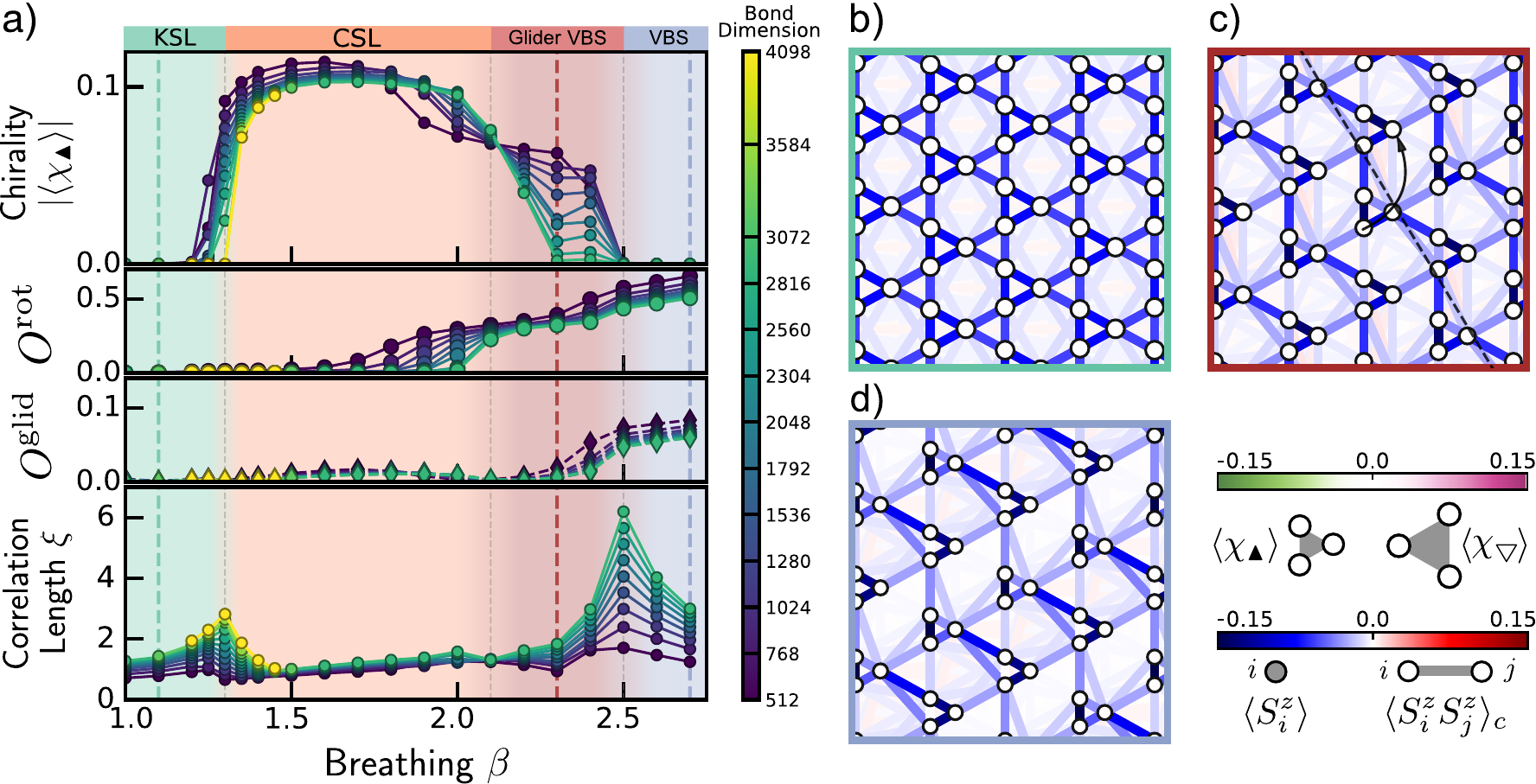}
    \caption{{\bf (a)} Phase diagram with respect to the breathing parameter $\beta$ as measured by the (1) chirality $\langle \chi_\blacktriangle \rangle$, (2) rotation and glider symmetry breaking order parameter $O^{\text{rot}}$ and $O^{\text{glid}}$, and (3) correlation length $\xi$. 
    The nature of the transition is determined by the behavior of the order parameter $\chi_\blacktriangle$ and $\xi$. 
    The transition from the KSL into the CSL at $\beta=1.3$ appears second order owing to the continuous behavior of the chirality and the associated diverging correlation length across the transition. 
    These considerations also apply to the valence-bond solid (VBS) phases at the Glider-VBS to VBS transition at $\beta=2.5$. 
    The transition from the CSL into the Glider-VBS pattern appears first order due to the discontinuous change of the chirality and the lack of a diverging correlation length.
    {\bf (b)} Correlations for the different observed ground states at the $\beta$ values indicated by the colored vertical lines: {\bf (b)} the Kagome Spin Liquid (KSL), {\bf (c)}  Glider-VBS, and {\bf (d)} VBS phases.
    The KSL exhibits homogeneous spin-spin correlations, without breaking any spin rotation or lattice translation.
    In the Glider-VBS phase, the correlations no longer obey translation symmetry, but instead obey a \textit{glider} symmetry composed of a translation and a mirror transformation (dashed line). 
    In the VBS state, this glider symmetry is also broken.
    }
    \label{fig:CSL_breathing}
\end{figure*}

We conclude by remarking on an intriguing feature of the present calculation: in conducting the above analysis for $L_y=6$ we find that to obtain the correct modular transformations, the vacuum and semionic ground states must be identified opposite to their entanglement spectrum expectation. More precisely, the state with double degeneracy in the lowest entanglement spectrum is identified as the vacuum state, where the single degenerate state is identified as the semionic state.
This might arise from the underlying crystalline symmetry structure of the state, which leads to the rotation $\mathcal{R}_{2\pi/3}$ acting non-trivially.

\subsection{Breathing $\beta$ phase diagram} 
\label{sec:PhaseDiagram}

We extend our analysis to the full range of the breathing parameter, from the Kagome point at $\beta=1$ to the limit of weakly interacting triangles at very large $\beta\gtrsim 3$.

We begin by analyzing the phase diagram near the $\beta=1.5$ point studied in the previous section.
Across the region $\beta \in [1.3 ,2.1]$, we observe a robust spontaneous breaking of TRS, suggesting the presence of the CSL in a large extended region.
Concomitantly, we also observe the topological features of the CSL, notably the presence of chiral modes in the entanglement spectrum~\cite{SM}.

As we approach the Kagome point $\beta \le 1.3 = \beta^c_{KSL-CSL}$, TRS is restored once again and the chirality operator no longer exhibits long-range order~[Fig.~\ref{fig:CSL_breathing}(a)]
Curiously, much like in the CSL state, the spin-spin correlations are featureless and exhibit no symmetry breaking pattern (in either spin or real space), suggesting the presence of a \emph{distinct} spin liquid state [Fig.~\ref{fig:CSL_breathing}(b)].
This observation is consistent with previous literature reporting a gapless \emph{Kagome spin liquid} (KSL) in the Heisenberg antiferromagnet in the Kagome lattice \cite{he_signatures_2017}, which is not destabilized by the \emph{long-range} XY nature of our interactions~\cite{bintz2024DSL}.

Zooming in on the transition between the CSL and the KSL, our numerical exploration indicates the presence of a second-order phase transition.
This is shown by two observations. 
First, around the transition $\beta\approx \beta^c_{\text{KSL-CSL}}$ we observe a divergence of the correlation length that is limited by the bond dimension of the iMPS wave function~[Fig.~\ref{fig:CSL_breathing}(a)], consistent with a continuous phase transition.
This also agrees with previous studies, which observed a second-order transition and whose understanding arises from the opening of the KSL spinon Dirac points~\cite{he_signatures_2017, bintz2024DSL}.
Second, the expectation value of the chirality smoothly interpolates from zero to its maximal value by $\beta \approx 1.5$.

On the other side of the CSL phase in the large breathing limit, we find evidence for two distinct valence-bond solid (VBS) phases which preserve spin-rotation symmetry. When $2.1 \lesssim \beta \lesssim 2.5$, we find that the system breaks rotation and translation symmetries---captured by the non-zero value of $O^{\text{rot}}$ (see SM for details \cite{SM})---while preserving a glide transformation [combination of translation and mirror transformation whose axis is the dashed line in Fig.~\ref{fig:CSL_breathing}(c)]-- we term this phase the Glider-VBS. We also observe that the chirality decays to zero with increasing bond dimension, suggesting that the true ground state in this regime respects TRS.

In the second phase for $\beta \gtrsim 2.5$, the gliding symmetry is also broken.
We introduce $O^{\text{glid}}$ (see SM for details~\cite{SM})) as an order parameter to distinguish these two phases, and observe it taking a non-zero value only in the VBS phase.

Given the interplay of the cylinder boundary conditions, which can both weakly break certain symmetries and frustrate certain orders, a precise determination of the nature of the transitions is difficult. 
In Fig.~\ref{fig:CSL_breathing}, we focus on one particular geometry (YC8) and range of interactions (R1), and leave further discussion to the SM.
The CSL-Glider VBS transition appears to be discontinuous and first-order, as expected for a non fine-tuned transition across which one symmetry is restored (TRS) and another is broken (translation).
At the same time, the transition into the VBS state appears to be continuous and second-order, consistent with only a single new symmetry being broken.

While the large breathing limit offers an intriguing and highly frustrated regime that warrants further study, we emphasize that the observation of the CSL at intermediate breathings is robust for all geometries (width and boundary conditions) and range of interactions considered.
For additional details, we refer the reader to the Supplementary Materials~\cite{SM}.

\subsection{Building an effective spin-chirality model}
\label{sec:2SpinModel}
The breathing parameter $\beta$ also provides a new perspective into the nature of the chiral spin liquid state.
This perspective arises from the identification of the relevant degrees of freedom in the large breathing limit.
For $\beta \to \infty$, the dominant energy scale $\approx (J/a^3)$ arises from the interactions between spins in the same triangle, while the long-range inter-triangle interaction is weak $\approx (J/a^3) \beta^{-3}$.
In this limit, %
one can build an effective description of the system in terms of the low-energy subspace of each triangle.
Surprisingly, this approach is able to capture the full phase diagram of the model, even at small $\beta$, suggesting its validity lies beyond a perturbative analysis around $\beta \to \infty$.
At the same time, these effective degrees of freedom offer a new language with which to interpret the emergence of TRS-breaking, and offer a path towards understanding the mechanisms generate the spin liquid state.

We start by calculating the low-energy manifold of a single triangle where the Hamiltonian can be written in terms of the total spin operator of the three constituent spins, $\vec{S}_{\text{\text{tot}}} = \sum_{i\in \blacktriangle} \vec{S}^i$:
\begin{align}
  \frac{H_\blacktriangle}{(J/a)^3} &= \sum_{i,j\in \blacktriangle}\hat{S}^x_i\hat{S}^x_j + \hat{S}^y_i\hat{S}^y_j = (\vec{S}_{\tot})^2 - (S^z_\tot)^2 - \frac{3}{2}, \notag
\end{align}

The sole dependence on the total spin of the triangle reflects the permutation symmetry of $H_\blacktriangle$, while the antiferromagnetic interactions favor states with low total spin.
Owing to the rules of spin addition, $\frac{1}{2}\otimes \frac{1}{2}\otimes \frac{1}{2} = \frac{1}{2} \oplus \frac{1}{2} \oplus \frac{3}{2}$, 
where the two spin-$1/2$ degrees of freedom make up a four-fold degenerate ground state manifold.

\begin{figure*}[t]
    \includegraphics[width = \textwidth]{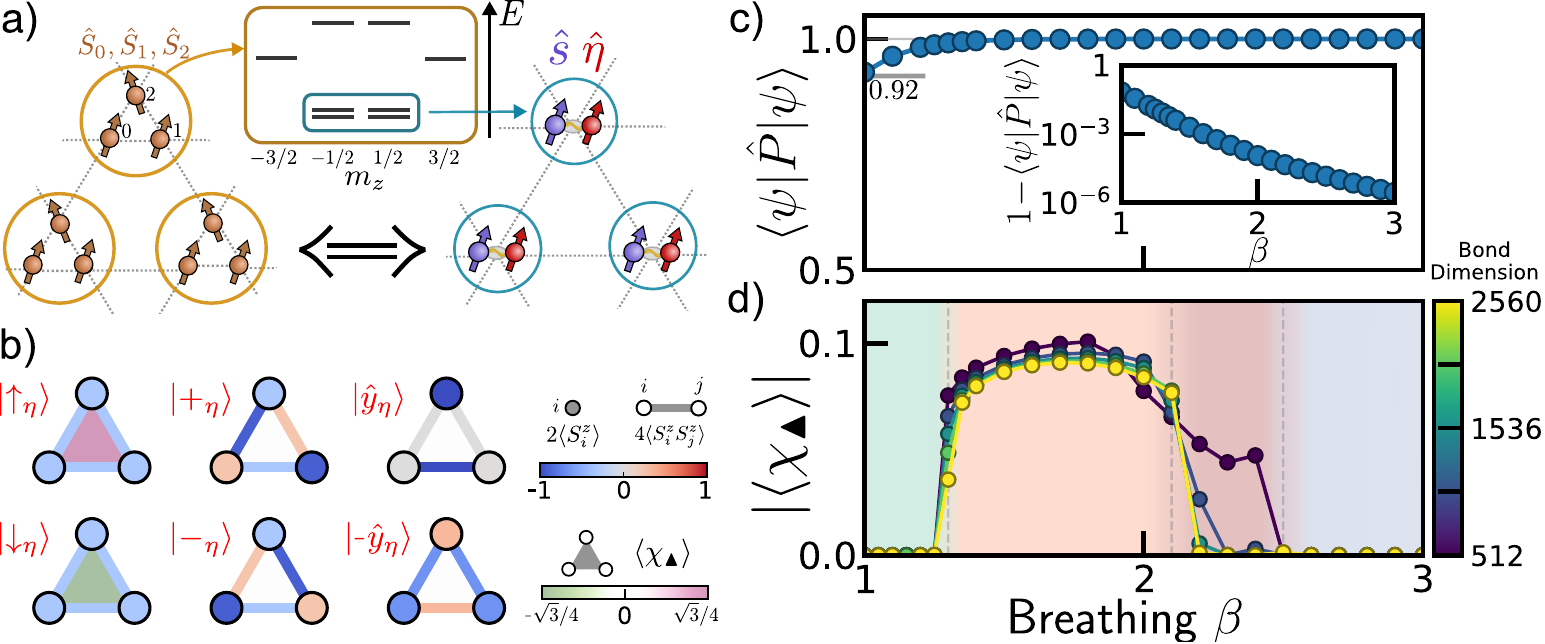} 
    \caption{
    {\bf (a)} Schematic of the basis we consider for our $\beta \to \infty$ perturbative effective model.
    For each triangle composed of 3 spins, we project into the 4 lowest-lying, degenerate states, which can be described by 2 spin-$1/2$ d.o.f.s, $\hat{s}$ and $\hat{\eta}$.
    $\hat{s}$ encodes the spin rotation properties of this manifold, while $\hat{\eta}$ encodes the chirality and the planar distribution of correlations in the triangle.
    In our effective model, each pair of spins lives on the vertex of a triangular lattice.
    {\bf (b)} Spin correlations for different $\vec{\eta}$ states (with total spin $m_s=-1/2$) in the low-energy manifold.
    {\bf (c)} Projection of the full wavefunction into this low-energy subspace. Even away from the $\beta \rightarrow \infty$ limit, the overlap of the wavefunction with this subspace remains high ($\gtrsim  92 \%$ at the Kagome point $\beta=1$).  
    {\bf (d)} Using our second-order perturbative effective Hamiltonian $H_{\eff}$, the phase diagram (with respect to $\beta$) of our model closely matches that of the full DMRG calculation for YC8 cylinders with R1 cutoff. 
    This underscores that the $\hat{s}$ and $\eta$ spins are the relevant degrees of freedom that inform the physics across all values of $\beta$.
    }
    \label{fig:2Spin}  
\end{figure*}

We can write the two distinct spin-1/2 degrees of freedom as a spin $\vec{s}_\blacktriangle$ and a chirality $\vec{\eta}_\blacktriangle$~\cite{repellin:2017}.
Whereas $\vec{s}_\blacktriangle$ encodes the total spin $\vec{S}_{\text{tot}}$ in this manifold; $\vec{\eta}_\blacktriangle$ encodes the spatial spin-correlations within each triangle~[Fig.~\ref{fig:2Spin}(a)].
More specifically, $\hat{\eta}_\blacktriangle^z$ measures the chirality of the three spins in the triangle, while $\hat{\eta}_\blacktriangle^{x/y}$ measures the spatial imbalance of correlations [Fig.~\ref{fig:2Spin}(b)]:
\begin{align}
  \eta^z_\blacktriangle &=  \frac{4}{\sqrt{3}} \chi_\blacktriangle,\\
  \eta^x_\blacktriangle &= \frac{2}{\sqrt{3}} \left[\vec{S}_b\cdot \vec{S}_c - \vec{S}_a \cdot \vec{S}_c \right],\\
  \eta^y_\blacktriangle &= \frac{2}{3} \left[ \vec{S}_a\cdot \vec{S}_c + \vec{S}_b\cdot \vec{S}_c - 2 \vec{S}_a \cdot \vec{S}_b  \right],
\end{align}
where the right-hand side is interpreted before the projection into the appropriate ground state manifold~\cite{SM}.

One might worry that these effective degrees of freedom are only relevant for very large $\beta$ where the system spontaneously breaks translation symmetry.
This is not the case: for all $\beta > 1$, the ground state's population in this subspace (per triangle) is $\gtrsim 92\%$ \emph{even at the Kagome point $\beta=1.0$}~[Fig.~\ref{fig:2Spin}(b)].
This suggests that these are the relevant low-energy degrees of freedom for describing the underlying phase diagram and phase transitions.

Reframing of the original spin-1/2 system into $\vec{s}$ and $\vec{\eta}$ degrees of freedom offers two key advantages: (i) it provides a more effective basis that improves the convergence of iDMRG calculations, and (ii) it offers a new language to probe the mechanism that stabilizes the CSL.

To directly compute the correlated ground state in the $\vec{s}$ and $\vec{\eta}$ basis, we must first calculate their effective interactions, $H_{\eff}$.
We start by recognizing which terms are not allowed by symmetry: the combination of U(1) spin conservation and $\mathbb{Z}_2$ spin flip symmetry preclude any single spin $s^\alpha_\blacktriangle$ terms, while time reversal symmetry also precludes any individual $\eta^z_\blacktriangle$~\cite{SM}.
More broadly, because time reversal flips both $\vec{s}$ and $\eta^z$, any interactions must have an even number of such terms.
By contrast, single- and two-body $\eta^{x/y}_\blacktriangle$ terms are not constrained by the onsite symmetries, but rather the lattice symmetries.
Interactions between different pairs of sites are related by the rotation and mirror symmetries that define the $C_{3v}$ symmetry group~\cite{SM}.

To zero-th order, $H_{\text{eff}}$ is simply given by the projection $\hat{P}$ of the full system Hamiltonian to the low-energy subspace, $H^{(0)}_{\eff} = \hat{P} H \hat{P}$.
Crucially, we find that $H^{(0)}_\eff$ is insufficient to capture the phase diagram of our model as it contains only a small subset of all possible terms. 
Higher-order corrections arise from virtual processes that are mediated by the eliminated states; using a Schrieffer–Wolff transformation~\cite{SM} we obtain an order-by-order perturbative effective Hamiltonian $H_\eff = H^{(0)}_{\eff} + H^{(1)}_{\eff}+\ldots$:
\begin{align*}
  \label{eq:Heff}
  H_{\eff} = \sum_{i<j}&\bigg[ (s^x_{\blacktriangle_i} s^x_{\blacktriangle_j} + s^y_{\blacktriangle_i} s^y_{\blacktriangle_j}) f_{1}(\vec{\eta}_{\blacktriangle_i}, \vec{\eta}_{\blacktriangle_j}) \\
  &+ s^z_{\blacktriangle_i} s^z_{\blacktriangle_j} f_{2}(\vec{\eta}_{\blacktriangle_i}, \vec{\eta}_{\blacktriangle_j}) + f_3(\vec{\eta}_{\blacktriangle_i}, \vec{\eta}_{\blacktriangle_j})\bigg],
\end{align*}
where $f_\alpha(\vec{\eta}_{\blacktriangle_i}, \vec{\eta}_{\blacktriangle_j})$ are different functions of the chirality interaction terms between triangles~\cite{SM}.

Defining a new two-spin model that includes terms through $H_{\eff}^{(2)}$ and performing iDMRG calculations on this effective model [Fig.~\ref{fig:2Spin}(d)], we find that the resulting phase diagram exhibits good quantitative agreement with the original system, demonstrating the predictive power of this perturbative approach across all values of $\beta$.
Notably, this approach allows us to capture the first-order phase transition into the Glider VBS with smaller bond dimension; local updates on the iDMRG optimization affect tensors corresponding to six spins $\vec{S}$, helping to establish the Glider VBS pattern of correlations more effectively.

More broadly, the separation of the local Hilbert space between the spin $\vec{s}_\blacktriangle$ and the chirality $\vec{\eta}_\blacktriangle$ degrees of freedom offers a simple, yet powerful picture for understanding the spontaneous time-reversal symmetry breaking.
Three observations are important.
First, time-reversal symmetry breaking can be simply recast as the formation of a long-range ordered Ising state in $\eta^z$.
Second, the nature of the long-range order in $\chi_\blacktriangle$ and $\chi_\bigtriangledown$ is very distinct.
For $\chi_\blacktriangle$, it arises from ordering of the local chiral $\eta^z$ d.o.f.s, while for $\chi_\bigtriangledown$, 
it arises from ordering of complex three-body spin correlations across different triangles $\chi_\bigtriangledown\sim \vec{s}_\blacktriangle \cdot (\vec{s}_{\blacktriangle'} \times \vec{s}_{\blacktriangle''})$.
Third, from symmetry, $H_{\eff}$ can contain a term proportional to  $\vec{s}_\blacktriangle \cdot (\vec{s}_{\blacktriangle'} \times \vec{s}_{\blacktriangle''})$.
While this term is time reversal symmetric, when $\eta^z$ become ordered, this term induces a simple spin chirality term in the Hamiltonian---this term is known to stabilize the CSL in frustrated magnets~\cite{bauer:2014}.
Indeed, by extending our Schrieffer-Wolff approach to three unit-cells, we observe this term in $H_\eff$.

However, the mechanism driving the CSL state appears to be more robust than expected.
While we observe the Ising order of $\eta^z$, we find that the effective Hamiltonian \emph{does not} ferromagnetically couple $\eta^z\eta^z$, but rather anti-ferromagnetically couples them.
This remains true even when considering terms mediated by spin-spin correlations (e.g. $(s^x_{\blacktriangle}s^x_{\blacktriangle'} +s^y_{\blacktriangle}s^y_{\blacktriangle'}  )\eta^z_{\blacktriangle} \eta^z_{\blacktriangle'}$).
Even more surprisingly, while the $\vec{s}_\blacktriangle \cdot (\vec{s}_{\blacktriangle'} \times \vec{s}_{\blacktriangle''})$ term offers a natural mechanism for the CSL state, we find that the CSL remains robust even when this term is not included.
Indeed, the iDMRG calculations of Fig.~\ref{fig:2Spin} only take into account two-triangle interactions, and thus, do not include this three-triangle interaction.
These two facts highlight how the CSL state is much more robust than expected and driven by a more nuanced mechanism.
Overall, we hope that the $\vec{s}$ and $\vec{\eta}$ degrees of freedom provide a novel perspective and help to better study its origin.

\section{Experimental Considerations}
\label{sec:ExperimentalConsiderations}

In this section we explore how the ingredients of modern AMO platforms---namely the ability to design the system's interaction, geometry, and apply generic fields---can be leveraged to better stabilize, prepare and probe the CSL.

\subsection{Stability}

\begin{figure}[t]
\centering
    \includegraphics[width = 3.2in]{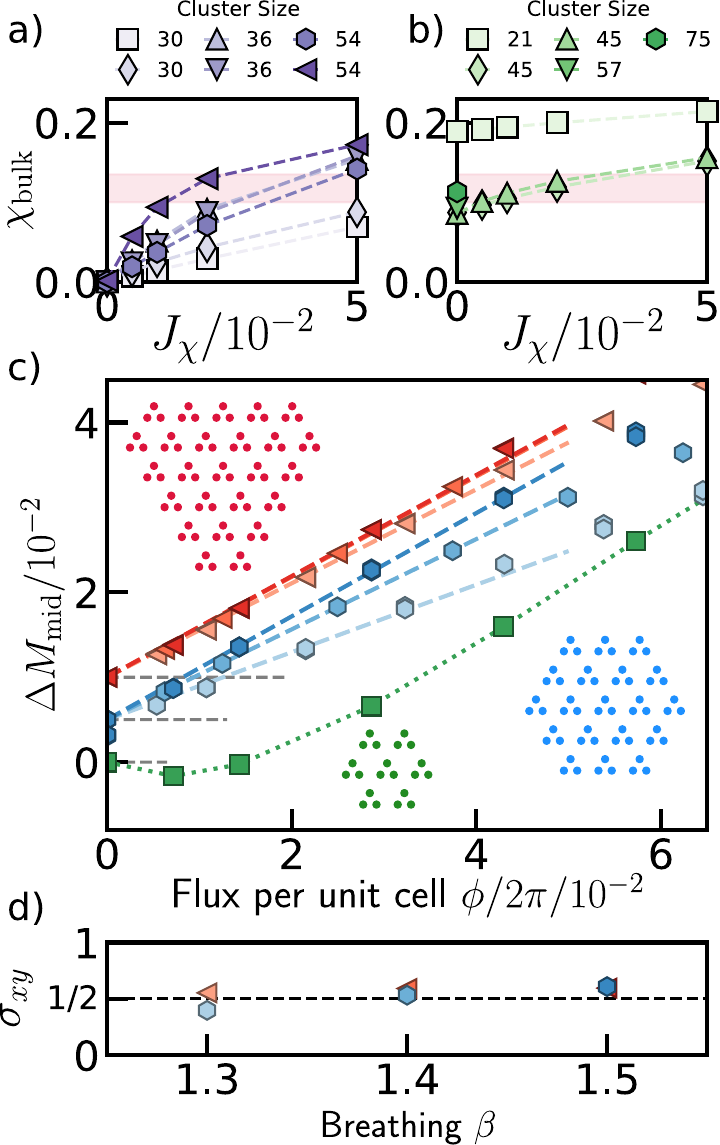}
    \caption{{\bf (a[b])} Average chirality of cluster as a function of a symmetry breaking field $J_\chi$ for different even[odd] cluster sizes.
      The even clusters display large susceptibility, while the odd clusters display symmetry breaking at $J_\chi=0$.
      {\bf (c)} When applying a flux to the system (see~Eq.~\ref{eq:Streada}), the magnetization in the middle unit-cell of the cluster increases linearly for medium sized clusters (red and blue) and for different breathings (different brightnesses).
      By contrast, the small 21 spin cluster (green) exhibits TRS-breaking but does not have the correct response.
      {\bf (d)} Extracted transverse conductivity $\sigma_{\text{xy}}$ from the linear growth in {\bf (c)} agrees with expected fractional response.
    }
    \label{fig:DiffClusters}
\end{figure}

So far, we have only studied the stability of the CSL state in an infinite geometry, using dipolar interactions.
To connect with a suitable experimental realization, we need to consider the stability of the phase in \emph{finite-sized} clusters.
On the one hand, such investigation provides a direct measure of the extent of the experiment required to see this topological state [Fig.~\ref{fig:DiffClusters}].
On the other hand, it offers the opportunity to directly explore the edge mode via spatial correlations rather than through the entanglement spectrum of in the infinite cylinder geometry. 

While finite system sizes can hurt, engineering different interactions using Floquet engineering offers a path towards further stabilizing the CSL state \cite{choi:2020,scholl:2022,koyluoglu:2024}.
To this end, we explore the broader breathing phase diagram upon the inclusion of a long-range dipole Ising term [Fig.~\ref{fig:Heisenberg}].

\subsubsection*{Finite Clusters}

Much like in the infinite geometry state, we begin by studying the stability of the CSL state in finite clusters by searching for evidence of TRS-breaking.
To this end, we study the response of the bulk chirality to an applied TRS-breaking global field $H_\chi = J_\chi \sum_i \chi_{\blacktriangle,i}$ [Figs.~\ref{fig:DiffClusters}(a,b)]. 
For clusters with an even number of spins, we observe that the bulk chirality is zero at $J_\chi=0$, but exhibits a large, system-size dependent susceptibility.
The odd clusters behave drastically differently: for all $J_\chi$, the bulk chirality remains nonzero, and exhibits a weak, size-independent response to the $J_\chi$ field.

Even though they manifest very differently, both responses show evidence of TRS-breaking---the origin of this difference is the nature of the ground state manifold.
For even-sized clusters, the ground state is unique and corresponds to a superposition of both symmetry breaking configurations:
at $J_\chi=0$ the symmetry is respected and the bulk chirality is zero.
When $J_\chi$ is added, this state hybridizes with the other orthogonal macroscopic superposition state whose energy gap is exponentially controlled by the size of the cluster. 

For odd-sized clusters, the total magnetization cannot be zero; the closest it can be is $\pm 1/2$.
This excess magnetization acts as an excitation on top of the CSL state, but because of the gapped bulk and gapless edge, this excitation will populate one of the chiral edge modes.
By populating this edge mode, the ground state at non-zero magnetization has a non-zero angular momentum: this forces the ground state manifold to be exactly two-fold degenerate (for fixed magnetization $\langle \sum_i S^z_i \rangle = 1/2$).
The DMRG calculation thus converges to one of the two TRS-breaking states.

We confirm this interpretation in multiple ways~\cite{SM}: 
by directly computing the angular momentum of the cluster and observing its flip when the other chiral ground state is prepared, and by observing the accumulation of magnetization and spin current at the edge of the cluster.

Together, these observations highlight that moderately sized clusters $N\gtrsim 21$ already display evidence of TRS-breaking and thus, might exhibit features of the CSL.
In what follows, we restrict ourselves to the odd-sized clusters, where the preparation of a symmetry breaking state is easier and the responses are easier to analyze.

As we saw in Section~\ref{sec:CSL_beta1.5}, the TRS-breaking cannot unambiguously characterize the CSL; one must also probe nonlocal quantities. 
One way to probe whether the system exhibits features of the CSL is to measure its conductivity, or response in a Quantum Hall-like setup.
The ratio between the changing electronic density $n$ and magnetic flux (per unit cell) is given by the transverse conductivity of the state---this is encoded in Streda's formula~\cite{Streda1982}:
\begin{equation} \label{eq:Streada}
    \frac{\Delta n}{\phi/2\pi} = \sigma
\end{equation}
The magnetic flux can be engineered by including a complex phase in the hopping term related to the starting and ending locations of the excitation $S^+_iS^-_j \to S^+_iS^-_je^{i\phi_{ij}}$; by choosing the phases appropriately, one can simulate the effect of an homogenous magnetic field of flux $\phi$~\cite{SM}.

Focusing on the magnetization at the center of the cluster, we observe that the linear relationship of Eq.~\ref{eq:Streada} only occurs for large enough clusters, $N\gtrsim 57$.
The slope of this linear increase is also in agreement with our expectations; 
the extracted spin transverse conductivity $\sigma_{xy}^{N=57} \in [0.39(1), 0.69(1)]$ for $\beta\in [1.3, 1.5]$ is in agreement with the value of $\sigma_{xy} = 1/2$ expected for the CSL state [Fig.~\ref{fig:DiffClusters}(d)].
Upon increasing the size of the cluster from 57 to 75, we observe a much more robust response; this highlights that while the observed conductivities are close to the expected value, there are still finite-sized effects that can be ameliorated by increasing the system size.

A few remarks are in order.
First, because our simulations are performed in an isolated cluster with fixed total magnetization, the linear change in magnetization density with flux is only present for a small range of $\phi$.
For large enough $\phi$, the edge is no longer an adequate bath of magnetization to the bulk and the magnetization eventually organizes itself onto a larger pattern---at this point, the CSL is destroyed~\cite{SM}.

Second, the observation of the linear relationship between $\phi$ and $\Delta n$ only occurs for clusters where the magnetization and spin current profiles also suggest the presence of a chiral edge mode in the system.
More specifically, for the smaller cluster of size $N=21$, although we observe finite bulk chirality, there is no accumulation of magnetization at the edge~\cite{SM}.
The resulting magnetic flux response is much more complex~[Fig.~\ref{fig:DiffClusters}].

\subsubsection*{Ising coupling $\Delta$}
Ising interactions arise in dipolar XY tweezer arrays in two ways: (i) the natural van der Waals $1/r^6$ coupling and (ii) Floquet engineering. 
This expands the landscape of possible Hamiltonians one can consider by carefully designing the absolute length scale of the Kagome lattice, or periodically modulating the dipolar system.

We now investigate how this term affects the stability of the CSL by considering a variety of cuts in the phase diagram with fixed $\beta$ or $\Delta$.
Due to the size of the associated phase space, we restrict our numerics to a fixed geometry (YC8-0) and a fixed interaction cutoff (R1).

\begin{figure}[t]
    \includegraphics[width = \textwidth]{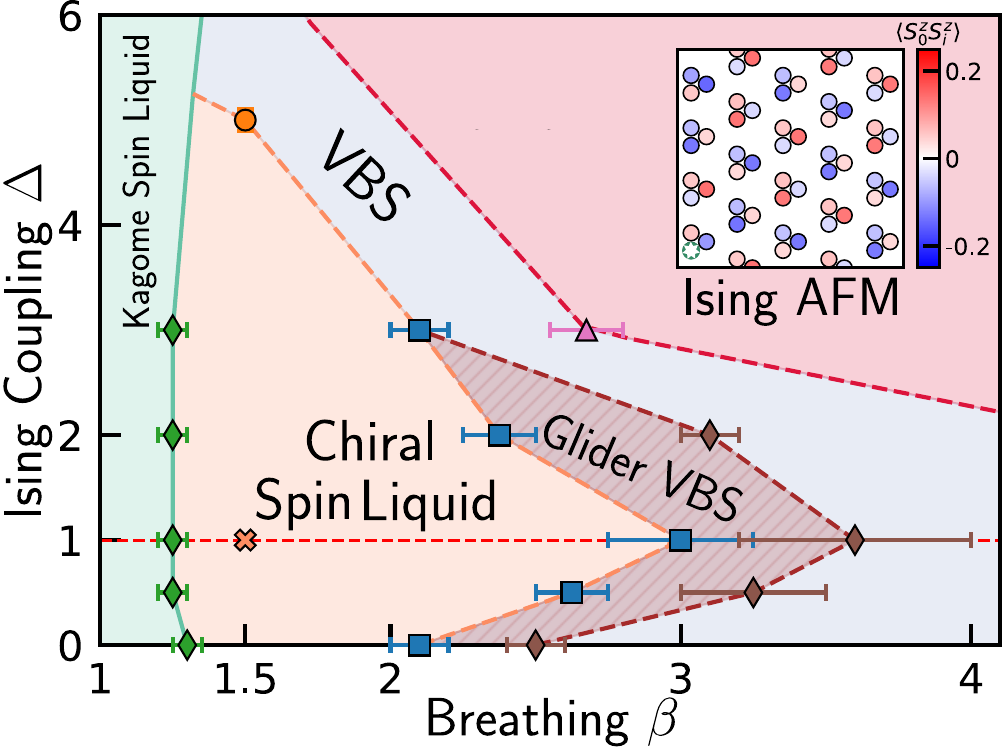}
    \caption{
      Phase diagram with respect to breathing $\beta$ and Ising coupling strength $\Delta$.
      All numerics were performed on a YC8 cylinder geometry with R1 interaction cutoff.
      We observe a robust region of the CSL up to large values of $\Delta$, and for large $\Delta$ and $\beta$, we see the emergence of a new ordered phase characterized by Ising AFM order which breaks translation symmetry.
      {\bf (Inset) } The two-point $\langle S^z_0S^z_i\rangle$ correlation function, with respect to a fixed site (chosen to be 0, bottom left white site), highlights the presence of long-range Ising AFM order.
    } 
    \label{fig:Heisenberg}
\end{figure}

The resulting phase diagram [Fig.~\ref{fig:Heisenberg}] presents a few intriguing features.
First, the CSL phase remains robust across a large area of parameter space. 
At the Heisenberg point $\Delta=1$, the CSL is further stabilized, with the critical breathing increasing from $\beta_{c}^{\text{XY}} = 2.1(2)$ to  $\beta_{c}^{\text{Heis}} = 3.0(2)$.
Experimentally, this suggests that the stability of the CSL is maintained and perhaps even improved when considering additional Ising van der Waals interactions~\cite{browaeys:2020}. 

Second, the Kagome spin liquid \emph{also} appears to be quite robust.
For all Ising couplings $\Delta$, we continue to observe signatures of the KSL at small breathings; spin-spin correlations remain homogeneous without exhibiting signatures of either long-range order or lattice symmetry breaking (translations or rotations). 
This observation connects to the expectation in the literature that the easy-axis limit $\Delta \gg 1$ is a Kagome Spin Liquid~\cite{he_kagome_2015}.
Beyond $\Delta \gtrsim 5$, the phase diagram exhibits a direct transition between the KSL and the translation symmetry broken state.
While an intriguing observation, some caveats are important. 
The KSL is believed to be a gapless critical theory and thus is very sensitive to finite size and interaction range effects. 
In particular, in finite width cylinders, we expect the underlying gapless mode to become gapped, which might stabilize the phase beyond what occurs in the true thermodynamic behavior.
A conclusive understanding of this limit requires a more in-depth and careful study.

Third, when the Ising interaction dominates, the nature of the translation symmetry broken phases is different.
At large $\beta$ and $\Delta$, we observe a 
new antiferromagnetic phase (AFM-Ising)~[Fig.~\ref{fig:Heisenberg}(inset)].

\subsection{Preparation}

Having established the stability of the CSL state, we now address its preparation in a quantum simulator.
Because the CSL is a topological state, preparation using unitary dynamics must scale with the system size~\cite{bravyi:2006}.
However, unlike other topological states previously studied in these devices, the lack of a zero-correlation-length reference state for this phase prevents the development of digital approaches~\cite{satzinger:2021}.

To this end, we consider the ingredients for the adiabatic preparation of the CSL state.
Such approach offers guarantees for the preparation of the ground state supported by the long-range interactions considered in our work, provided we can: (i) prepare a Hamiltonian and its ground state with high fidelity, and (ii) continuously deform the Hamiltonian to the CSL Hamiltonian~\cite{Hamma_2008}.

Crucially, these conditions are not met by our breathing phase diagram. 
Preparing from the Kagome Spin liquid state requires the preparation of this complex state---an outstanding challenge in its own right~\cite{bornet:2026, geim:2026}.
At the same time, preparing from the valence bond solid is blocked by the presence of a \emph{discontinuous} transition which precludes a smooth deformation of the ground state across the phase transition.

As a result, we focus on the effect of global and local fields on the breathing phase diagram of the form:
\begin{equation} \label{eq:FieldDefinition}
    H_{\text{global}} = h_x\sum_iS^x_i \quad 
    \text{and} \quad H_{\text{local}} = h_z \sum_i \lambda_i S^z_i,
\end{equation}
where $\lambda_i = \pm 1$ determines the pattern of the local perturbation.

These perturbations are particularly well suited for adiabatic preparation protocols.
Single-site fields are simple to implement and can even be changed dynamically during the experiment, and at large field strengths, the system's ground state is an easily preparable paramagnetic state, which serves as an ideal starting point for an adiabatic ramp into the phase of interest~\cite{semeghini:2021}.

\begin{figure}[t]
    \includegraphics[width = \textwidth]{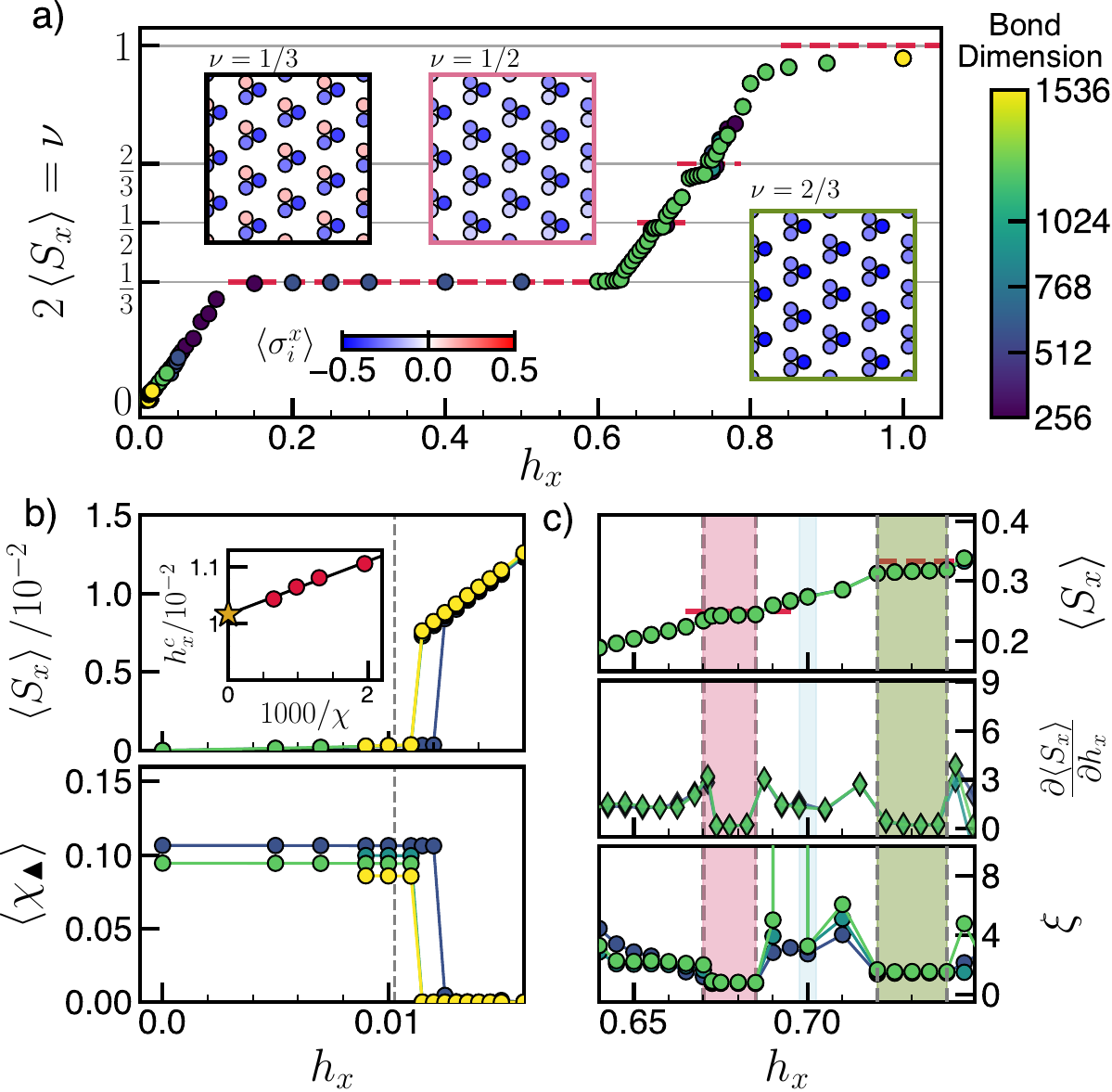}
    \caption{Phase diagram with respect to a global $\hat{x}$ field of strength $h_x$, computed for  $\beta = 1.5$ on a YC8 cylinder geometry with R1 interaction cutoff.
    {\bf (a)} The average magnetization $\expval{S_{\text{ave}}^x}$ as a function of $h_x$ displays a series of plateaus consistent with incompressible states at fractional magnetization density. The inset plots display the spatial values for $\langle \sigma_i^x \rangle$ values for the $\nu = 1/3, 1/2, \text{ and } 2/3$ phases, from left to right. 
    {\bf (b)} The transition out of the CSL at small $h_x$ is first-order, as the observables $\expval{S_x}$ and $\langle \chi_\blacktriangle\rangle$ jump discontinuously.
    This location of this discontinuity depends on the bond dimension considered; upon a finite-size bond dimension scaling analysis, we estimate the true transition to occur at the critical point $h_x^c \approx 0.0102$, demonstrating the general robustness of the CSL state (inset).
    {\bf (c)} At large $h_x$ we verify the rigid nature of the $\nu = 1/2$ and $\nu = 2/3$ states by their (approximately) zero susceptibility and finite correlation length.
    Between them we also observe a reduction in the susceptibility and correlation length, compatible with a more intricate $\nu=7/12$ state. 
    }
    \label{fig:hx} 
\end{figure}

\subsubsection*{Global field $h_x$}

We first consider the effect of a global field by computing the system's phase diagram at breathing $\beta = 1.5$ for varying $h_x$, keeping $\Delta = h_z = 0$.
Because the CSL is a gapped topological state, we expect it to remain stable for small yet non-zero $h_x$.
Indeed, we observe the persistence of finite chirality and nearly zero $\hat{x}$ magnetization of the state up to $h^c_x \approx 0.01$ [Fig.~\ref{fig:hx}(b)].
At $h^c_x$, the system undergoes a transition: the chirality suddenly becomes zero, indicating the restoration of time reversal symmetry, while the $\hat{x}$ magnetization jumps and linearly increases.
These discontinuities suggest a first-order transition out of the CSL state.

While we observe a flow of the transition location with increasing bond dimension in the numerics, the flow behavior is consistent with a finite $h_x$ transition~[Fig.~\ref{fig:hx}(b, inset)].
We estimate the location of this transition by performing a finite bond dimension scaling~[dashed line in Fig.~\ref{fig:hx}(b)]~\cite{SM}.

As $h_x$ further increases the phase diagram is quite rich.
The average magnetization $\expval{S^x}$ exhibits a staircase-like behavior with pseudo-plateaus near fractional magnetizations ($1/6$, $1/4$, $2/6$,...), Fig~\ref{fig:hx}(a)
(we do not expect exact quantization of the plateaus as $S^x$ is not conserved). 
The most prominent magnetization pseudo-plateau occurs at $\expval{S^x} = 1/6$.
Intuitively, the robustness of this phase comes from the existence of an energetically favorable and unit-cell separable configuration: one spin in the triangular unit-cell is pinned by the external field, while the other two spins in the triangle form a singlet state, minimizing the interaction energy.
Since 1 out of 3 spins are aligned with the applied field, we refer to this as the $\nu=1/3$ phase. 

We see evidence for similar such states at $\nu=1/2$ and $\nu =2/3$. In Fig.~\ref{fig:hx}(c), we identify the phases by measuring a sharp drop of $\pdv{\langle S_x\rangle }{h_x}$ and drop in correlation length $\xi$, consistent with a gapped phase of matter. The onset of the paramagnetic phase $\nu=1$ occurs when $h_x \approx J$, or when all the spins are aligned with the field.

More generally, we might expect pseudo-plateaus at each possible ``filling''. 
The MPS ansatz is a finite unit-cell with 12 spins (YC8), so we might expect to observe the subset of states with magnetization $\langle S^x \rangle = n/24$, for $n\in \mathbb{Z}_{12}$. 
Though we see hints of the other fillings in our data--- for example, the correlation length dips at $h_x \approx 0.7$ which would correspond to the $\nu = 7/12$ plateau---the careful characterization of these states is outside the scope of this work. 

Ultimately, from a preparation perspective, the global $\hat{x}$ field would not allow for adiabatic preparation due to the lack of a direct continuous transition from the trivial paramagnet to the CSL.

\subsubsection*{Local field $h_z$}

\begin{figure}[t]
  \includegraphics[width = 3.4in]{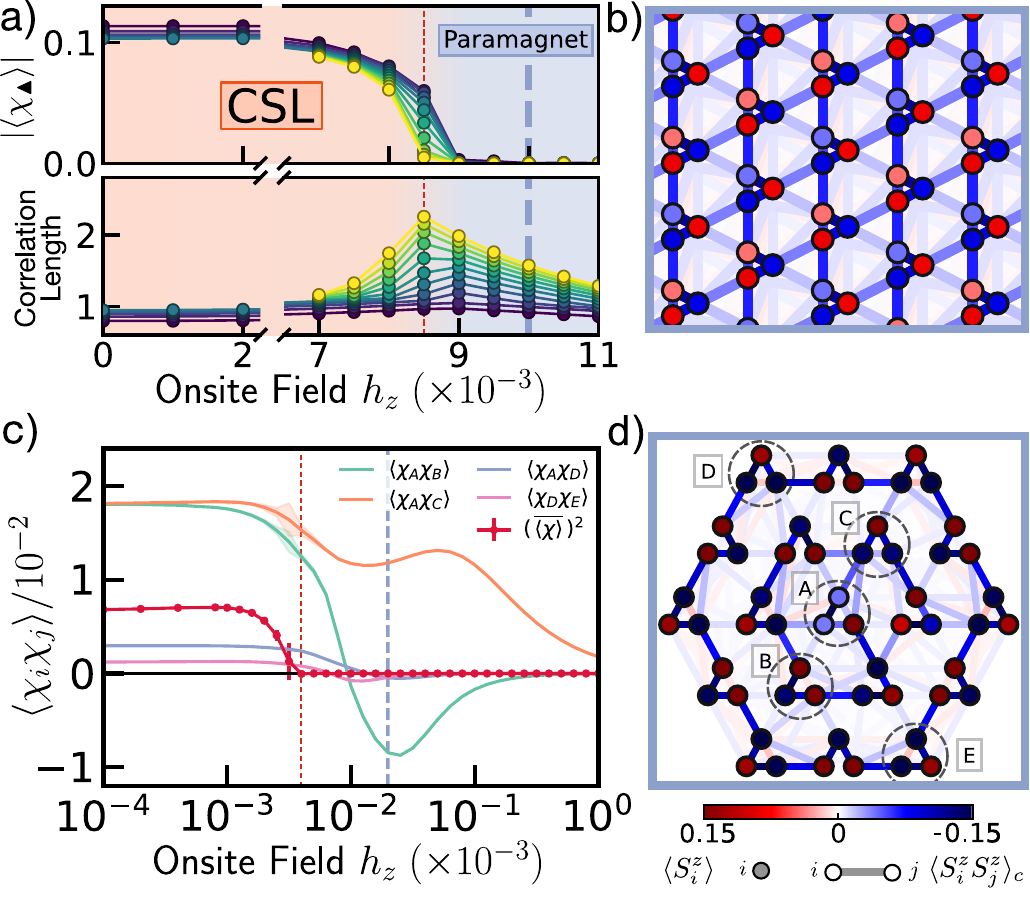}  
    \caption{Phase diagram with respect to a locally varying magnetic field pattern with strength $h_z$.
    \textbf{(a)} For small $h_z$, the system remains in the CSL phase; as $h_z$ increases, $\langle \chi_\blacktriangle \rangle$ continuously decays, reaching zero at the critical point, $h_z^c \approx 8.5 \times 10^{-3}$.
    \textbf{(b)} Concomitantly, the correlation length $\xi$ diverges with bond dimension, indicating a second-order transition.
    Beyond this transition, the system is in a trivial paramagnetic phase. 
    Vertical dashed line denotes the parameter utilized in the paramagnetic state illustrated in Fig.~\ref{fig:schematic}(d).
    \textbf{(c)} Chiral-chiral correlations between selected pairs of triangles in an $N=57$ spin cluster.
    The correlations remain negligible until $h_z \lesssim 0.2$, when some pairs start developing correlations. Below $h_z \lesssim 7 \times 10^{-3}$, all correlations become ferromagnetic, which we use as a proxy for long-range order in a finite system. The red line shows the average chirality over all triangles in the system, which also becomes non-negligible at this field strength. 
    \textbf{(d)} Spatial correlations for the cluster at the red dashed line-- the spins mostly align with the field. The letters that label the triangles indicate which chiral-chiral correlations we consider in (a).
    } 
    \label{fig:OnsiteField}
\end{figure}

In general, many different patterns can be achieved by changing the values of $\lambda_i$ in Eq.~\ref{eq:FieldDefinition}.
Here, we focus on a specific one---corresponding to the paramagnetic state in Fig~\ref{fig:OnsiteField}(b)---which minimizes the total magnetization in each triangle, and globally favors the state of zero magnetization.
Since the Hamiltonian has a $U(1)$ spin rotation symmetry, magnetization is conserved as we adiabatically ramp down the strength of $h_z$, connecting the initial paramagnetic state to the previously studied zero magnetization CSL state.

The symmetry properties of the local pattern also gives insight into the nature of the transition. 
Previously, we understood the origin of the first order transition from the CSL to Glider-VBS pattern by the change of two symmetries in the system---restoring the TRS and breaking a lattice symmetry.
By contrast, the local pattern we consider (and any other pattern that has odd magnetization per unit-cell and zero total magnetization) will immediately break rotation and translation symmetry, while maintaining TRS~\footnote{While the TRS defines as $\Theta = Ke^{i\pi\sum_jS^y_j}$ is not respected because of the presence of the field, the Hamiltonian remain purely real, so one can define a spinless time reversal symmetry $\Theta' = K$ which remains respected}.
As a result, at non-zero $h_z$, the lattice symmetries are already broken and the transition out of the CSL must only recover the TRS, so we expect this transition to be second-order.

Indeed, this expectation agrees with our numerical observations.
Much like the $h_x$ perturbation, as we increase the applied field $h_z$ to a small non-zero value, we observe a region where the CSL state remains the ground state of the system.
However, unlike the $h_x$ perturbation, we observe that the transition to the paramagnetic state exhibits both a smooth decay of the chirality $\langle \chi_\blacktriangle\rangle$ as well as a divergence of the system's correlation length.
This provides evidence for a single second-order transition that connects the paramagnetic state to the CSL state at $h_z^c \sim 8.5\times 10^{-3}$. 

The location of this transition depends on the breathing $\beta$ since it determines the strength of the inter-triangle interactions that compete with the applied field~[Fig.~\ref{fig:schematic}(d)]. 
By increasing the inter-triangle interactions (reducing $\beta$) the transition occurs at a higher critical value of $h_z$, allowing one to optimize the preparation of the CSL.

\subsubsection*{Preparing Finite Clusters}

We now turn to considering the preparation requirements of the CSL state by leveraging the second-order phase transition from a staggered local magnetic field~[Fig.~\ref{fig:OnsiteField}].
To this end, we begin by considering the ground state for the 57 spin cluster as a function of the applied field $h_z$, monitoring the emergence of long-range order of $\chi_\blacktriangle$ as a proxy for the onset of the CSL topological order.

Starting at large applied field $h_z \gtrsim 1$, the ground state of the system falls squarely in the paramagnetic phase, with all the spins aligned with the chosen pattern and exhibiting negligible spin-spin and $\langle \chi_i\chi_j\rangle$ correlations [Fig.~\ref{fig:OnsiteField}(c)].
As the local field is reduced to $0.01 \lesssim h_z \lesssim 0.2$, the cluster begins developing correlations~[Fig.~\ref{fig:OnsiteField}(d)], until all $\langle \chi_i\chi_j\rangle$ become ferromagnetic for $h_z \lesssim 7 \times 10^{-3}$. 
This is in good agreement with the infinite cylinder geometry value $h_{z,c}^{\text{cyl}}/J = 8.5 \times 10^{-3}$.
In this region, even the most distant triangles (i.e. $\langle \chi_D\chi_E\rangle$) exhibit small but non-zero chiral-chiral correlations, signaling the onset of TRS-breaking order.

The lack of any sharp features as a function of $h_z$ arises from finite-size effects;
while in the thermodynamic limit, we expect a true divergence of the correlation length, such behavior is ``truncated'' by finite size effects which imbue the system with a non-zero energy gap for all $h_z$.
Although this gap disappears in the thermodynamic limit, it does so polynomially with system size, suggesting that an adiabatic preparation is possible even for moderately sized clusters.
While this sets the scaling of the gap closing, it is important to note that the overall value of the gap is determined by the small energy scales associated with the frustrated spin system: this is based on the observation that the magnitude of the critical field $h_{z,c}$ is $\sim 7\%$ of the nearest triangle-triangle coupling of the system.

This trend for $h_{z,c}$ is observed across the entire phase diagram~[Fig.~\ref{fig:schematic}].
The breathing parameter $\beta$ thus offers a trade-off between the size of the energy gap and the size of the TRS-breaking signal of the CSL phase.
More specifically, a smaller breathing parameter within the CSL phase leads to a sharply increased interaction strength between triangles $J_{ij} \sim (2\beta-1)^{-3}$, which directly increases the critical field, although at the cost of a smaller value of $\langle \chi_i\chi_j\rangle$.

\subsection{Experimental Signatures}

While we characterized the nature of the CSL phase by its response to an applied magnetic flux, such an experiment is quite challenging in current AMO platforms due to the need to implement artificial gauge fields.
Although some strategies have been developed~\cite{AidelsburgerHofstadter2013, Cesa2013, Aidelsburger2022ColdAtomsLGT}, it is important to explore alternative signatures to verify the topological properties of the spin liquid state.
In particular, we will focus on four distinct signatures and how they can be probed using individual spin control.

\subsubsection*{Observing transverse conductivity}

As previously discussed, one of the defining features of the CSL state is the presence of a topologically protected chiral edge mode that is responsible for the presence of a quantized transverse conductivity.
The conventional solid-state approach to extract conductivity leverages the ability to connect leads to a material and apply a small chemical potential difference $V$.
The relationship between the \emph{steady state} current $I$ and strength of the applied perturbation $V$ directly defines the conductance of the system.
Unfortunately, the isolated nature of an AMO system and the difficulty of engineering suitable spin baths precludes this approach (even if some remarkable experiments have been able to probe particle conductance~\cite{Krinner_2014}).
However, even without the system reaching a true steady state regime, we now demonstrate that it is possible to observe signatures of the transverse conductivity.

\begin{figure}[t!]
\centering
    \includegraphics[width = 3.4in]{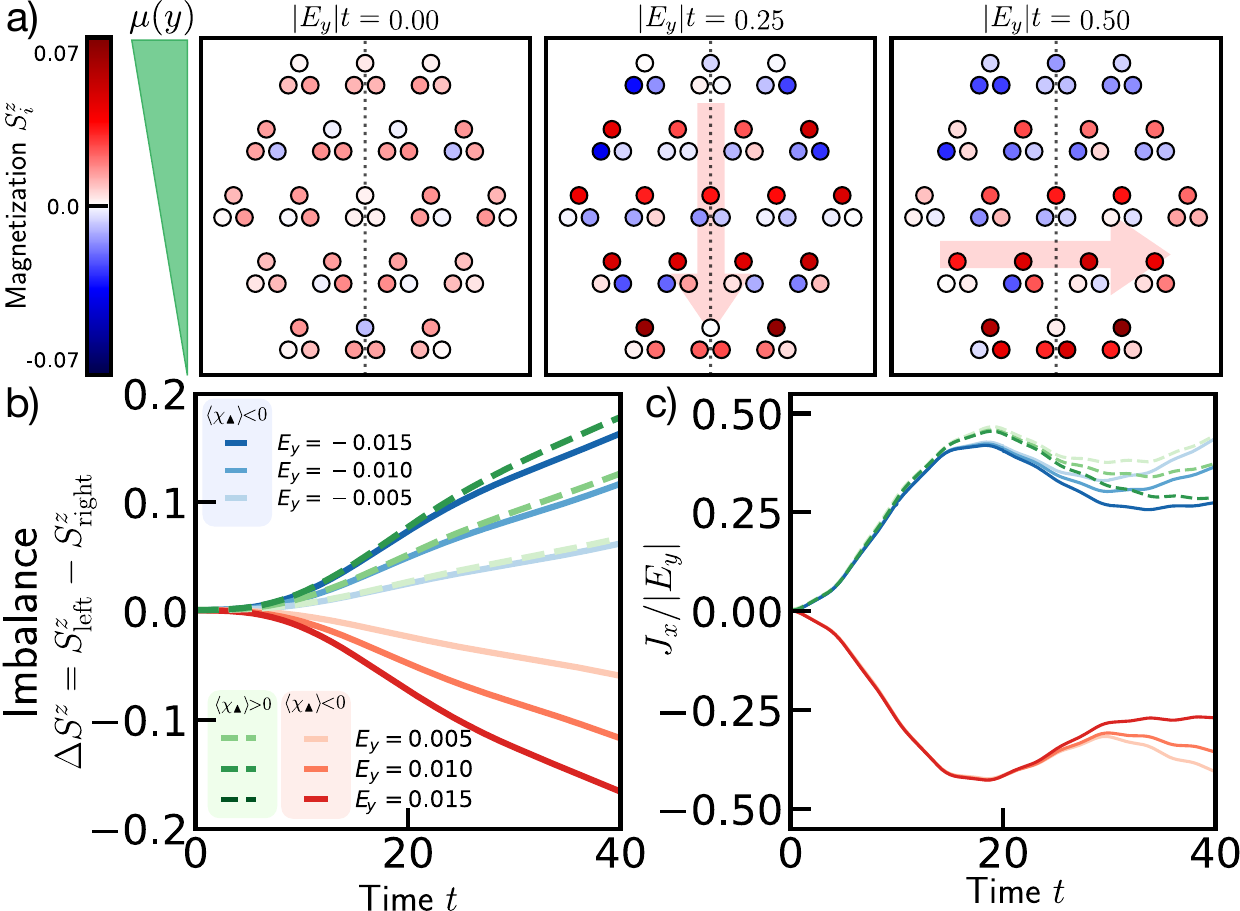}
    \caption{
      {\bf (a)} Dynamical response of the cluster to a quenched chemical potential gradient $\mu(y)$ in the vertical direction.
      Throughout the evolution, the magnetization dynamics follows the chemical potential gradient as expected. However, at later times, we can also discern a \emph{transverse} motion of the magnetization, in agreement with a non-zero transverse conductivity.
      {\bf (b)} To quantify this transverse response we compute the total spin imbalance $\Delta S^z$ between the left and right sides of the cluster (red).
      Crucially, the direction of this motion reverses when either we consider the opposite TRS-breaking ground state (blue), or we reverse the sign of the chemical potential gradient (dashed line).
      {\bf (c)} By rescaling the transverse spin current $J_x$ by the strength of the gradient $E_y$, we observe a scaling collapse of the early-time dynamics confirming we are studying the linear response of the system.
    }
    \label{fig:Cluster_Dynamics}
\end{figure}

Our main insight is to leverage the ability of AMO platforms to investigate the real-time dynamics of the magnetization motion (i.e. current) upon the application of a chemical potential imbalance.
We consider the following setting:
starting from the ground state of the $N=57$ cluster, we apply a chemical potential  gradient $ \delta H_{\text{grad}}= \sum_i \mu(y_i) S^z_i$ with $\mu(y_i) = E_y y_i$ and study the resulting quench spin dynamics~[Fig.~\ref{fig:Cluster_Dynamics}(a)].
Starting from a three-fold rotationally symmetric spin distribution, two distinct features become clear with the subsequent dynamics.
First, the presence of a vertical chemical potential induces a vertical imbalance of magnetization, with magnetization moving towards the bottom of the cluster where the chemical potential is smaller.
Second, even though the left and the right side of the cluster have no chemical potential imbalance, the cluster's magnetization displays a tendency to move to the right.
Indeed, by quantifying this imbalance by computing $\Delta S^z = \langle S^z_{\text{left}}\rangle - \langle S^z_{\text{right}}\rangle$, we observe a systematic increase in the strength of the imbalance~[Fig.~\ref{fig:Cluster_Dynamics}(b)].
This transverse spin motion highlights the presence of a transverse response, whereby a vertical chemical potential translates into a horizontal flow of magnetization.

While our observations occur before the system reaches steady state, they capture important features of the system's equilibrium linear response.
First, the strength of the observed spin current, $I_x = \frac{d \Delta S^z}{dt}$, is \emph{proportional} to the strength of the applied chemical potential difference, $I_x \propto E_y$.
This is in agreement with the system being in a linear response regime (i.e. a non-zero transverse conductivity), rather than dominated by incoherent processes--i.e. Fermi's Golden rule.
Indeed, in this setting, the presence of low-lying excited states arising from the edge mode should preclude such an analysis.
Second, the direction of the transverse spin current is controlled by both the sign of the chemical potential gradient $E_y$, and also the sign of the spontaneous TRS-breaking of the cluster $\overline{\langle \chi \rangle}$ [Fig.~\ref{fig:Cluster_Dynamics}(c)]; flipping either one reverses the transverse motion of the magnetization.

\subsubsection*{Dynamics of a chiral edge mode}

\begin{figure}[t]
    \centering
    \includegraphics[width=\linewidth]{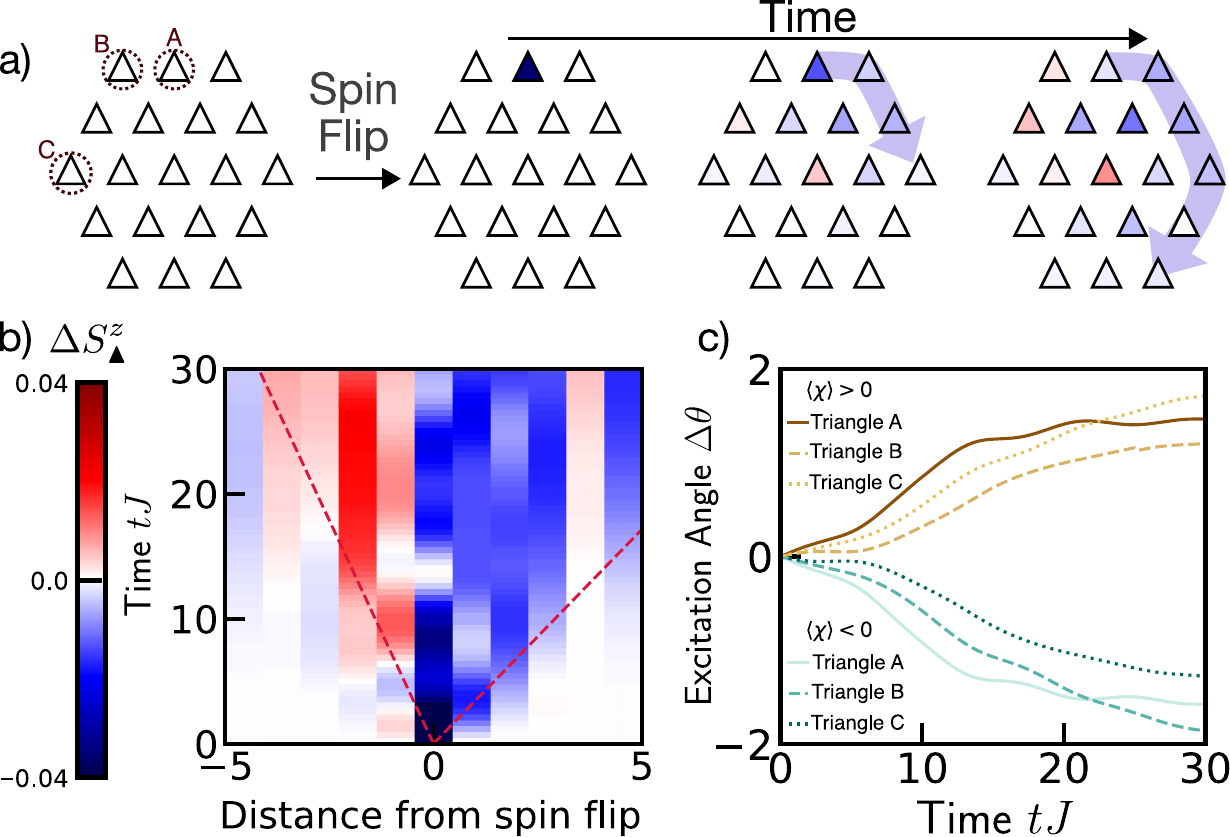}
    \caption{{\bf (a)} Edge dynamics upon flipping the spins in a single unit cell.
      We study the dynamics of the unit-cell magnetization \textit{difference} $\Delta S^z_\blacktriangle$ (compared to the initial ground state) observing a robust clockwise propagation along edge of the cluster. 
    {\bf (b)} Considering the dynamics of only the edge unit cells highlights the chiral, ballistic nature of the dynamics. 
    {\bf (c)} The chiral edge mode is robust to flipping different unit cells on the edge, and the propagation direction is controlled by the direction of the TRS-breaking order of the cluster.}
    \label{fig:Cluster_Dynamics_Chiral}
\end{figure}
While the magnetization transport dynamics directly connects to the expected spin conductivity of the CSL, the ability to measure each individual spin in the system offers an entirely novel approach to probing this phase of matter.
We highlight this point by considering an experiment that allows one to directly probe the chiral edge mode.
To achieve this, we must consider a low-energy spinful excitation near the edge of the cluster.
This populates the edge mode with a pair of semions, which then propagate in a well-defined direction.

We again start with the ground state of our Hamiltonian in a finite cluster, perform a spin-flip within the low-energy manifold of a single unit cell and study the subsequent dynamics [Fig.~\ref{fig:Cluster_Dynamics_Chiral}(a)].
Experimentally, this can be done by resonantly driving the $\left|\uparrow \right\rangle \leftrightarrow \left|\downarrow  \right\rangle$ transition with a Rabi frequency that is much smaller than the strong intra-triangle coupling $J$, allowing one to only address the low-lying manifold of states~[Fig.~\ref{fig:2Spin}(a)].
Numerically, we leverage our spin-chirality model to simulate this process: since the chiral and total spin degrees of freedom are already decoupled, the initialization is much simpler and the subsequent dynamics are easier to parse.

Our observations are summarized in Fig.~\ref{fig:Cluster_Dynamics_Chiral}(a-c), focusing on the magnetization dynamics measured with respect to the initial ground state: $\Delta S^z_\triangle(t) = \langle S^z_\triangle(t) \rangle_{\text{dyn}} - \langle S^z_\triangle \rangle_{\text{gs}}$.
The magnetization is initially confined to a single triangle, then subsequently propagates \emph{clockwise}, highlighting the chiral nature of the edge mode.
While the edge mode is not exactly contained within the outer layer of triangles, most of the magnetization remains far from the center of the cluster.

To quantify this chiral propagation, we consider the dynamics of $\Delta S^z_\triangle(t)$ as a function of the distance from the initial perturbation along the edge of the cluster~[Fig.~\ref{fig:Cluster_Dynamics_Chiral}(b)].
This further emphasizes the chiral nature of the dynamics, with the magnetization moving only to the right.
Crucially, it also highlights that the motion is ballistic, with the magnetization propagation mirroring a one-sided light-cone.
Curiously, towards the opposite direction we observe a ``wake'' of opposite magnetization, propagating at a much lower velocity.

Finally, let us note that the observed dynamics are not restricted to a particular choice of spin flip. 
By varying the location of the edge spin-flip, we can observe similar chiral movement, as illustrated by the angular displacement of the magnetization center of mass, $\vec{r}_{\text{CM}} = \sum_i \vec{r}_{\blacktriangle_i} \langle S^z_{\blacktriangle_i}\rangle / \sum_i \langle S^z_{\blacktriangle_i}\rangle$.
At the same time, by reversing the chirality of the ground state considered (i.e. considering the other spontaneous symmetry broken sector), the direction of the propagating edge mode is correspondingly flipped.
These observations stand in contrast to the dynamics of a spin flip in the bulk, whereby the magnetization excitation remains in the same position~\cite{SM}.

\subsubsection*{Trapping a semion}

\begin{figure}[t]
\centering
    \includegraphics[width = 3.4in]{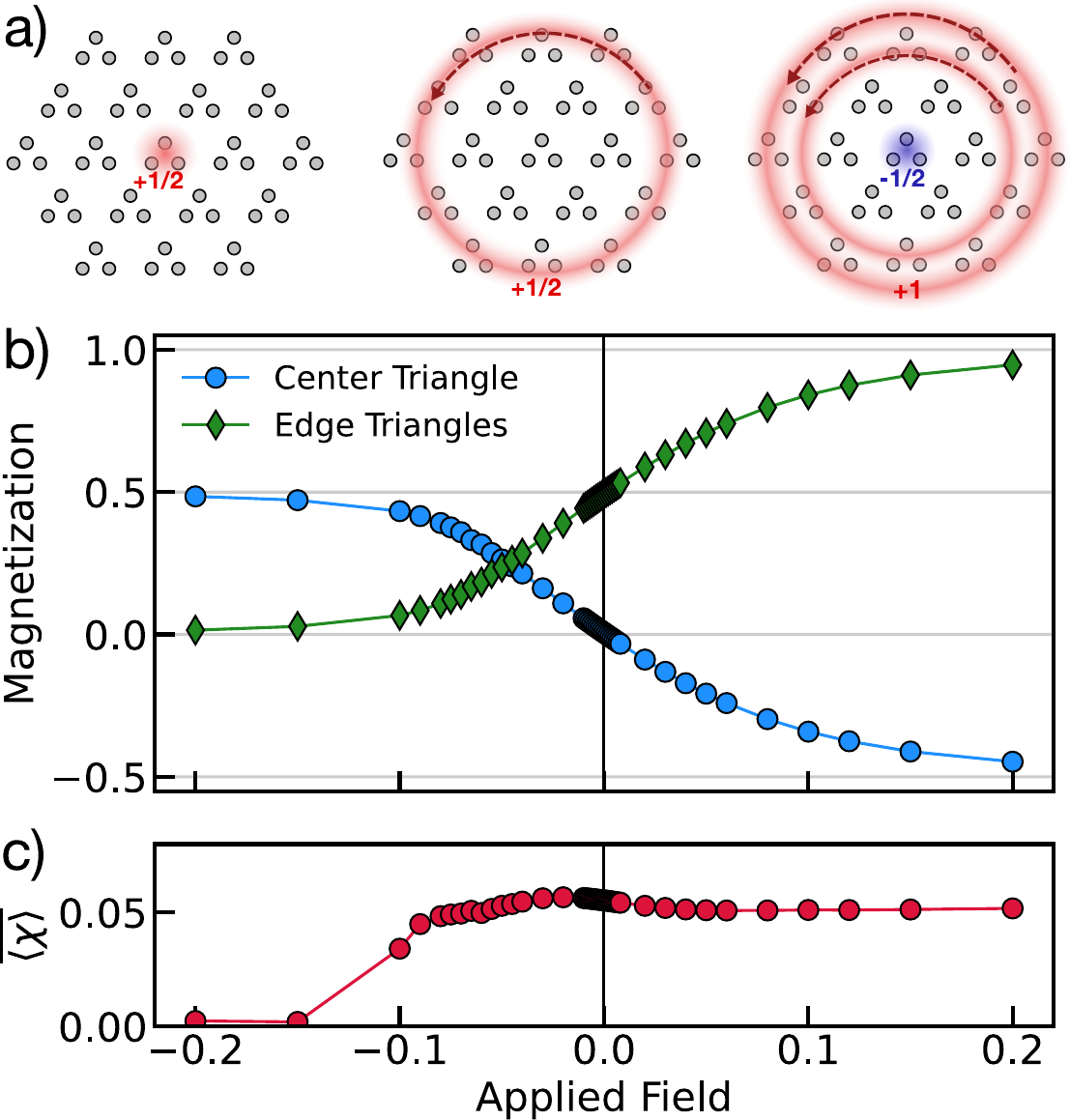}
    \caption{
      {\bf (a)} In an odd cluster, the additional $1/2$ magnetization is carried by an excitation on top of the CSL ground state.
      By applying a local field on the center of the cluster (colored sites), we can change the nature of this excitation.
      When the field is negative, the central triangle will trap a localized semion.
      When the field is zero, the magnetization lives on the edge, occupying a chiral gapless mode.
      When the field is positive, a semion of opposite magnetization is trapped, and two edge modes are occupied.
      {\bf (b)} The magnetization of the center and edge triangles. At no applied field, there is a net $1/2$ magnetization at the edge triangles, consistent with an anyon on the edge. When increasing the applied field, the edge picks up another net $1/2$ magnetization while the center has a $-1/2$ magnetization. At negative applied field, the net magnetization travels from the edge to the center.
    {\bf (c)} The average chirality $\bar{\langle \chi \rangle}$ as a function of the applied field. When the field is very positive, the system remains in the CSL state by adding another excitation to the edge. When the field is very negative, the TRS symmetry is restored and the system is no longer in the CSL state.
    }
    \label{fig:TrappedSemion}
\end{figure}

We now discuss how a semion might be trapped using local controls.
The key insight is that, because the semion is a half-spin excitation, its chemical potential is controllable using local applied fields.
This offers a path towards controlling the dynamics of the semionic excitations, using the same tools as used when preparing the CSL.
More specifically, we focus on the effect of a locally changing potential at the center of our cluster: $H_{\text{trap}} = h_z \sum_{i\in \text{center}} S^z_i$.

Because we are considering a cluster with total magnetization of $+1/2$, the observations are very different for positive or negative $h_z$.
When $h_z$ is made more positive, any spin magnetization near the center moves towards the edge of the cluster, populating \emph{another} chiral edge mode.
As a result, the ground state manifold is again composed of states with non-zero angular momentum, ensuring that it is exactly two-fold degenerate.
When $h_z \gtrsim 0.1$, the entire system continues to display strong correlations, with the chirality of the center triangle aligned with the chiralities of the other triangles in the cluster~\cite{SM}.
This suggests that the CSL remains stable across the entire cluster, with the excess opposite magnetization in the middle corresponding to the trapped semionic excitation.

The story is more complicated when $h_z < 0$.
As $h_z$ becomes more negative, the magnetization initially located near the edge of the cluster becomes concentrated near the center.
For $h_z\lesssim -0.15$, the middle triangle contains almost all of the cluster's magnetization, and its correlations with neighboring spins are reduced~\cite{SM}.
As such, the cluster is now effectively an even cluster with a hole in the middle and no occupied edge modes. 
Thus, this finite-sized even cluster exhibits no spontaneous TRS-breaking as previously discussed.

This proof-of-principle exploration highlights that semions can be manipulated using local controls. 
Using these tools to engineer the braiding of semions, or help with more complex interferometry experiments, would provide direct evidence of the anyonic statistics of these excitations.

\subsubsection*{Detecting anyonic flux through Wilson loops}

Access to local observables also potentiates the study of more complex correlation functions. 
This opens up the opportunity to study the Wilson loop associated with the CSL state to probe the number of anyons.

To investigate this operator as probe of the CSL state, we first return to the infinite cylinder states obtained in the Sec.~\ref{sec:Model}, where the Wilson loop \emph{around} the cylinder helps to identify the topological sector.
Building on previous work~\cite{he_kagome_2015}, we define a candidate Wilson loop along a path $\ell = (a_1,b_1,a_2,b_2,\ldots)$ as the product of sequential raising and lowering operators: 
\begin{equation}\label{eq:Wilson}
    W_\ell = \prod_{i=1}^{|\ell|/2} S^+_{a_i}S^-_{b_i}~.
\end{equation}
For a fixed point wavefunction, it is expected that $|\langle W_\ell\rangle|=1$, but this is not the case for ground states with a finite correlation length, where the Wilson loop is ``dressed'' and thus will always decay with the increasing length of the path (regardless of the nature of the state).
One way to circumvent this is by explicitly projecting the wavefunction to the relevant constrained configuration space. 
However, due to the computational difficulty of this procedure (and the ambiguity of the correct constrained space) we can achieve a similar effect by renormalizing the Wilson loop as follows~\cite{he_kagome_2015}:
$$
\langle W_\ell \rangle_{\text{ren}} = \frac{\langle W_\ell\rangle}{\langle W_\ell^\dag W_\ell\rangle} = W_0 e^{i\alpha}
$$
where $W_0$ denotes the magnitude of the Wilson loop, and $\alpha$ encodes the phase which determines the flux information.
Crucially, we observe that $|\langle W_\ell \rangle_{\text{ren}}| \gtrsim 0.3$ for all paths considered, suggesting the system is indeed in the deconfined regime.
Moreover, the phase information of $\langle W_\ell \rangle_{\text{ren}}$ directly measures the anyonic flux in the system: 
computing it for the two wavefunctions at $\theta=0$ and $\theta=2\pi$, we get symmetric values, $\langle W_\ell \rangle_{\text{ren}}^{\theta=0} = 0.81$ and $\langle W_\ell \rangle_{\text{ren}}^{\theta=2\pi} = -0.80$, corresponding to a flux of $\alpha^{\theta=0}=0$ and $\alpha^{\theta=2\pi} = \pi$, respectively. 
Note that while there is a phase ambiguity in the definition of $W_\ell$, this does not change the phase difference $\Delta \alpha$ between the two states which differentiates between the two topological sectors.

To translate these observations in into an experimental probe, we must ensure that the Wilson loop remains a valid measure, and that its numerical value is large enough to be measured in experiments.
In the case of a finite-sized cluster, $W_\ell$ does not determine the topological sector of the state but can count the number of anyonic excitations within it.
Computing $W_\ell$ around the center unit cell we observe a change of sign between finite cluster with zero and large trapping fields, corroborating the previous interpretation of the trapping of a semion.
However, the structure of the $W_\ell$ for different regions becomes more complex, suggesting that larger system size calculations are important to better understand the robustness of this observation.
Moreover, we find that the non-normalized Wilson loop has a magnitude of $\approx 0.3$ per unit cell, enabling the measurement of the relative phase of $W_\ell$ for loops interfacing with 4 unit-cells.

\subsection{Experimental Platforms}

We now turn to a discussion of experimental AMO platforms where the CSL state is expected to naturally arise: 
Rydberg atoms and polar molecules in tweezer arrays 
and magnetic atoms in optical lattices.

\subsubsection*{Tweezer Arrays: Neutral Atoms and Molecules}

Initially utilized to trap and control microscopic particles~\cite{Ashkin1986}, 
optical tweezers were then extended with great success to trap and control individual atoms and molecules~\cite{Chu1986TrappedAtoms}. 
More recently, the ability to prepare scalable defect-free arrays~\cite{Barredo_2016}, as well as leverage long-range and strong interactions, has enabled tweezers to flourish as a platform for studying many-body quantum dynamics and perform quantum computational tasks~\cite{Bernien_2017, Bluvstein_2023, Bluvstein2026FaultTolerantNeutralAtom}.

Within the context of our proposal, one of the most appealing features of tweezer arrays is arbitrarily dictating the geometry of the lattice of tweezers~\cite{Schlosser_2023}.
This directly enables continuous geometric control over features such as the breathing parameter studied in our work.
At the same time, this technique is not restricted to only a few tweezers; recent technical improvements have enabled the demonstration of 1000s of tweezers within a single experiment, enabling the study of many-body phenomena near the thermodynamic limit~\cite{Pichard_2024}.
Finally, the tweezer technology itself can be leveraged beyond trapping. 
The ability to control the strength and position of a secondary optical tweezer array has been leveraged to dynamically move particles~\cite{Bluvstein_2023}, apply local control gates~\cite{Bluvstein_2023}, and engineer a local Hamiltonian~\cite{chen_continuous_2022}.
In conjunction with global addressing and readout, this has enabled new preparation schemes~\cite{chen_continuous_2022}, mid-circuit readout protocols \cite{Bluvstein_2023}, and the direct measurement of multi-body, multi-basis observables~\cite{Bornet2024enhancing}.

\emph{Rydberg atoms}---Many of the above techniques have first been demonstrated in the context of Rydberg atoms~\cite{Bluvstein_2023,chen_continuous_2022}.
In this particular approach, strong, long-range interactions can be generated by exciting the trapped atom into a Rydberg state of high principal quantum number $n$.
In such a state, the electronic wavefunction becomes highly delocalized, increasing its susceptibility and dipole moment.
As a result, neutral atoms can exhibit very strong interactions even when separated by nominally large distances (i.e.$~\mathrm{MHz}$ at $\sim10s ~\mathrm{\mu m}$).
This interaction has gained prominence as the key step in ``Rydberg blockade'' gates, where the very strong van der Waals $\sim 1/r^6$ interaction prevents two nearby atoms from both occupying the Rydberg state, enabling robust digital quantum gates~\cite{Jaksch2000,Lukin2001}.
At the same time, it naturally corresponds to a long-range Ising-like interaction $\sim S^zS^z/r^6$, of interest to multiple many-body quantum simulation experiments~\cite{Bernien_2017,Ebadi_2021,semeghini:2021}.

Extending this type of interaction to an XY-type model requires a different scheme, where the qubit is encoded in two \textbf{distinct} Rydberg levels. 
When the levels are dipole-coupled, the atoms can coherently exchange their states, which serves as the basis for implementing the dipolar XY model ($\sim (S^xS^x+S^yS^y)/r^3$)~\cite{Barredo_2016}. 
This interaction is characterized by a different power-law and symmetry, and has enabled the observation of a variety of novel phenomena, including spin squeezing, continuous symmetry-breaking phases, and  bosonic SPT phases \cite{de_L_s_leuc_2019,chen_continuous_2022,Bornet_squeezing_2023}.

Indeed, current dipolar XY Rydberg experiments already have all the ingredients necessary to adiabatically prepare the CSL discussed, from the Hamiltonian to the use of a secondary optical tweezer array to engineer the necessary local magnetic field patterns and measure multi-basis observables~\cite{Bornet2024enhancing}. Furthermore, the proposed dynamical experiments (Figs.~\ref{fig:Cluster_Dynamics} and \ref{fig:Cluster_Dynamics_Chiral}) measure topological signatures of the CSL (conductivity and the edge mode) are also quite natural and within the scope of current experiments. 

While all the experimental capabilities have been demonstrated, it is important to highlight some current limitations and promising future directions.
Below, we briefly discuss three of these points and future directions of exploration.

First, the coherence time of the experiment is limited by the inability to trap the atoms while in the Rydberg manifold.
Rydberg states for alkali atoms are anti-trapped for red-detuned dipole traps, which can be understood by considering the highly excited Rydberg electron as approximately ``free''. 
Free electrons have a negative polarizability and are thus repelled by the tweezer light~\cite{Wilson_alkEarth_2022}.
To prevent the atoms from being pushed apart, one must turn off the traps during the course of the experiment; during this time, the atoms are free to expand owing to the velocity distribution set by the initial temperature, limiting the interaction times to $\sim 10-20~\mu s$.~\cite{chen_continuous_2022}
Because of this random, thermal noise, for longer times, different shots of the experiment experience different dynamics and thus the system decoheres.
This limits the speed of the adiabatic preparation and thus the fidelity of the final CSL state. 
The importance of this positional randomness is made more poignant due to the frustrated nature of the CSL, as positional disorder destabilizes the frustration necessary to drive the system into a spin liquid state. 
In our system, we have quantified the noise tolerance by studying the 57 spin cluster with random onsite (static) positional disorder sampled from a Gaussian distribution. For this system size, we find that the CSL is stable up to positional disorder of $\lesssim 0.5\%$ of $a$ for $\beta=1.5$.

By contrast, alkaline earth atoms such as ytterbium and strontium have been experimentally shown to have a significant positive polarizability due to their ion cores, so they are attracted to intensity maxima and can be trapped in their Rydberg states~\cite{Wilson_alkEarth_2022}.
``Magic'', or state-insensitive, trapping conditions can be achieved by tuning parameters such as the Rydberg state $n$ and the size of the beam waist. Alkaline earth atoms have significantly longer lifetimes $\gtrsim 100~\mu s$ that more closely approach the typical room temperature lifetime, allowing for longer coherence times and less positional disorder~\cite{Wilson_alkEarth_2022}. Thus, keeping the traps on during the experiment would allow one to get closer to the ultra-low entropy and frustrated regime necessary to observe clear signatures of spin liquids.

Second, there exist certain subtleties when preparing and measuring a TRS-breaking phase such as the CSL. 
During the adiabatic ramp, the time evolution breaks TRS but only perturbatively, so the state will still spontaneously break into each TRS sector approximately half of the time. 
As a result, we cannot simply measure $\expval{\chi_i}$ but need to measure $\expval{\chi_i \chi_j}$. 
To measure the full six-body observable, one can design a more complicated experiment, such as using three SLMs for the tweezer array, initial state preparation, and readout ~\cite{Bornet2024enhancing}. 
Another approach is to explicitly break TRS, which would lower one chirality state in energy over the other. 
This would reduce the complexity of the measurement scheme from six to three-body observables, as well as increase the robustness of the preparation scheme, ensuring a higher fidelity of the ground state.

One promising route to adding a TRS-breaking term in the Hamiltonian comes from using three Rydberg levels to engineer terms with complex phases~\cite{weber_experimentally_2022}.
The presence of a third level enables one to directly leverage the $\Delta m=\pm 2$ dipolar interaction to imbue a complex phase into the Hamiltonian.
Making one of the states off-resonant enables this complex phase to be mapped into the dynamics of a two-level system.
Alternatively, one can leverage additional auxiliary qubits to generate a homogeneous magnetic flux in a rectangular lattice~\cite{eix2024homogeneousmagneticfluxrydberg}.

Finally, one of our observations is that the CSL is further stabilized with antiferromagnetic Heisenberg-like interactions (see Figure~\ref{fig:Heisenberg}). 
In the long term, Rydberg atoms may be able to explore the aforementioned Ising phase diagram by using Floquet engineering~\cite{scholl:2022}.

\emph{Polar molecules}---While more nascent, polar molecules trapped in tweezers are another promising experimental platform for realizing a CSL~\cite{Anderegg_2019,KaufmanNi2021Review}.
In these systems, the two-level qubit is encoded in different molecular rotational states, which enables one to make use of the large permanent electric dipole moment of the polar molecule. 
When combined with an applied electric field, the strength and angle between the field and dipole moment tunes the strength of the dipolar coupling~\cite{Micheli_2006}.

Similarly to Rydberg atoms, polar molecules have all the ingredients necessary to realize the CSL--- 
(1) arbitrary geometries, local Hamiltonians, and multi-basis measurements from the tweezer technology and 
(2) the antiferromagnetic dipolar XXZ model. 
At the same time, there are magic trapping conditions between the two lowest rotational states that allow for state-independent trapping, and thus enable a longer coherence time~\cite{Seesselberg2018Molecule,Gregory2024}.

While the complexity of the internal molecular structure makes certain experimental aspects more challenging, there has been rapid development in the preparation and manipulation of ultracold molecules in recent years, making this platform a promising future direction~\cite{KaufmanNi2021Review, lu:2026}.

\subsubsection*{Optical lattices}

More speculatively, ultracold atoms in optical lattices potentially provide another strategy towards preparing and measuring a CSL. While not able to produce arbitrary geometries, optical lattices can generate the breathing Kagome lattice by overlaying two commensurate triangular lattices with different spacings~\cite{Barter_2020}. 
The use of additional standing waves can be used to create staggered potentials, allowing for local Hamiltonian engineering.
This can be leveraged for multi-basis measurements~\cite{Impertro2024OpticalLatticeLocalControl}, time-dependent potentials  (for generating TRS-breaking terms), or even artificial gauge fields~\cite{AidelsburgerHofstadter2013}. 
Finally, optical lattices in a quantum gas microscope also enable site-resolved measurements of the lattice, providing the main necessary experimental components of our proposed experiment. 

\section{Discussion and Outlook}

In this work, we demonstrate how the continuous control of lattice geometry is a novel and powerful tool to prepare and analyze frustrated magnetism, including spin liquids. 
Through extensive DMRG simulations, we show the existence of a robust chiral spin liquid phase for the long-ranged, dipolar XY model on the \textit{breathed} Kagome lattice, and provide pathways to directly prepare and probe this phase on near-term experiments. 
In addition, we introduce the effective \textit{spin-chirality} model, which not only reproduces the phase diagram but also provides a deeper understanding of the relevant degrees of freedom.

Looking forward, there are a number of interesting experimental and theoretical directions to explore. 
On the experimental side, we have already shown how local control allows us to perform dynamical experiments that directly probe the transverse conductivity, chiral edge mode, and localization of semionic excitations. 
With the ability to trap and control semions, the outstanding goal would be to exchange semions and measure their anyonic statistics.
Indeed, the measurement and interferometry of the topologically protected chiral edge mode can be utilized in certain robust quantum state transfer protocols~\cite{yao_topologically_2013}.
On the practical side, we note that the adiabatic preparation timescale always scales with the size of the system. 
Developing alternative preparation methods is crucial for the high-fidelity dynamical characterization of the CSL in large system sizes.

On the theoretical side, the paradigm of tunable geometry opens the doors to a much broader search for spin liquid states in AMO settings. 
Our spin-chirality model provides a new microscopic framework to consider controlling and probing the CSL or even other types of spin liquids. 
For example, although we are motivated by dipolar interactions, our spin-chirality model informs us that the necessary interactions are only nearest-neighbor on a triangular lattice, which could potentially be implemented on other experimental platforms without long-range interactions. 
In addition, one can consider generalizations of higher spin or even more complex geometries, potentially leading to more exotic topological phases, such as non-Abelian spin liquids.

\emph{Acknowledgements}--- We gratefully acknowledge the insights of M.~Bintz, E.~Demler, J.~Hauschild, L.~Homeier,  R.~Mong, L.~Pollet, and R.~Verresen. FM acknowledges support from the NSF through a grant for ITAMP at Harvard University, as well as from the Netherlands Organisation for Scientific Research (NWO/OCW), as part of Quantum Limits (project number SUMMIT.1.1016).
SC acknowledges support from the National Science Foundation Graduate Research Fellowship under Grant No. DGE 2140743. MZ was supported by the U.S. Department of Energy, Office of Science, Basic Energy Sciences, under Early Career Award No. DE-SC0022716.

\bibliography{refs}%

\end{document}


\title{Supplementary Information: \\
A Dipolar Chiral Spin Liquid on the Breathed Kagome Lattice}

\author{Francisco Machado}
\thanks{These authors contributed equally to this work.}
\affiliation{ITAMP, Harvard-Smithsonian Center for Astrophysics, Cambridge, MA 02138, USA}
\affiliation{Department of Physics, Harvard University, Cambridge, MA 02138, USA}
\affiliation{QuTech, Delft University of Technology, PO Box 5046, 2600 GA Delft, The Netherlands}

\author{Sabrina Chern}
\thanks{These authors contributed equally to this work.}
\affiliation{Department of Physics, Harvard University, Cambridge, MA 02138, USA}

\author{Michael P. Zaletel}
\affiliation{Department of Physics, University of California, Berkeley, CA 94720, USA}
\affiliation{Material Science Division, Lawrence Berkeley National Laboratory, Berkeley, CA 94720, USA}

\author{Norman Y. Yao}
\affiliation{Department of Physics, Harvard University, Cambridge, MA 02138, USA}
\affiliation{Harvard Quantum Initiative, Harvard University, Cambridge, MA 02138, USA}

\maketitle

\section{Convergence and robustness of the Chiral Spin Liquid}

In this section, we complement the numerics of the main text with additional evidence for the convergence of our calculation to the chiral spin liquid state, as well as the robustness of the phase with regards to the choices of interaction distance cutoff and simulation geometry.
We restrict ourselves to the pure dipolar XY model at $\beta=1.5$ for this benchmark.

{\bf Geometry---}We consider two compactifications of the two-dimensional lattice into an infinite cylinder.
In YCN, the lattice is wrapped along one of the directions defined by the basis vector $\vec{a}_1$ such that unit cells separated by $\vec{a}_1 \frac{N}{2}$ are identified to be the same.
In this geometry, each $N/2$ unit-cells along the compactified direction form the unit cell for the iMPS state, which is infinitely repeating.
In YCN-2, the two dimensional lattice is compactified with a different identification, namely that the unit cell at $\vec{a}_1 \frac{N}{2}$ is identified with the unit cell at $\vec{a}_2$.
Since the iMPS is defined to run across all unit-cells along $\vec{a}_1$ before moving in the $\vec{a}_2$ direction, this compactification imbues the system with an additional translation symmetry by a single unit-cell translation along the iMPS direction. %
This leads to a much smaller unit-cell for the iMPS, reducing the computational cost. 
The two compactifications become equivalent in the thermodynamic limit.

{\bf Interaction cutoff---}As noted in the main text, the structure of each unit-cell provides important clues for the structure of the chiral spin liquid state and the underlying phase diagram. To this end, when choosing how to cutoff the long-range dipolar interactions, we do not focus on the distance as measured by either the spatial distance (which changes considerably as we change the breathing $\beta$) nor the distance as measured by graph distance within the original Kagome lattice (which does not accurately capture the significance of the different interactions terms as one changes $\beta$).
Instead, we perform the interaction cut-off as a function of the lattice distance between \emph{unit-cells}.
More specifically, we consider the interaction cut-off R1,R2, and R3 to incorporate all spin-spin interactions across nearest, next-nearest, and next-next-nearest unit cells.

\begin{figure}[h]
    \centering
    \includegraphics[width=0.75\linewidth]{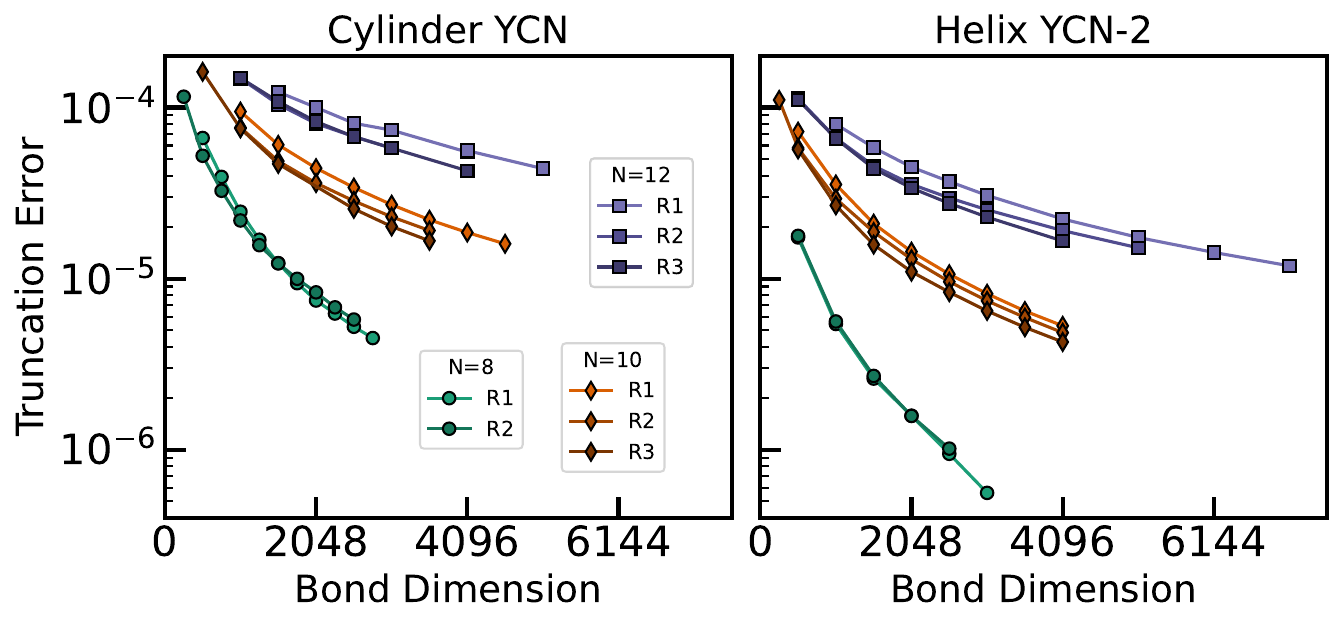}
    \caption{Truncation error as function of bond dimension for the different geometries (YCN and YCN-2) as well as different interaction cutoffs (R1,R2, and R3).}
    \label{fig:ErrorVsTruncation}
\end{figure}

Summarized in Figure~\ref{fig:ErrorVsTruncation}, we highlight the evolution of the truncation error with increasing bond dimension across these two choices. 
We observe similar levels of truncation error when varying the interaction cutoff across different geometries, while wider cylinders exhibit a much slower decay of truncation error (as expected from the exponentially large representation with increasing cylinder width).
For the YCN-2 geometry, the error is well below or within the $10^{-5}$ regime, while for the more computationally costly YCN geometry, we achieve this small error for $N=8,10$ with $N=12$ exhibiting truncation error closer to $4\times 10^{-5}$.

\begin{figure}[t]
    \centering
    \includegraphics[width=0.75\linewidth]{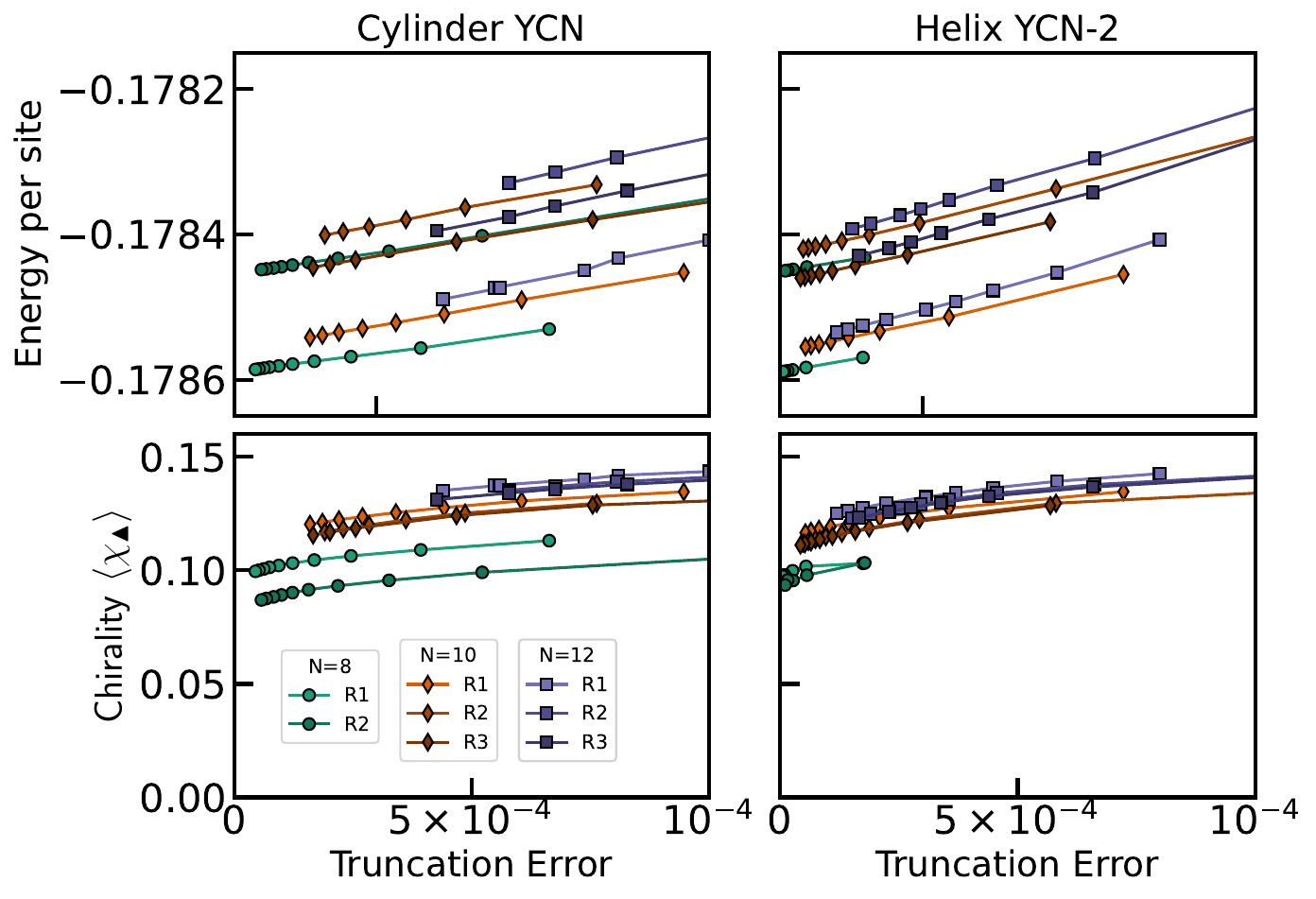}
    \caption{Convergence of energy density and chirality with respect to truncation error for varying geometry and interaction cutoffs. 
    For small truncation error, we observe an approximate linear dependence that highlights the robustness of our numerical approach and, consequently, the robustness of the time reversal symmetry breaking characteristic of the CSL.
    }
    \label{fig:ConvergenceAndRadii}
\end{figure}

To better understand the convergence properties of the iDMRG calculation, we directly consider the convergence of both the energy density and the chirality as a function of the truncation error, Figure~\ref{fig:ConvergenceAndRadii}.
For both observables, across geometry and interaction cutoff, we observe similar trends with the convergence approaching a linear regime for small ($\lesssim 3\times 10^{-5}$) truncation error.
Crucially, the observed chirality approaches a sizeable, non-zero value, highlighting robust time reversal symmetry breaking.
We note that with increasing system size, we observe a further stabilization of the TRS-breaking behavior, with the magnitude of the chirality $\langle \chi_{\blacktriangle}\rangle$ increasing with cylinder width $L_y = N/2$.

\section{Breathing Phase Diagram}

In the large breathing limit, we consider two different operators to capture different possible symmetry breaking patterns, notably different ways that the rotation symmetry can be broken.
The first is an operator that only acts \textit{within} each triangle, and captures breaking of rotation symmetry:
\begin{align}
    O^{\text{(rot)}}_{\blacktriangle} = &\left[ S_{\blacktriangle,a}^zS_{\blacktriangle,b}^z + S_{\blacktriangle,b}^zS_{\blacktriangle,c}^z -  2S_{\blacktriangle,a}^zS_{\blacktriangle,c}^z \right]~.
\end{align}
Indeed, this operator is related to $\eta_y$ (the $\hat{z}$-component of those correlations).
In general, our iDMRG tensor network breaks the rotation symmetry, ensuring that the correlations are not always exactly rotationally symmetric. 
To this end, and given that the patterns of correlations observed break both rotation and translation symmetries, we consider the \emph{difference} between $O^{\text{(rot)}}_{\blacktriangle}$ across triangles separated by $\vec{a}_1$. This defines our observable that captures the onset of the different VBS phases:
\begin{equation}
    O^{\text{rot}}  = \langle O^{\text{(rot)}}_{\blacktriangle_{\vec{r}}} \rangle - \langle O^{\text{(rot)}}_{\blacktriangle_{\vec{r}+\vec{a}_1}} \rangle 
\end{equation}

To differentiate between the VBS and the Glider VBS phases, we need an observable that is sensitive to the breaking of the Glider symmetry. 
To this end, we consider the glider symmetry highlighted in the main text, we have that the pair of unit-cells at $(\vec{r}, \vec{r}+\vec{a}_2 - \vec{a}_1)$ gets mapped to unit cells $(\vec{r'}, \vec{r'}-\vec{a}_1)$, where $\vec{r}$ and $\vec{r}'$ are related by the glide operation.
When the unit cell is near the glide plane (on the left side), we have $\vec{r}' = \vec{r} + \vec{a}_1$, which implies that the following pairs of triangles develop the same correlations:
\begin{equation}
(\vec{r}, \vec{r}+\vec{a}_2 - \vec{a}_1) \quad \text{and} \quad (\vec{r}, \vec{r}+\vec{a}_1)
\end{equation}
As such we can construct the following operator that is sensitive to the breaking of the glider symmetry:
\begin{align}
    O^{(\text{glid})}_{\vec{r}}= & \left[ \left(\sum_i S_{\blacktriangle{\vec{r}},i}^z\right) \left(\sum_i S_{\blacktriangle{\vec{r}+\vec{a}_1},i}^z\right) - \left(\sum_i S_{\blacktriangle{\vec{r}},i}^z\right) \left(\sum_i S_{\blacktriangle{\vec{r}+\vec{a}_2-\vec{a}_1},i}^z \right) \right] 
\end{align}
where the spin indices $a,b,c$ as well as the basis vectors $\vec{a}_j$ are defined in Fig.~$1$ in the main text.

Notably, if the system has broken translation symmetry but preserved glider symmetry, then there are two possible configurations of the glider symmetric correlations, seperated by $\vec{a}_1$.
Since each configuration is equally likely when performing iDMRG, we consider as a witness of the glider symmetry breaking the combination of the $O^{(\text{glid})}_{\vec{r}}$ across these two sites.
The resulting order parameter is given by:
\begin{equation}
 O^{\text{glid}} = \sqrt{ \langle O^{(\text{glid})}_{\vec{r}}\rangle^2  + \langle O^{(\text{glid})}_{\vec{r}+\vec{a}_1}\rangle^2}
\end{equation}

When $\beta \gtrsim 2.5$ at large breathings, we find that the VBS phase breaks both translation and rotational symmetry, and both $O^{\text{glid}}$ and $O^{\text{rot}}$ are robustly non-zero.

At slightly smaller breathings, $2.1 < \beta <2.5$, we observe the intermediate Glider-VBS regime where $O^{\text{rot}}$ remains non-zero, yet $O^{\text{glid}}$ becomes zero. 
This signals the presence of a different translation symmetry breaking state, that remains symmetric under a glide transformation.

\begin{figure}[t]
    \centering
    \includegraphics[width=1\linewidth]{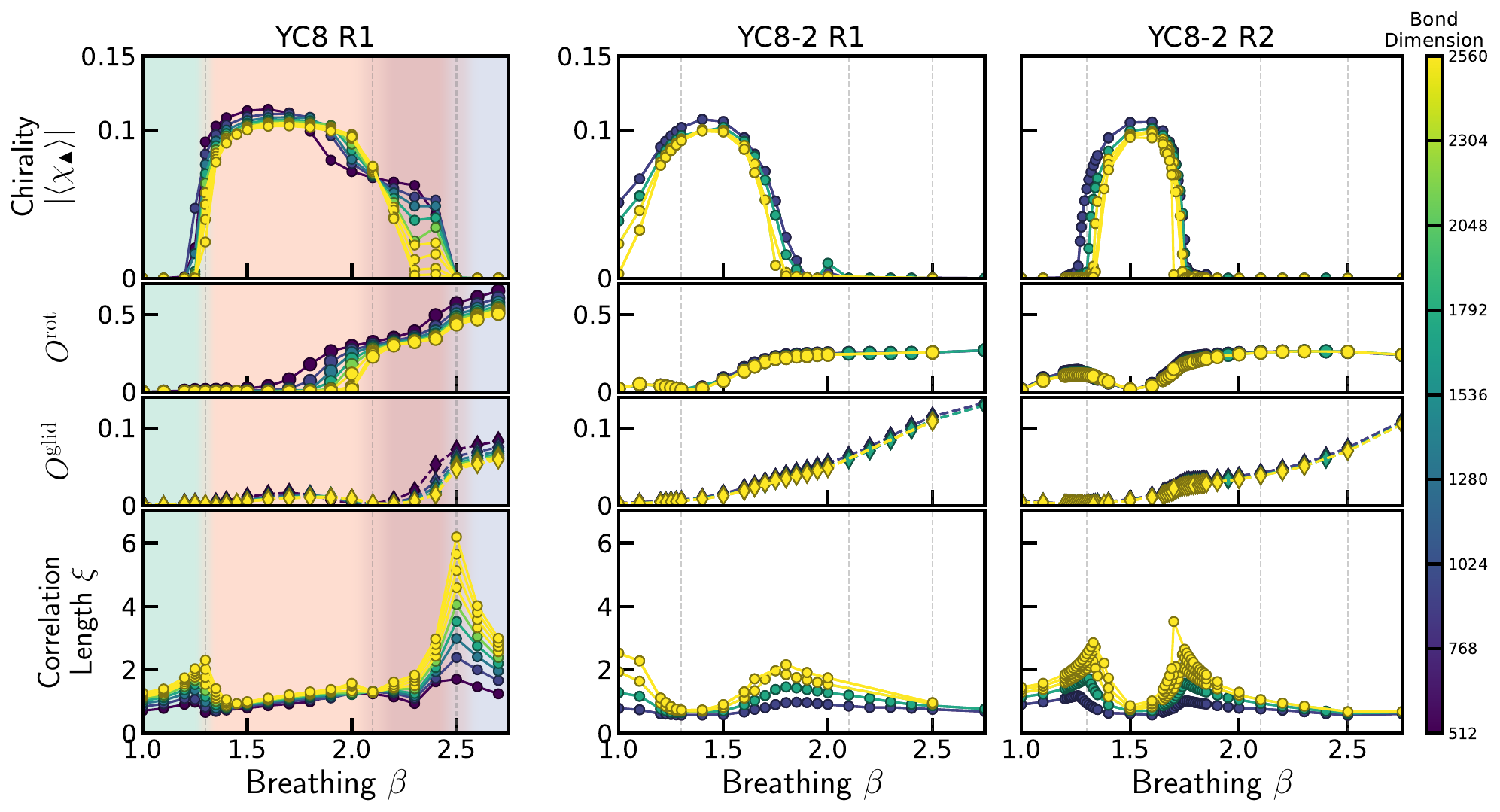}
    \caption{Phase diagram as a function of breathing $\beta$ for different geometry and interaction cutoff. 
    In the YCN-2 geometry, we observe that $O^{\text{rot}}$ and $O^{\text{glid}}$ can develop a non-zero expectation value even within the CSL phase. 
    While the origin of this behavior is not fully clear, the explicit breaking of mirror symmetry in the YCN-2 compactification may enable the iDMRG to converge to a symmetry broken state more robustly than in the YCN compactification. For this reason, in the main text we only consider the YCN compactification.}
    \label{fig:breathingDiagram_Helix}
\end{figure}

The nature of the transition between the Glider-VBS and the CSL is difficult to determine due to the boundary conditions and range of interactions in our numerics.
In the main text, we focused on a YC8 cylinder geometry with periodic boundary conditions and interactions truncated to the first ring of triangles (R1).
When considering the helix geometry, the CSL phase is surrounded by two continuous transitions, and the phase diagram appears to exhibit only a single VBS phase at large $\beta$, Figure~\ref{fig:breathingDiagram_Helix}.
We believe this arises from the reduced symmetry in the helix (YCN-2) geometry, which allows the state to smoothly connect to a spatial symmetry broken state. The only transition is out of the topologically ordered CSL state which appears to be continuous.

\section{Emergent SU(2) symmetry in the entanglement spectrum}

\begin{figure}[t]
    \centering
    \includegraphics[width=1\linewidth]{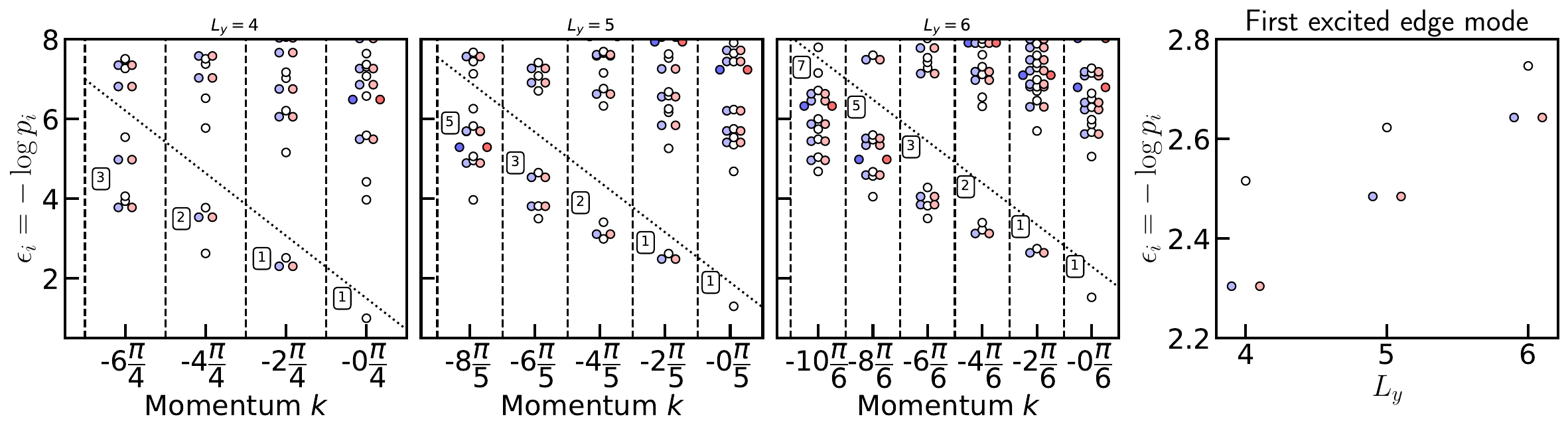}
    \caption{Entanglement spectrum for different YC($2L_y$). Across all system sizes we observe the gapless mode, with the number of modes characteristic of a SU(2)$_1$ theory. Notably, while the underlying model has only U(1) symmetry, we observe evidence of an emergent SU(2) symmetry of the edge mode---with increasing system size, the gap between the modes with different spin quantum number reduces.}
    \label{fig:convergenceofEmergentSU(2)}
\end{figure}

While our model strongly breaks SU(2) spin rotation symmetry, we observe that the entanglement spectrum exhibits an approximate SU(2) symmetry.
For the same momentum, modes with different spin quantum number are close to one another, approaching the spectrum in SU(2) symmetric models considered in the literature~\cite{bauer_chiral_2014, hu:2015, Kuhlenkamp2024}.
Notably, this symmetry improves with increasing system size (Figure~\ref{fig:convergenceofEmergentSU(2)}), suggesting that the edge mode becomes SU(2) symmetric in the thermodynamic limit.

\section{Extracting the $h_x$ transition location}

When adding a global $\hat{x}$ field, the transition out of the CSL is of a first-order nature. 
This field breaks the original U(1) spin rotation symmetry so we expect the system to transition into a state with non-zero $\expval{S_x}$.
Indeed, this is what we observe, with the system undergoing a discontinuous jump in the magnetization value [see Fig.~6(b) in the main text].

\begin{figure}
    \centering
    \includegraphics[width=\linewidth]{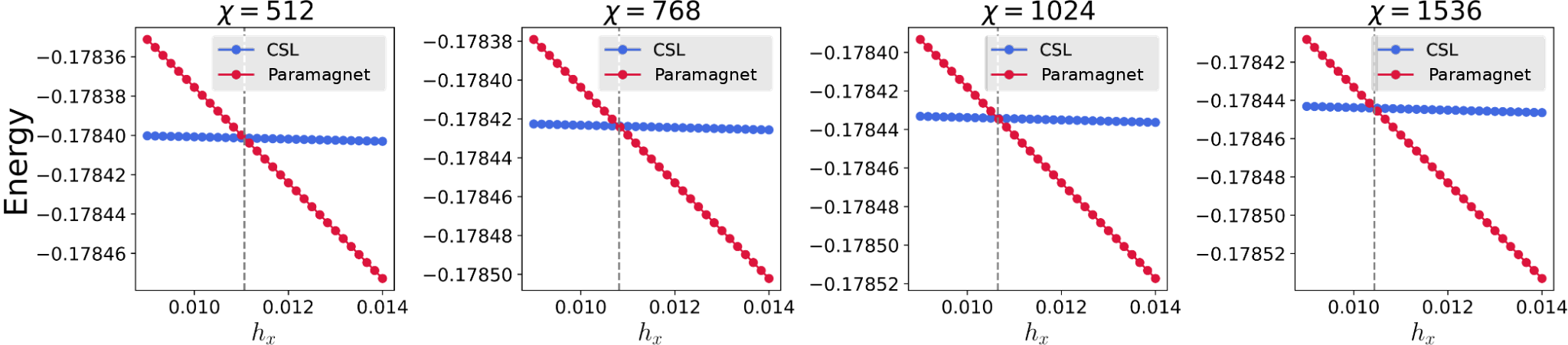}
    \caption{Energy as a function of the field for the CSL and paramagnetic state for different bond dimension of the MPS. We observe a modest shift of the crossing energy, which we utilize to extrapolate the location of the transition in the large bond dimension limit, Fig.~\ref{fig:supp_hx_FSS}.}
    \label{fig:supp_hx_energy}

    \centering
    \includegraphics[width=0.5\linewidth]{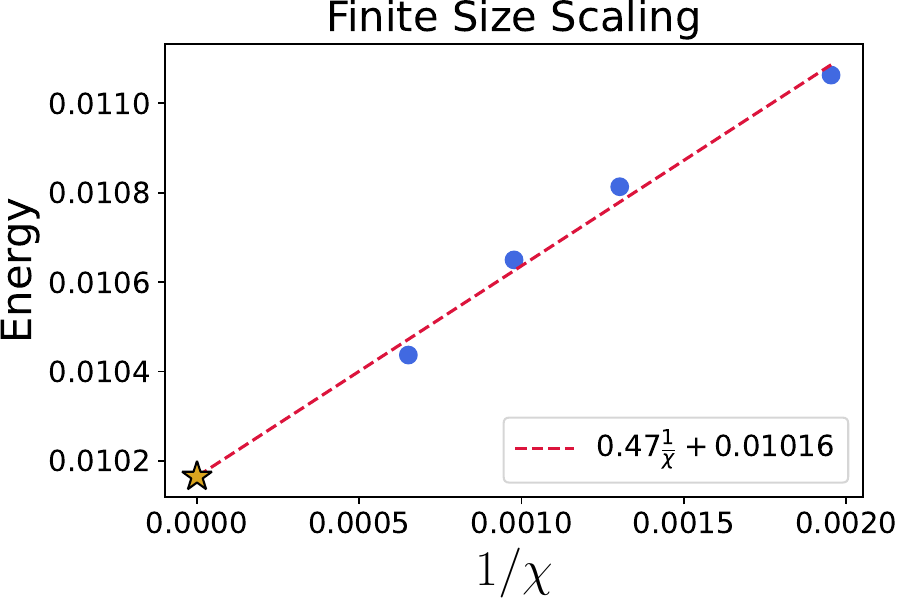}
    \caption{Finite size scaling for the $h_x$ transition location, demonstrating that the CSL is stable under the small but non-zero applied field.}
    \label{fig:supp_hx_FSS}
\end{figure}

To extract the location of this transition, we take two states: the identified CSL state at $h_x = 0.009$ and the paramagnetic state at $h_x = 0.014$, and compute the energy of each with varying $h_x$ strengths. A hallmark of a first-order phase transition is an energy crossing, and in Fig.~\ref{fig:supp_hx_energy}, we observe a crossing of the energies at a particular value of $h_x$. We then extract this crossing a function of bond dimension and perform a finite-size scaling analysis to approximate the true location of the transition at $h_x^{crit} = 0.0102$ (star in Fig.~\ref{fig:supp_hx_FSS}). 

Importantly, we note that the estimated transition does \textit{not} go to zero as bond dimension is increased, but rather crosses at a finite, non-zero value, corroborating the existence of a finite width CSL phase away from the $h_x = 0.0$ point, as expected from the topological nature of the CSL state, which is robust to any (small) perturbation.

\section{Computation of the Modular matrices}

In this section, we provide more detail on the Monte Carlo sampling used to compute the modular matrices and also comment on the finite size scaling performed for the central charge. 
The modular matrices $\mathcal{U}$ and $\mathcal{S}$ correspond to the generators of modular transformations, which map the torus to itself.
On a lattice specified by the basis vectors $\{\vec{a}_1,\vec{a}_2\}$, these transformations act by integer linear
recombinations of the basis vectors that preserve the oriented area (and hence the boundary-condition identifications).

Because any transformation that leaves the lattice constant can be recast in terms of $\mathcal{U}$ and $\mathcal{S}$, we chose to consider a $2\pi/3$ rotation, which is a symmetry of the lattice and the topological state.
Writing the basis vectors in terms of exponential numbers ($\vec{b_j} \to w_j = x_j + iy_j$), the action of the modular matrices can be simply recast into two simple linear transformation:
$$
\mathcal{S} : \begin{bmatrix} w_1\\w_2 \end{bmatrix} \to \begin{bmatrix} w_2\\-w_1 \end{bmatrix} = \mqty[0 & 1 \\-1 & 0] \mqty[w_1\\w_2]\quad, \quad \mathcal{U}: \begin{bmatrix} w_1\\w_2 \end{bmatrix} \to \begin{bmatrix} w_1+w_2\\w_2 \end{bmatrix}= \mqty[1 & 1 \\0 & 1] \mqty[w_1\\w_2]
$$
At the same time, the rotation corresponds to:
$$
\mathcal{R}_{2\pi/3} : \begin{bmatrix} w_1\\w_2 \end{bmatrix} \to \begin{bmatrix} w_2-w_1\\-w_1 \end{bmatrix} = \mqty[-1 & 1 \\-1 & 0] \mqty[w_1\\w_2] = \mqty[1 & 1 \\0 & 1] \mqty[0 & 1 \\-1 & 0] \mqty[w_1\\w_2] = u s \mqty[w_1\\w_2]
$$
where $u$ and $s$ are the linear representation of the $\mathcal{U}$ and $\mathcal{S}$ transformations.
This implies that $\mathcal{R}_{2\pi/3} = \mathcal{U}\mathcal{S}$~\cite{zhang:2012}.
Expanding in terms of the action of the $\mathcal{U}$ and $\mathcal{S}$ into the ground state manifold:
\begin{align}
  &\mathcal{U} = e^{-i \frac{2 \pi}{24} c} \mqty(\theta_1 & 0 \\ 0 & \theta_s),\quad \mathcal{S} = \mqty(S_{11} & S_{1s} \\ S_{s1} & S_{ss}) \\
  &\Rightarrow R_{2\pi/3} =  e^{-i \frac{2 \pi}{24} c} \mqty(e^{-i\phi_1} & 0\\ 0 & e^{-i\phi_2})\mqty(\theta_1 & 0 \\ 0 & \theta_s)\mqty(S_{11} & S_{1s} \\ S_{s1} & S_{ss})
\mqty(e^{i\phi_1} & 0\\ 0 & e^{i\phi_2})\\
&\boxed{R_{2\pi/3} = e^{-i \frac{2 \pi}{24} c} \mqty(S_{11} & S_{1s} e^{-i(\phi_1-\phi_2)}\\ \theta_s S_{s1} e^{i(\phi_1-\phi_2)} & \theta_s S_{ss})} = \begin{bmatrix} V_{11} & V_{1s} \\ V_{s1} & V_{ss} \end{bmatrix}\label{eq:RTransformation}
\end{align}

The actions of the modular matrices encode the mutual and self statistics of the anyons in the model. 
Entries $S_{ij}$ correspond to the phase picked up by anyon $i$ being encircled with anyon of type $j$, $c$ is the central charge, and $\theta_i$ is the phase picked up by swapping an anyon with another of the same type. 
We simplified the above expression by using the fact that $\theta_1 = 1$ for the vacuum topological sector. 

Importantly, the form of Eq.~\ref{eq:RTransformation} allows the extraction of all the self and mutual statistics as well as the quantum dimension of the state~\cite{cincio:2013}. The final step is the actual computation of the overlaps under the spatial rotation of the wavefunction.

\begin{figure}[t]
  \centering
  \includegraphics[width = 0.6\textwidth]{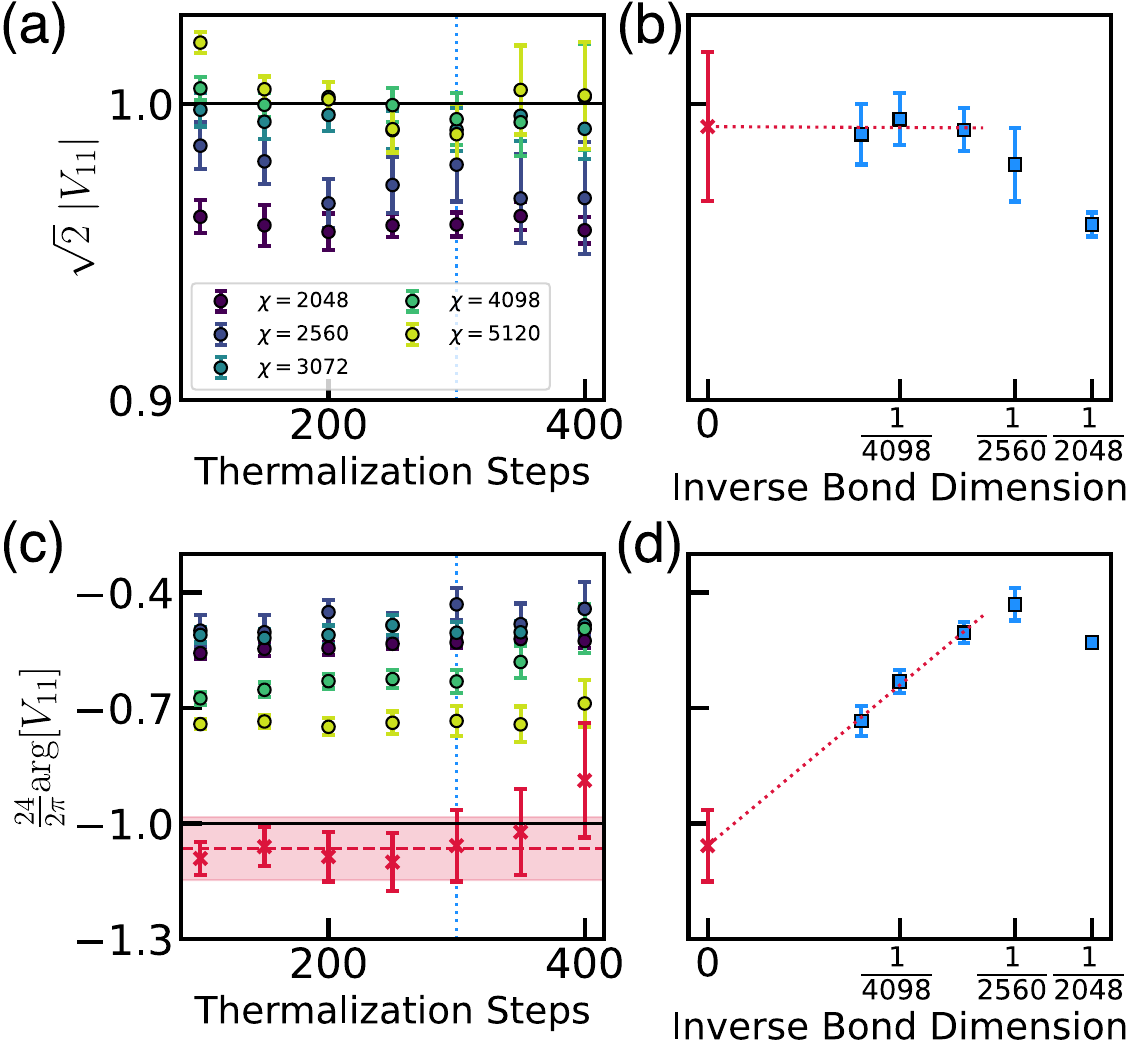}
  \caption{  Convergence of the modular matrix diagonal term $V_{11}$ with increasing bond dimension. {\bf (a-b)} For large bond dimension, $\chi\gtrsim 2560$, we observe an accurate calculation of magnitude of $V_{11}$ provided the number of initial thermalization steps is large enough ($\gtrsim 150$).
  {\bf (c-d)} By increasing the bond dimension of the iMPS, we observe a convergence of the central charge of the theory to its expected value of $c=-1$, provided we perform an inverse bond dimension extrapolation. This observation is robust to the number of thermalization steps considered in our Monte Carlo.}
  \label{fig:sup_modularMatrix}
\end{figure}

We follow the prescription described in~\cite{cincio:2013}. From our iDMRG, we obtain an infinite wave function, which can be used to compute the torus wave function~\cite{zaletel_flux_2014}. Building the wavefunction on a $L\times L$ torus then enables us to compute $R_{2\pi/3}$---a rotation is obtained by relabeling the sites of the MPS and then computing the overlap between it and the unrotated wavefunction.
Unfortunately, this overlap is, in general, exponentially hard to compute since the contraction cannot be simplified. To overcome this challenge we recast the problem of computing the overlap into a Monte Carlo sampling problem~\cite{cincio:2013}:
\begin{align}
  \langle \psi_A | R_{2\pi/3} |\psi_B\rangle &= \sum_{\sigma_A, \sigma_B} \langle \sigma_A | R_{2\pi/3} | \sigma_B \rangle c^*_{\sigma_A} c_{\sigma_B} = \sum_{\sigma_B}  \sum_{\sigma_A} \delta \left(\sigma_A = R_{2\pi/3}\sigma_B\right) c^*_{A,\sigma_A} c_{B,\sigma_B} \notag\\
  &= \sum_{\sigma} c^*_{A, R_{2\pi/3}\sigma} c_{B,\sigma} = \sum_{\sigma} \frac{c^*_{A, R_{2\pi/3}\sigma}}{c_{B,\sigma}^*} c_{B,\sigma}^*c_{B,\sigma} = \sum_{\sigma} \frac{c^*_{A, R_{2\pi/3}\sigma}}{c_{B,\sigma}^*} P_{B,\sigma}
\end{align}
where $\sigma$ corresponds to a simple product state configuration and $c_{i,\sigma} = \langle \sigma | \psi_i\rangle$ is the value of the wavefunction. Since $P_{B,\sigma} = |c_{B,\sigma}|^2$ corresponds to a probability distribution over the space of configurations, the matrix element can be easily computable using a Monte Carlo procedure, leveraging the Metropolis algorithm for generating and accepting new configurations $\sigma \to \sigma'$.
At the same time, this also enables the calculation to be parallelizable. 

We compute the modular matrices in a $6\times 6$ torus, using the wavefunctions resulting from our YC12 geometry.
We begin by focusing on the upper left matrix element $R_{2\pi/3, 11}$ as it encodes information of both the quantum dimension of the ground state as well as the central charge.

In Fig.~\ref{fig:sup_modularMatrix}(a-b), we see that the magnitude of $\langle \psi_A | R_{2\pi/3} |\psi_B\rangle_{11}$, which should encode $S_{11}$, approaches 1 as we increase the number of thermalization steps and increase the bond dimension. 
When plotted on an inverse bond dimension scale, we see that the value plateaus and predict a finite size scaled value that is consistent with $|c| = 1$.

In Fig.~\ref{fig:sup_modularMatrix}(c-d), the complex argument encodes the central charge of the model, which is $\pm 1$. While the value of the central charge at our available bond dimensions does not reach 1, from doing another finite size scaling at each thermalization step, we see that the value also approaches $-1$ within error bars.

\section{Defining the Two Spin model}

The goal of this section is to summarize and provide more details on the idea behind the two-spin model. This effective model arises from the observation that, in the large breathing limit, the low energy manifold of each triangle forms a 4-fold degenerate ground state which can be encoded with two spin-$\frac{1}{2}$ degrees of freedom.

Our starting point is a single isosceles triangle composed of three spin 1/2 interacting via an XY coupling.
The spins are placed a distance $a$ from one another and interact with energy $J$.
We will use a normalization $a=J=\hbar = 1$.

The triangle's Hamiltonian is given by:
\begin{align}
    H_\blacktriangle &= J \sum_{i<j} (\sigma_x^i\sigma_x^j + \sigma_y^i\sigma_y^j) = J \sum_{\alpha\in\{x,y\}} \left\{\frac{1}{2} \left[\sum_i \sigma^\alpha_i \right]\left[\sum_j \sigma^\alpha_j \right] - \frac{3}{2}\right\}\\
    &=\frac{J}{2} \left( 4[S^{\text{tot}}_\blacktriangle]_x [S^{
    \text{tot}}_\blacktriangle]_x + 4[S^{\text{tot}}_\blacktriangle]_y [S^{\text{tot}}_\blacktriangle]_y - 3\right) = \\ 
    &=\boxed{ 2J\left( [S^{\text{tot}}_\blacktriangle]^2 - [S^{\text{tot}}_\blacktriangle]_z^2- \frac{3}{4}\right) }
\end{align}
Where $[S^{\text{tot}}_\blacktriangle]_\alpha = \frac{1}{2}\sum_i \sigma^\alpha_i$. For the rest of this section, we will disregard the constant factor of $3J/2$.

As a result, the level spectrum of the triangle is best understood in terms of the total spin of the full system, with the spectrum taking a very simple form.
From spin addition rules, we can also get the degeneracy of the different states and their energies:
$$
\frac{1}{2} \otimes \frac{1}{2} \otimes\frac{1}{2} = \frac{3}{2} \oplus \frac{1}{2} \oplus \frac{1}{2} \quad \begin{cases}
    \frac{3}{2} \Rightarrow & \frac{15J}{2} - 2Jm_z^2 \in \{3J, 7J\}\\
    \frac{1}{2} \Rightarrow & \frac{3J}{2} - 2m_z^2 \in \{J\}\\
\end{cases}
$$

The lowest states correspond to the two spin-$\frac{1}{2}$ manifold, where all 4 states are fully degenerate.

We can explicitly write down these states as follows:
\begin{align}
|\chi_+, \frac{1}{2}\rangle &= \frac{i}{\sqrt{3}} \left[ |\uparrow \uparrow \downarrow\rangle + \omega  |\downarrow \uparrow \uparrow\rangle + \omega^\dag | \uparrow \downarrow \uparrow\rangle\right]\\
|\chi_+, -\frac{1}{2}\rangle &= \frac{i}{\sqrt{3}} \left[ |\downarrow \downarrow \uparrow\rangle + \omega  |\uparrow \downarrow \downarrow\rangle + \omega^\dag | \downarrow \uparrow \downarrow\rangle\right]\\
|\chi_-, \frac{1}{2}\rangle &= \frac{1}{\sqrt{3}} \left[ |\uparrow \uparrow \downarrow\rangle + \omega^\dag  |\downarrow \uparrow \uparrow\rangle + \omega | \uparrow \downarrow \uparrow\rangle\right]\\
|\chi_-, -\frac{1}{2}\rangle &= \frac{1}{\sqrt{3}} \left[ |\downarrow \downarrow \uparrow\rangle + \omega^\dag  |\uparrow \downarrow \downarrow\rangle + \omega| \downarrow \uparrow \downarrow\rangle\right]
\end{align}
with $\omega = e^{i 2\pi/3}$ and where the order of the spins are chosen to be anti-clockwise around the triangle.

First, the choice of the phase $\omega$ ensures that states are orthogonal to the $m_z= \pm 1/2$ with $S_{\text{tot}}=3/2$, which correspond to permutation symmetric combinations of the spin configurations:
$$
|3/2,1/2\rangle = \frac{1}{\sqrt{3}} \left[ |\uparrow \uparrow \downarrow\rangle + |\downarrow \uparrow \uparrow\rangle +  | \uparrow \downarrow \uparrow\rangle\right]\quad \text{and} \quad
|3/2,-1/2\rangle = \frac{1}{\sqrt{3}} \left[ |\downarrow \downarrow \uparrow\rangle +  |\uparrow \downarrow \downarrow\rangle + | \downarrow \uparrow \downarrow\rangle\right]
$$

Second, these states have well defined chirality $\vec{S}_i\cdot(\vec{S}_j \times \vec{S}_k)$ which can be written as:

\begin{align}
    \vec{S}_i\cdot(\vec{S}_j \times \vec{S}_k) = \frac{1}{8} \epsilon_{ijk} \sigma^x_i\sigma^y_j\sigma^z_k = \frac{1}{8} \epsilon_{ijk} \frac{1}{i}(\sigma^+_i+\sigma^-_i)(\sigma^+_j-\sigma^-_j)\sigma^z_k = \frac{-i}{4} \epsilon_{ijk} \sigma^+_i\sigma^-_j\sigma^z_k
\end{align}

Using this form, it becomes easy to check  $|\chi_+, \frac{1}{2}\rangle$ is an eigenstate of the chirality operator (the order of the operators matches the order of the spins in the ket): 
\begin{align}
    &\frac{-i}{4} \epsilon_{ijk} \sigma^+_i\sigma^-_j\sigma^z_k |\chi_+, \frac{1}{2}\rangle \\
    &= \frac{i}{4\sqrt{3}} \left[( \sigma^-\sigma^z \sigma^+ - \sigma^z\sigma^- \sigma^+) |\uparrow\uparrow\downarrow\rangle + \omega( \sigma^+\sigma^-\sigma^z  - \sigma^+\sigma^z\sigma^- ) |\downarrow\uparrow\uparrow\rangle) + \omega^\dag ( \sigma^z\sigma^+\sigma^-  - \sigma^-\sigma^+\sigma^z )|\uparrow\downarrow\uparrow\rangle\right]\\
    &=\frac{-i}{4\sqrt{3}} \left[ (\omega^\dag-\omega) |\uparrow\uparrow\downarrow\rangle + (1-\omega^\dag) |\downarrow\uparrow\uparrow\rangle +  (\omega-1) |\uparrow\downarrow\uparrow\rangle\right]\\
    &=\frac{-i}{4\sqrt{3}} (\omega^\dag-\omega) \left[  |\uparrow\uparrow\downarrow\rangle + \omega|\downarrow\uparrow\uparrow\rangle +  \omega^\dag |\uparrow\downarrow\uparrow\rangle \right]= \boxed{\frac{\sqrt{3}}{4}|\chi_+, \frac{1}{2}\rangle}
\end{align}

Analogous calculations hold for the other states.
Within each $m_z$ subspace, we can then identify $\vec{S}_i\cdot(\vec{S}_j \times \vec{S}_k) = \frac{4}{\sqrt{3}} \eta^z$, where $\eta^z$, or the \textit{chirality} of the triangle, is the first of the Pauli operators used to define a new spin-1/2-like degree of freedom (we drop the $\blacktriangle$ notation for simplicity).

The $\hat{\eta}^{x/y}$ can then be defined in terms of the transitions between states with positive and negative chirality at fixed $m_z$.
This approach is cumbersome and not particularly enlightening; instead we can find the correct form of the operator by making an informed guess and confirming our result.

From the discussion so far, we already know some properties of this low-energy subspace and so we can infer properties on $\eta^{x/y}$. In particular, both operators must flip the chirality, without modifying the total $m_z$ or total spin quantum numbers; $\eta^{x/y}$ must be sums of spin rotation invariant terms:
\begin{equation}
  \eta^{x/y} = \alpha (\vec{S}_0\cdot \vec{S}_1) + \beta (\vec{S}_1\cdot \vec{S}_2) + \gamma (\vec{S}_0\cdot \vec{S}_2)
\end{equation}
Finally, we can note that $\alpha=\beta=\gamma$ is exactly the Hamiltonian within the low-energy subspace (up to constant factors), so the $\eta^{x/y}$ have to be orthogonal (i.e. $\alpha+\beta+\gamma =0$).
This gives us two independent solutions:
\begin{align}
  \eta^{x} &\propto (\vec{S}_1\cdot \vec{S}_2) - (\vec{S}_0\cdot \vec{S}_2) = \frac{\sigma^z_1\sigma^z_2 + 2(\sigma^+_1\sigma^-_2+\sigma^-_1\sigma^+)}{4} - \frac{\sigma^z_0\sigma^z_2 + 2(\sigma^+_0\sigma^-_2+\sigma^-_0\sigma^+_2)}{4} \\
  \eta^y &\propto (\vec{S}_1\cdot \vec{S}_2) + (\vec{S}_0\cdot \vec{S}_2) - 2(\vec{S}_0\cdot \vec{S}_1)
\end{align}

We can explicitly carry out the calculation for $\eta^x$:
\begin{align}
  &\left[(\vec{S}_1\cdot \vec{S}_2) - (\vec{S}_0\cdot \vec{S}_2)\right]\ket{\chi_+, \frac{1}{2}} =\\
  &= \left[\frac{\sigma^z_1\sigma^z_2 + 2(\sigma^+_1\sigma^-_2+\sigma^-_1\sigma^+)}{4} - \frac{\sigma^z_0\sigma^z_2 + 2(\sigma^+_0\sigma^-_2+\sigma^-_0\sigma^+_2)}{4}\right]\frac{i}{\sqrt{3}} \left[ |\uparrow \uparrow \downarrow\rangle + \omega  |\downarrow \uparrow \uparrow\rangle + \omega^\dag | \uparrow \downarrow \uparrow\rangle\right]\\
  &= \frac{i}{4\sqrt{3}}  \left\{\left[ -1+1 + 2\omega^\dag -2\omega \right] |\uparrow \uparrow \downarrow\rangle +   \left(\omega + \omega- 2\right)|\downarrow \uparrow \uparrow\rangle + \left[-\omega^\dag - \omega^\dag + 2\right] | \uparrow \downarrow \uparrow\rangle\right\}\\
  &= \frac{i}{4\sqrt{3}}  \left\{\left[  -2i\sqrt{3}  \right] |\uparrow \uparrow \downarrow\rangle +   2\left(\omega - 1\right)|\downarrow \uparrow \uparrow\rangle + 2\left[1-\omega^\dag\right] | \uparrow \downarrow \uparrow\rangle\right\}\\
  &= \frac{1}{2}  \left\{|\uparrow \uparrow \downarrow\rangle + \omega^\dag|\downarrow \uparrow \uparrow\rangle + \omega | \uparrow \downarrow \uparrow\rangle\right\}  = \frac{\sqrt{3}}{2} \ket{\chi_-, \frac{1}{2} } \\
\end{align}

A similar calculation can be done for $\eta^y$:
The final set of operators are then:
\begin{align*}
  \eta^x &= \frac{2}{\sqrt{3}} \left[(\vec{S}_1\cdot \vec{S}_2) - (\vec{S}_0\cdot \vec{S}_2)\right]\\
  \eta^y &= 2\left\{(\vec{S}_1\cdot \vec{S}_2) + (\vec{S}_0\cdot \vec{S}_2) - 2(\vec{S}_0\cdot \vec{S}_1)\right\}\\
  \eta^z &= \frac{4}{\sqrt{3}}   \vec{S}_i\cdot(\vec{S}_j \times \vec{S}_k)
\end{align*}

\section{Developing interactions between this reduced Hilbert Space}

Having developed the translation between the original degrees of freedom and the low-energy manifold of each triangle, we now develop the theory of the effective Hamiltonian $H_{\eff}$ that governs these degrees of freedom.

In mapping the breathed Kagome lattice of the spin to our effective low-energy description, the two-spins now live on a triangular lattice with unit length $2\beta$, where $\beta$ is the breathing of the system.

\subsection{Symmetry constrains the allowed terms}

The first important consideration is how the original symmetries of the system translate to these effective degrees of freedom, as that immediately constrains the terms of $H_{\eff}$.

\begin{itemize}
    \item {\bf U(1) Spin Rotation}: This symmetry makes the total $S^z$ a good quantum number, action of this symmetry rotates $S^x$ and $S^y$ into one another.
    In the original d.o.fs it prevents terms proportional to $S^x$ or $S^y$; since $s^x$ and $s^y$ corresponds to similar operators in a restricted Hilbert space, the same constraints apply. 
    \item {\bf Spin Reversal symmetry} $S^z \leftrightarrow -S^z$: Generated by $\prod_i \sigma^x_i$ (or $\prod_i \sigma^y_i$), this symmetry prevents single terms proportional to $S^z$.
    In the effective Hilbert space, it takes $s^z\leftrightarrow -s^z$, but also one of the other spin rotation terms to minus itself. 
    As a result, there are either Ising-like $s^z s^z$ with an even number of operators, or spin rotation combinations of $s^x,s^y,s^z$ (like the chirality) or their product. 
  \item {\bf Time Reversal Symmetry} (TRS): Since $s_\alpha$ refer to the total spin operators in the $S_{\text{tot}}^2 = 3/4$ subspace, TRS still maps $s_\alpha$ to $-s_\alpha$. On the $\eta_\alpha$ spin, TRS does not affect $\eta_{x/y}$ but reverses $\eta^z$ (the chirality).
    To this end, each term of the Hamiltonian has to be composed of an even number of $s_\alpha,\eta^z$ operators.

\noindent{{\bf Note} that the constraints above already preclude any single-site term (either $s^\alpha$ or $\eta^\alpha$) from emerging in our effective Hamiltonian.}
    
  \item {\bf Lattice symmetries ($C_{3v}$ rotations and reflections)}: Lattice transformations do not affect $s_\alpha$, but affect $\eta_{x/y}$. In particular, $\eta_{x,y}$ transform like the two-dimensional irreducible representation of $C_{3v}$ (i.e. how two dimensional vectors rotate under $z$-rotations).
    Analogously, we find that $\eta_{x/y}$ is odd/even under vertical reflections.
    These conditions imply that single $\eta_{x/y}$ are forbidden by symmetry, but quadratic terms or bond-dependent terms are allowed.

    Having built these considerations we can now write down how the interactions between two triangles are related to one another when doing the lattice transformations:
    \begin{equation}
    h_{R\vec{a},R\vec{b}} = \hat{R}h_{\vec{a},\vec{b}}\hat{R}^\dag
    \end{equation}
    where $R$ is the lattice transformation on the positions of the triangles and $\hat{R}$ is the corresponding transformation on the Hilbert space of the system. $\hat{R}$ does not just relabel the spins, but also acts on the operators. In the table below we summarize the action of the different relevant transformations.
    Different terms form either one or two dimensional irreducible representations, which connects how the coefficients of the different interactions terms relate to one another.
    \begin{table}
      \begin{tabular}{|c|c|c|}
      \hline
        Operator & $\hat{R} = R_{2\pi/3}$ Rotation by $2\pi/3$ & $\hat{R} = \sigma_v$ Vertical Reflection\\ \hline \hline
        $\eta^x$ & $(-\eta^x + \sqrt{3}\eta^y)/2$ & $-\eta^x$\\
        $\eta^y$ & $(-\eta^y - \sqrt{3}\eta^x)/2$ & $\eta^y$\\
        \hline
        $\eta^z$ & $\eta^z$ & $\eta^z$\\\hline\hline
        $\eta^x_\blacktriangle\eta^x_{\blacktriangle'} + \eta^y_\blacktriangle\eta^y_{\blacktriangle'}$ & $\eta^x_\blacktriangle\eta^x_{\blacktriangle'} + \eta^y_\blacktriangle\eta^y_{\blacktriangle'}$ & $\eta^x_\blacktriangle\eta^x_{\blacktriangle'} + \eta^y_\blacktriangle\eta^y_{\blacktriangle'}$ \\
        $\eta^x_\blacktriangle\eta^y_{\blacktriangle'} - \eta^y_\blacktriangle\eta^x_{\blacktriangle'}$ & $\eta^x_\blacktriangle\eta^y_{\blacktriangle'} - \eta^y_\blacktriangle\eta^x_{\blacktriangle'}$ & $-(\eta^x_\blacktriangle\eta^y_{\blacktriangle'} - \eta^y_\blacktriangle\eta^x_{\blacktriangle'})$\\
        $\eta^z_\blacktriangle\eta^z_{\blacktriangle'}$ & $\eta^z_\blacktriangle\eta^z_{\blacktriangle'}$ & $\eta^z_\blacktriangle\eta^z_{\blacktriangle'}$\\
        \hline 
        $\eta^x_\blacktriangle\eta^x_{\blacktriangle'} - \eta^y_\blacktriangle\eta^y_{\blacktriangle'}$ & $\left[-(\eta^x_\blacktriangle\eta^x_{\blacktriangle'} - \eta^y_\blacktriangle\eta^y_{\blacktriangle'}) + \sqrt{3}(\eta^x_\blacktriangle\eta^y_{\blacktriangle'} - \eta^y_\blacktriangle\eta^x_{\blacktriangle'}) \right]/2$ & $\eta^x_\blacktriangle\eta^x_{\blacktriangle'} - \eta^y_\blacktriangle\eta^y_{\blacktriangle'}$ \\
        $\eta^x_\blacktriangle\eta^y_{\blacktriangle'} + \eta^y_\blacktriangle\eta^x_{\blacktriangle'}$ & $\left[-(\eta^x_\blacktriangle\eta^y_{\blacktriangle'} + \eta^y_\blacktriangle\eta^x_{\blacktriangle'}) - \sqrt{3}(\eta^x_\blacktriangle\eta^x_{\blacktriangle'} + \eta^y_\blacktriangle\eta^y_{\blacktriangle'})\right]/2$ & $-(\eta^x_\blacktriangle\eta^y_{\blacktriangle'} + \eta^y_\blacktriangle\eta^x_{\blacktriangle'})$\\
         \hline
      \end{tabular}
    \end{table}

    For two-site interactions the system will then take the form of:
    \begin{align}
      H_\eff^{i,j} &= A_{zz} s^z_is^z_j + A_{xx} (s^x_is^x_j +s^y_is^y_j)+ \sum_{\alpha,\beta \in \{x,y\}}B_{\alpha\beta}\eta^\alpha_i\eta^\beta_j + B_{zz} \eta^z_i\eta^z_j + \\  
        &+  a_{xx} (s^x_is^x_j +s^y_is^y_j)\times \left[b_x(\eta^x_i-\eta^x_j) + b_y (\eta^y_i+\eta^y_j) + \sum_{\alpha,\beta \in \{x,y\}}b_{\alpha\beta}\eta^\alpha_i\eta^\beta_j + b_{zz} \eta^z_i\eta^z_j \right]\\
      &+ a_{zz} s^z_is^z_j \times \left[b_x'(\eta^x_i-\eta^x_j) + b_y' (\eta^y_i+\eta^y_j) + \sum_{\alpha,\beta \in \{x,y\}}b_{\alpha\beta}'\eta^\alpha_i\eta^\beta_j + b_{zz}' \eta^z_i\eta^z_j \right]  
    \end{align}
    
    Finally, let us note that a similar analysis can be extended to 3-triangle interactions (which emerge at higher order in perturbation theory, which we will get to in the next section). 
    The analysis of these terms is even more burdensome so we do not carry it out in this note.
    However, it is important to note that one term that is naturally permitted by the symmetries of the system is $\eta^z_as^z_bs^x_cs^y_d$. This term directly couples the chirality degrees of freedom $\eta^z$, to the chirality of the ``leftover" spin degrees of freedom $\vec{s}_b\cdot(\vec{s}_c\times \vec{s}_d)$.
    
    \item {\bf Translation Symmetry}: Trivially acts to ensure that the interactions in the effective Hamiltonian remain the same if both triangles are translated by lattice spacing.
\end{itemize}

\subsection{Computing $H_{\text{eff}}$}
The symmetry discussion allows us to determine which terms are zero, but does not provide insight into the strength and sign of the nonzero interaction terms. 
In this subsection we describe how to compute it, first at a zeroth order approximation and then through a systematic perturbative expansion that highlights some terms that we failed to capture through the symmetry analysis. 

\subsubsection{Projecting into the low-energy subspace}

To zeroth order, the dynamics of these effective degrees of freedom can be obtained by projecting the full Hamiltonian of the system into our low-energy subspace.
To this end, it is insightful to write the different spin operator $S^\alpha_i$ projected into the $s^\alpha$ and $\eta^\alpha$ basis:
\begin{align*}
  S^\alpha_{\blacktriangle,i} = (-1)^{\delta_{\alpha x}} \frac{1}{3} s^\alpha_\blacktriangle \otimes \left[1_\blacktriangle - \eta^x_\blacktriangle 2\sin \frac{(i+1) 2\pi}{3} + \eta^y_\blacktriangle 2\cos \frac{(i+1) 2\pi}{3}\right]
\end{align*}
where we label the $i$-th spin according the Fig.~1 as follows ($a=0$, $b=1$, $c=2$).

From the above expression for the projected spin operators, it is easy to see that $H_\eff$ between two triangles the following simple form:
\begin{align}\label{eq:HeffProjection}
  H_\eff = \frac{s^x_\blacktriangle s^x_{\blacktriangle'}+s^y_\blacktriangle s^y_{\blacktriangle'}}{9} \otimes \sum_{i,j} &\left\{\frac{1}{r_{\blacktriangle i,\blacktriangle j}^3}\left[1_\blacktriangle - \eta^x_\blacktriangle 2\sin \frac{(i+1) 2\pi}{3} + \eta^y_\blacktriangle 2\cos \frac{(i+1) 2\pi}{3}\right] \times \right. \notag \\
  &\times\left.\left[1_{\blacktriangle'} - \eta^x_{\blacktriangle'} 2\sin \frac{(j+1) 2\pi}{3} + \eta^y_{\blacktriangle'} 2\cos \frac{(j+1) 2\pi}{3}\right] \right\}
\end{align}
where $r_{\blacktriangle i,\blacktriangle j}$ is the distance between spin $i$ in triangle $\blacktriangle$ and $j$ in triangle $\blacktriangle'$.

A few remarks are in order.
First, in the case of two very distant triangles $r_{\blacktriangle i, \blacktriangle' j } \approx r_{\blacktriangle, \blacktriangle'}$ and the only term that does not cancel is the anti-ferromagnetic $s^x_\blacktriangle s^x_{\blacktriangle'}+s^y_\blacktriangle s^y_{\blacktriangle'}$.
This highlights how the chirality terms are subleading contributions in the large breathing limit of the lattice.
Second, note that the Eq.~\ref{eq:HeffProjection} does not include all symmetry allowed interaction terms. This is a consequence of the projection we perform, that does not take into account virtual processes which go into the excited triangle states, but bring the system back to the low-energy manifold.
Indeed, we find that the lack of these other terms preclude the observation of the CSL phase.
To understand this discrepancy, we now turn to a Schrieffer-Wolff perturbative approach to better understand the type and magnitude of the missing terms.

\subsubsection{Building a perturbative expansion using the Schrieffer-Wolff transformation}
\label{sec:Wolff}

The simplest way of perturbatively computing the effective Hamiltonian is via a Schrieffer-Wolff transformation. The main idea is to apply a unitary transformation to the system, informed by the non-diagonal components, to better diagonalize the system.
This approach is particularly useful, when the off-diagonal components are perturbatively small, as it provides a way of organizing this rotation, order-by-order, which facilitates its computation and interpretation.

The starting point is the original Hamiltonian, which can be divided into diagonal $H_0$ and off-diagonal terms $V$ (here this does not have to refer to a single state, but to a subspace of the Hilbert space):
\begin{equation}
  H = H_0 + V
\end{equation}

We then want to find a rotation of the Hamiltonian such that the resulting Hamiltonian is diagonal:
\begin{align*}
  H' = e^S H e^{-S} = H + [S,H] + \frac{1}{2}[S,[S,H]] + \ldots = H_0 + \left\{ V + [S,H_0] \right\} + [S, V] + \ldots
\end{align*}
Taking the rotation to be perturbatively small, like the interaction, $|V|\sim|S|\sim \lambda$, we can choose the rotation $S$ to satisfy:
\begin{equation}
  V + [S,H_0] = 0 \Rightarrow \boxed{[S, H_0]= -V} \Rightarrow \bra{n}S H_0 - H_0 S\ket{m} = S_{nm} \left[ E_m-E_n\right] = -V_{mn} 
\end{equation}
where $\ket{n}$ is the n-th eigenstate of $H_0$ with energy $E_n$.
This ensures that the lowest order term of the expansion is zero, but it also simplifies many of the higher order terms, namely:
\begin{equation}
  [S,[S,H]] = [S,[S,H_0]] + [S,[S,V]] = -[S,V] + [S,[S,V]]
\end{equation}
Armed with this simplification we can obtain $H'$ to any higher order.

\begin{table}
  \begin{tabular}{c | c | c}
    Order & Expansion Term & Simplification\\\hline\hline
    0 & $H_0$ & $H_0$ \\
    1 & $V + [S,H_0]$ &  $0$\\
    2 & $[S,V] + \frac{1}{2}[S,[S,H_0]]$ & $\frac{1}{2}[S,V]$\\
    3 & $\frac{1}{2}[S,[S,V]] + \frac{1}{3!}[S,[S,[S,H_0]]]$ & $ \left[\frac{1}{2}-\frac{1}{6}\right] [S,[S,V]]$\\
    \ldots & \ldots & \ldots \\
    n & $\frac{1}{(n-1)!}\text{ad}_S^{n-1} V + \frac{1}{n!}\text{ad}_S^{n} H_0$ & $\frac{1}{(n-2)! n} \text{ad}_S^{n-1} V$
  \end{tabular}
\end{table}

\begin{figure}[t]
            \centering
            \includegraphics[width=0.8\linewidth]{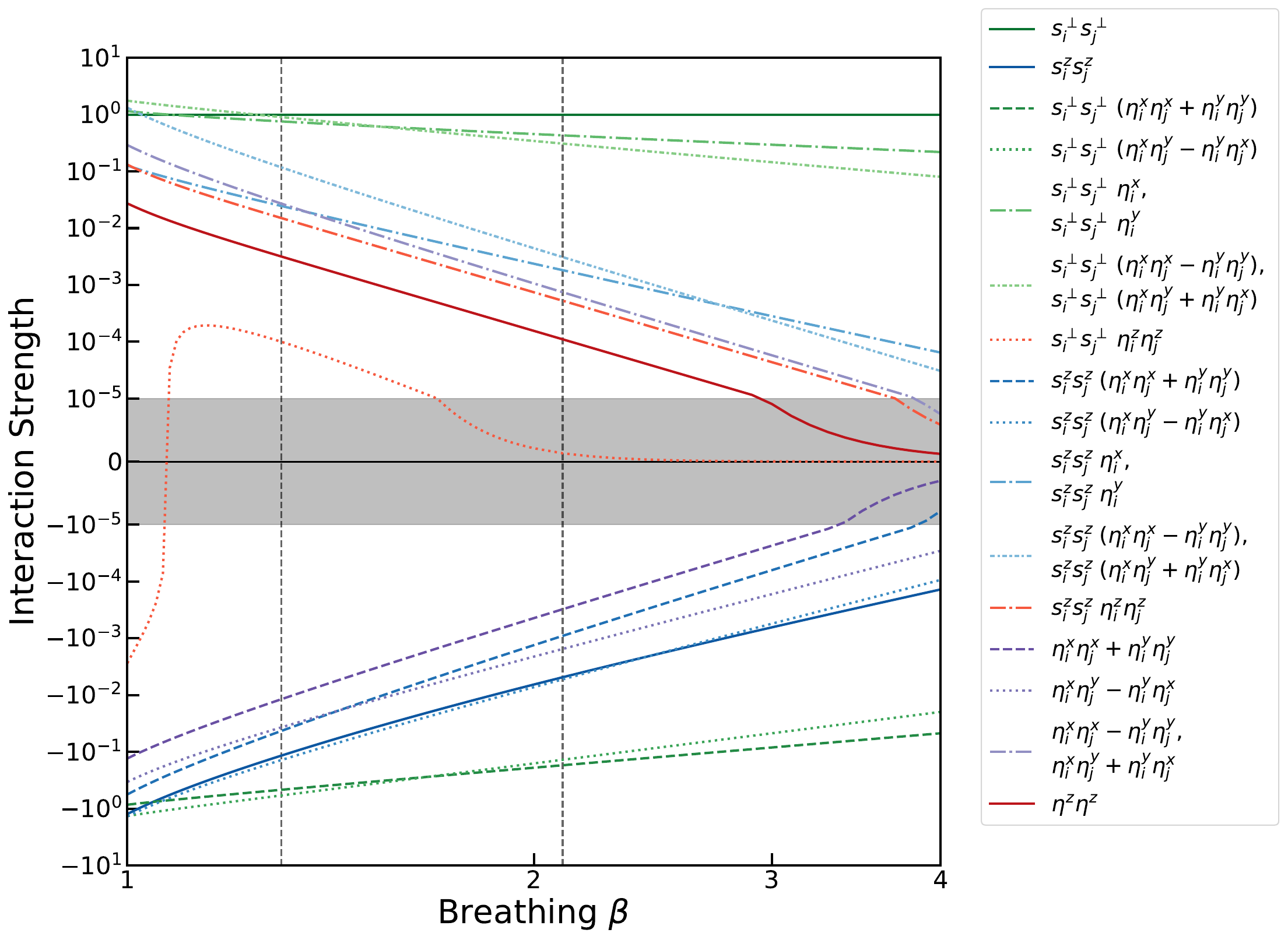}

                \caption{Parameters (normalized by the XY spin spin interaction) without and with second order perturbation theory, left and right respectively.
    The two dashed lines corresponds to the transitions out of the CSL state. In the grey shaded area, the y-range is linear. Breathing is plotted in logarithmic space to better emphasize the power-law behavior of the parameters.}
        \label{fig:PertTheoryParameters}

\end{figure}

Putting this approach together we can easily compute the effective two-spin Hamiltonian to second order perturbation theory, uncovering all the symmetry allowed terms, albeit with different breathing dependence, Figure~\ref{fig:PertTheoryParameters}.

Note that, since $V$ only connects the low-energy manifold to the excited states (and vice-versa), so does $S$. As a result any term composed of an odd number of $S$ and $V$ will not connect the low-energy manifold back to itself so we can immediately ignore the odd order terms of the above expansion.

Finally, let us note that, while we have been discussing the above construction in the context of understanding the interaction between two triangles, note that it is entirely independent of the size or form of the low/high energy states. As such it can be easily extended to understanding the nature of perturbative three-triangle terms.

Crucially, this suggests the mechanism that drives the system to exhibit the CSL ground state. Namely, from this analysis on the interactions between three triangles, interaction terms of the form $\eta^z_a (s^z_a s^x_b s^y_c- s^z_a s^y_b s^x_c)$ naturally emerge, corresponding to the coupling of the chirality degrees of freedom to the chirality of the remaining degrees of freedom.
In this language, as explained in the main text, the ferromagnetic ordering of the $\eta^z$ provides a background magnetic field for the $s^\alpha$ degrees of freedom which stabilizes the chiral spin liquid order.

\section{Finite Cluster}

\subsection{Symmetry properties of finite cluster ground states}

\begin{figure}[t]
            \centering
            \includegraphics[width=0.6\linewidth]{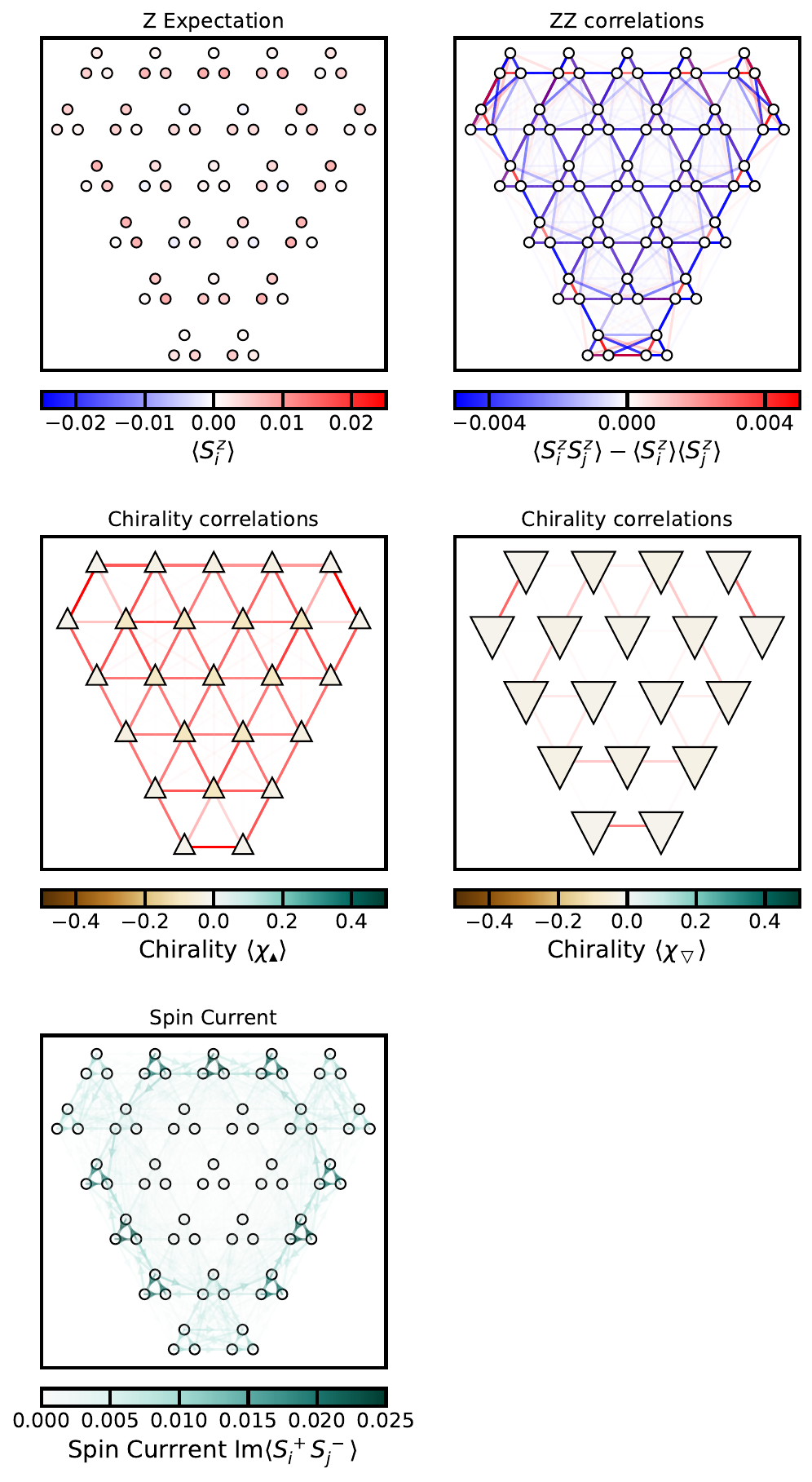}
                \caption{Different correlation functions of the 75 site cluster presented in Fig.~1.
                \bm{(a)} The cluster exhibits a higher density of onsite magnetization near the edge of the cluster, highlighting the presence of a spinful excitation in the edge.
                \bm{(b)} Spin-spin correlations demonstrate the liquidity of the cluster.
                \bm{(c)} Chirality and connected chirality-chirality correlations (lines) display a ferromagnetic behavior.
                \bm{(d)} Similar to (c), the chirality of the down facing triangles also display ferromagnetic-like correlations.
                \bm{(e)} Spin current---arrow direction points to direction--- highlight a pattern of a chiral current near the boundary of the cluster.
                }
        \label{fig:ClusterAllCorrelators}

\end{figure}

In odd-sized clusters, the results mirror the expectation from a first-order like transition.
Instead of observing the decay of $\chi_\text{bulk}$ to zero with decreasing $J_\chi$, $\chi_\text{bulk}$ approaches a non-zero constant at zero $J_\chi$.
This distinction is best understood in terms of the symmetry properties of the cluster and its eigenstates.
Because we focus on clusters with the same $C_{3v}$ symmetry of the full model, the ground state can either live on the non-degenerate (and thus symmetric) A$_1$ or A$_2$ irreducible representations, or the two-fold degenerate E irreducible representation.
While the even clusters' ground state is part of an A representation, we find that for odd sized clusters, the ground state manifold is always two-fold degenerate.
One way to intuit this feature is by connecting it to the physics of the CSL.
Because the E irreducible representation captures states with angular momentum of $\pm 1$, the presence of the additional unit of half-spin populates the chiral edge mode, imbuing the wavefunction with a well-defined non-zero angular momentum.
Under time-reversal, we must obtain the oppositely edge-populated state with opposite angular momentum and opposite chiralities.
Because of the non-zero angular momentum in both states, one cannot create symmetric state under rotations using these two wavefunctions: the resulting ground state manifold must be two-fold degenerate.
Besides the non-zero angular momentum, two other features corroborate this picture.
First, the presence of a large magnetization inhomogeneity, whereby the spin correlations are stronger at the edge~[Fig.~1(c) of the main text and Fig.~\ref{fig:ClusterAllCorrelators}].
Second, the presence of a chiral spin current, also confined to the edge of the cluster.
Computing the angular momentum of the wavefunction via the overlap of the cluster with its rotated version by $2\pi/3$, gives us a value of $e^{i2\pi/3}$.

\bibliography{refs.bib}